\documentclass[twocolumn, iop, numeberedappendix]{aastex6}

\usepackage{graphicx,amsmath,natbib, hyperref}
\hypersetup{colorlinks=true, urlcolor=black}
\usepackage{here}
\usepackage{epsf}
\usepackage{epstopdf}
\usepackage{xspace}

\usepackage[normalem]{ulem} 
\usepackage{cancel} 

\usepackage{color}

\shorttitle{ASPECS: [CII] line search in the UDF}
\shortauthors{Aravena et al.}

\def\Cii{[C\,{\sc ii}]}

\def\lsim{\mathrel{\rlap{\lower 3pt \hbox{$\sim$}} \raise 2.0pt \hbox{$<$}}}
\def\gsim{\mathrel{\rlap{\lower 3pt \hbox{$\sim$}} \raise 2.0pt \hbox{$>$}}}

\begin{document}

\title{The ALMA Spectroscopic Survey in the Hubble Ultra Deep Field: Search for \Cii{} line and dust emission in $6<z<8$ galaxies}

\author{
M. Aravena\altaffilmark{1},
R. Decarli\altaffilmark{2}, 
F. Walter\altaffilmark{2,3,4},
R. Bouwens\altaffilmark{5,6},
P.~A.~Oesch\altaffilmark{7},
C.~L.~Carilli\altaffilmark{5,8},
F.~E.~Bauer\altaffilmark{9,10,11}
E.~Da~Cunha\altaffilmark{12,13},
E. Daddi\altaffilmark{14},
J. G\'onzalez-L\'opez\altaffilmark{9},
R.\,J.~Ivison\altaffilmark{15,16},
D.~A.~Riechers\altaffilmark{17},
I. Smail\altaffilmark{18},
A.~M.~Swinbank\altaffilmark{18},
A. Weiss \altaffilmark{19},
T. Anguita\altaffilmark{20, 10},
R. Bacon\altaffilmark{21},
E. Bell\altaffilmark{22},
F. Bertoldi\altaffilmark{23},
P. Cortes\altaffilmark{24,4},
P. Cox\altaffilmark{24},
J. Hodge\altaffilmark{5},
E. Ibar\altaffilmark{25},
H. Inami\altaffilmark{21},
L. Infante\altaffilmark{9},
A. Karim\altaffilmark{23},
B. Magnelli\altaffilmark{23},
K. Ota\altaffilmark{26},
G. Popping\altaffilmark{15},
P.~van~der~Werf\altaffilmark{5},
J. Wagg\altaffilmark{27},
Y. Fudamoto\altaffilmark{15,28}
}
\altaffiltext{1}{N\'{u}cleo de Astronom\'{\i}a, Facultad de Ingenier\'{\i}a, Universidad Diego Portales, Av. Ej\'{e}rcito 441, Santiago, Chile. E-mail: {\sf manuel.aravenaa@mail.udp.cl}}
\altaffiltext{2}{Max-Planck Institut f\"{u}r Astronomie, K\"{o}nigstuhl 17, D-69117, Heidelberg, Germany.}
\altaffiltext{3}{Astronomy Department, California Institute of Technology, MC105-24, Pasadena, California 91125, USA}
\altaffiltext{4}{NRAO, Pete V.\,Domenici Array Science Center, P.O.\, Box O, Socorro, NM, 87801, USA}
\altaffiltext{5}{Leiden Observatory, Leiden University, PO Box 9513, NL2300 RA Leiden, The Netherland}
\altaffiltext{6}{UCO/Lick Observatory, University of California, Santa Cruz, CA 95064, USA}
\altaffiltext{7}{Astronomy Department, Yale University, New Haven, CT 06511, USA}
\altaffiltext{8}{Astrophysics Group, Cavendish Laboratory, J. J. Thomson Avenue, Cambridge CB3 0HE, UK}
\altaffiltext{9}{Instituto de Astrof\'{\i}sica, Facultad de F\'{\i}sica, Pontificia Universidad Cat\'olica de Chile Av. Vicu\~na Mackenna 4860, 782-0436 Macul, Santiago, Chile}
\altaffiltext{10}{Millennium Institute of Astrophysics, Chile}
\altaffiltext{11}{Space Science Institute, 4750 Walnut Street, Suite 205, Boulder, CO 80301, USA}
\altaffiltext{12}{Research School of Astronomy and Astrophysics, Australian National University, Canberra, ACT 2611, Australia}
\altaffiltext{13}{Centre for Astrophysics and Supercomputing, Swinburne University of Technology, Hawthorn, Victoria 3122, Australia}
\altaffiltext{14}{Laboratoire AIM, CEA/DSM-CNRS-Universite Paris Diderot, Irfu/Service d'Astrophysique, CEA Saclay, Orme des Merisiers, 91191 Gif-sur-Yvette cedex, France}
\altaffiltext{15}{European Southern Observatory, Karl-Schwarzschild Strasse 2, D-85748 Garching bei M\"unchen, Germany}
\altaffiltext{16}{Institute for Astronomy, University of Edinburgh, Blackford Hill, Edinburgh EH9 3HJ, UK}
\altaffiltext{17}{Cornell University, 220 Space Sciences Building, Ithaca, NY 14853, USA}
\altaffiltext{18}{Institute for Computational Cosmology, Durham University, South Road, Durham DH1 3LE, UK}
\altaffiltext{19}{Max-Planck-Institut f\"ur Radioastronomie, Auf dem H\"ugel 69, 53121 Bonn, Germany}
\altaffiltext{20}{Departamento de Ciencias F\'{\i}sicas, Universidad Andres Bello, Fernandez Concha 700, Las Condes, Santiago, Chile}
\altaffiltext{21}{Universit\'{e} Lyon 1, 9 Avenue Charles Andr\'{e}, 69561 Saint Genis Laval, France}
\altaffiltext{22}{Department of Astronomy, University of Michigan, 500 Church St, Ann Arbor, MI 48109, USA}
\altaffiltext{23}{Argelander Institute for Astronomy, University of Bonn, Auf dem H\"{u}gel 71, 53121 Bonn, Germany}
\altaffiltext{24}{Joint ALMA Observatory - ESO, Av. Alonso de C\'ordova, 3104, Santiago, Chile}
\altaffiltext{25}{Instituto de F\'{\i}sica y Astronom\'{\i}a, Universidad de Valparaiso, Avda. Gran Breta\~na 1111, Valparaiso, Chile}
\altaffiltext{26}{Kavli Institute for Cosmology, University of Cambridge, Madingley Road, Cambridge CB3 0HA, UK ; Cavendish Laboratory, University of Cambridge, 19 J.J. Thomson Avenue, Cambridge CB3 0HE, UK }
\altaffiltext{27}{SKA Organization, Lower Withington Macclesfield, Cheshire SK11 9DL, UK}
\altaffiltext{28}{Universit\"at-Sternwarte M\"unchen, Scheinerstr. 1, D-81679 M\"unchen, Germany}

\begin{abstract}

We present a search for [CII] line and dust continuum emission from optical dropout galaxies at $z>6$ using ASPECS, our ALMA Spectroscopic Survey in the Hubble Ultra--Deep Field (UDF). Our observations, which cover the frequency range $212-272$ GHz, encompass approximately the range $6<z<8$ for [CII] line emission and reach a limiting luminosity of L$_{\rm [CII]}\sim$(1.6-2.5)$\times$10$^{8}$\,L$_{\odot}$. We identify fourteen [CII] line emitting candidates in this redshift range with significances $>$4.5\,$\sigma$, two of which correspond to blind detections with no optical counterparts. At this significance level, our statistical analysis shows that about 60\% of our candidates are expected to be spurious. For one of our blindly selected [CII] line candidates, we tentatively detect the CO(6-5) line in our parallel 3-mm line scan. None of the line candidates are individually detected in the 1.2\,mm continuum. A stack of all [CII] candidates results in a tentative detection with $S_{1.2mm}=14\pm5\mu$Jy. This implies a dust--obscured star formation rate (SFR) of $(3\pm1)$\,M$_\odot$\,yr$^{-1}$. We find that the two highest--SFR objects have candidate [CII] lines with luminosities that are consistent with the low--redshift $L_{\rm [CII]}$ vs.\ SFR relation. The other candidates have significantly higher [CII] luminosities than expected from their UV--based SFR. At the current sensitivity it is unclear whether the majority of these sources are intrinsically bright [CII] emitters, or spurious sources. If only one of our line candidates was real (a scenario greatly favored by our statistical analysis), we find a source density for [CII] emitters at $6<z<8$ that is significantly higher than predicted by current models and some extrapolations from galaxies in the local universe.

\end{abstract}
\keywords{galaxies: evolution --- galaxies: ISM --- galaxies: star formation ---  galaxies: high-redshift --- submillimeter: galaxies --- instrumentation: interferometers}

\section{Introduction}

A key to understanding galaxy formation is to investigate the physical mechanisms that lead to the formation of the first galaxies and thereby to understand their role in the reionization of the Universe \citep{robertson10}. One of the main challenges in studying galaxies within the first Gigayear of the Universe (i.e. $z>6$) is that observations in the optical and near-infrared (NIR) regimes can only probe the high-resonance Ly-$\alpha$ line (rest wavelength: 1216\AA) and the faint underlying UV continuum. Both measurements are strongly affected by dust obscuration, and the Ly-$\alpha$ line is known to be hard to interpret due to the difficulties of radiative transfer modelling. Despite significant observational efforts, detection of Ly-$\alpha$ emission in non--quasar environments at $z>6$ has been very scarce. One interpretation of this finding is that the increasingly neutral intergalactic medium (IGM) that is surrounding galaxies during the epoch of reionization absorbs the Ly-$\alpha$ emission line through its damping wings. This in turn severely limits its usefulness as a spectroscopic redshift indicator \citep[e.g.,][]{schenker12,pentericci14}

Since the strength of the Ly-$\alpha$ line has been found to decline very rapidly beyond $z > 6$, optical spectroscopic confirmation has proven to be challenging with current facilities \citep[e.g.][]{treu13}. Moreover, at $z>6.5$, the Ly$\alpha$ line shifts into a range of the electromagnetic spectrum highly contaminated by sky lines, making the identification of $z\sim7$ sources even more challenging. Beyond that, the line enters the near--IR bands, that are limited in sensitivity through ground--based observations; this situation will not change until the launch of the {\it James Webb} Space Telescope ({\it JWST}) . As a consequence, only a handful of sources with spectroscopically confirmed redshift at $z > 6.5$ are known to date \citep[][]{oesch15}.

The far-infrared (FIR) fine-structure [CII]158$\mu$m emission line has been proposed as an alternative to Ly-$\alpha$ to study the first galaxies at $z>6$ \citep[e.g.][]{walter12}. The [CII] line is the dominant coolant of the interstellar medium (ISM) in star forming galaxies, making up 0.1-1\% of the integrated FIR luminosity of galaxies \citep[e.g. early work by ][]{stacey91}. This line appears to be ubiquitous in star forming galaxies, and at $z>6$ is redshifted into the accessible 1-mm band observable from the ground. Several studies have recently shown the power of this line to study objects in this redshift range \citep{carilli13}, but the majority of sources detected at z$>$6 are quasar host galaxies \citep[e.g.][]{maiolino05,walter09,venemans12,wang13,venemans16}. The current highest-redshift detections of [CII] emission in non--AGN--dominated galaxies, using ALMA, are at $z = 5.7-6.3$ \citep{riechers13,capak15,willott15,gullberg15,maiolino16,knudsen16}. The detection of such systems was not possible in the pre--ALMA era given the collecting area of the previous generation of millimeter interferometers.

The brightness of the [CII] line in principle makes it a unique tool to investigate the properties of galaxies well into the reionization epoch. Since this line is bright in typical star forming galaxies, it can be readily used as a direct way to identify and spectroscopically secure galaxies in blind millimeter spectroscopic surveys of the sky, in particular in galaxies that cannot be followed up spectroscopically in the optical/NIR (at least not before the launch if {\it JWST}). Compared to the more conventional tracer of the ISM, CO emission, the [CII] line is intrinsically much brighter. Given the rather high excitation temperature of $\sim$92\,K it also possible that the [CII] line is much less susceptible to the effects of the cosmic microwave background (CMB) than the CO emission \citep{dacunha13}, even though recent studies have suggested that in the cold neutral ISM, the spin temperature of the [CII] transition is almost coupled to the CMB temperature, and thus the [CII] luminosity could be as low as 20\% of the one that one could have inferred without taking into account of the CMB \citep{vallini15}.

Using the Atacama Large Millimeter/submillimeter Array (ALMA), we have  obtained the first full 1\,mm spectroscopic survey in a region of the cosmological deep field for which the deepest multi--wavelength data exist: the {\it Hubble} Ultra Deep Field \citep[UDF;][]{beckwith06}. This survey enables, for the first time, a blind search for [CII] emission in a redshift range $6<z<8$. The UDF is an ideal field to perform such a study, as this field contains a high density of dropout galaxies in this redshift range \citep[e.g.,][]{bouwens14,bouwens15,schenker12,mclure11, mclure13}. In this paper, we present the result of our search for [CII] line emission. This pathfinder study demonstrates that ALMA's tuning range and sensitivity is well--matched to detecting the normal galaxy population at $z>6$. The layout of this paper is as follows. In \S \ref{sect:obs}, we briefly describe our ALMA observations of the UDF and the multi--wavelength data of this field used in this study. In \S \ref{sect:results}, we introduce our methodology and algorithm to search for [CII] line emitters, present our list of candidate [CII] sources, and the level of contamination and completeness of our catalog. In this Section, we analyse the possibility that our line detections correspond to other redshift solutions based on the existence of other molecular line detections in our spectral scans. Here, we also present a blind [CII] line candidate based on possible detection of [CII] and CO(6-5) line emission. In \S \ref{sect:discussion}, we investigate the location of our [CII] line candidates in the SFR-[CII] luminosity plane and place first constraints on the [CII] luminosity function at $6<z<8$. Finally, in \S \ref{sect:conclusion}, we list the main conclusions of this study. We adopt a standard $\Lambda$CDM cosmology with $H_0=70$ km s$^{-1}$ Mpc$^{-1}$, $\Omega_{\Lambda}=0.7$ and $\Omega_{\rm M}=0.3$.

\section{Observations}
\label{sect:obs}
\subsection{ALMA UDF survey overview}

We used ALMA to conduct a spectroscopic survey of a region of the {\it Hubble} UDF, scanning essentially all of the 1-mm and 3-mm windows, corresponding to the ALMA bands 6 and 3, respectively.  Observations were performed in a single pointing in band 3, and using a 7-point mosaic in band 6, to map the same region in the sky, covering roughly a region of $\sim$1\,arcmin$^2$ of the HUDF. For details about our survey and blind line search, see \citet{walter16} -- Paper~I in this series. 

\begin{figure}[t]
\centering
\includegraphics[scale=0.45]{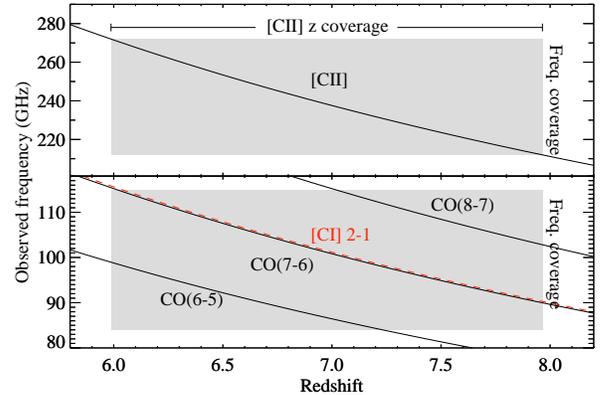}
\caption{Line redshift coverage for the ALMA millimeter line scans in band~3 and band~6, zooming in on the redshift range $6<z<8$ (for the full redshift coverage, see Fig.~1 in Paper~I). Only the redshifted CO, [CI] and [CII] lines are shown, as these lines are expected to be the brightest in this redshift range. The shaded area show the frequency coverage in each band, and the resulting redshift coverage for the [CII] , CO (and the fainter [CI]) lines.\label{fig:freqcov} }
\end{figure}

Our spectral line survey covers the frequency ranges $84.0-115.0$ GHz and $212.0-272.0$ GHz of the ALMA bands 3 and 6, respectively, thus encompassing the redshift range $0<z<6$ for the CO line emission (see Paper~I). Most importantly for this paper, the 1-mm frequency scan covers the redshift range $5.99<z<7.96$ for [CII] line emission. This redshift range is also covered by the 3-mm scan for the CO $J=6$ to 8 emission lines (see Fig. \ref{fig:freqcov}). 

\subsection{ALMA data and reduction}

\vspace{2mm}

The survey setup and data reduction steps are described in detail in \citet{walter16}. Here we repeat the most relevant information for the study presented here.

ALMA band 3 and band 6 observations were obtained during Cycle-2 as part of projects 2013.1.00146.S (PI: F. Walter) and 2013.1.00718.S (PI: M. Aravena). Observations in band-3 were conducted between July 01, 2014 to January 05, 2015, and observations were conducted between December 12, 2014 to April 21, 2015 under good weather conditions following observatory project execution restrictions. 

\begin{figure}[t]
\includegraphics[scale=0.3]{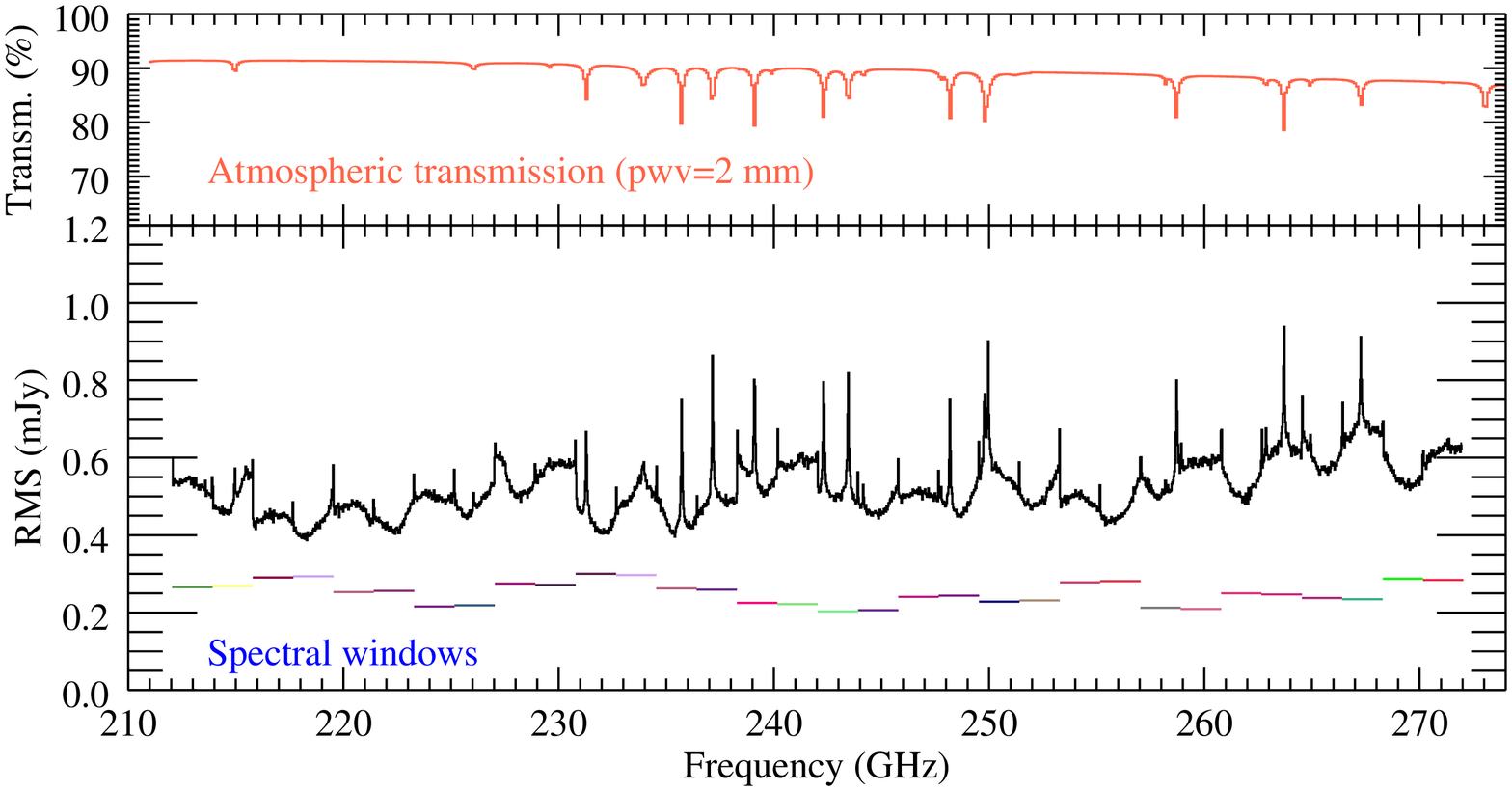}
\caption{{\em Lower panel:} RMS noise level as a function of frequency for the inner $30''$ of the ALMA 1-mm mosaic (not PB corrected). The RMS spectra is represented by the black curve and is shown in individual channels of 31.25 MHz ($\sim37.5$ km s$^{-1}$). The horizontal coloured lines show the location of the different tunings (SPWs) across the frequency axis. Colors indicate individual tunings: one tuning consists of a lower sideband and an upper sideband, separated by 12\,GHz. {\em Upper panel:} The red curve shows the atmospheric transmission for a precipitable water vapor (PWV) of 2 mm.  The majority of the RMS--peaks in the lower panel can be attributed to decreased atmospheric transmission at the respective frequencies.}
\label{fig_rms}
\end{figure}

Observations in band 3 were performed in a single pointing in spectral scan mode, using 5 frequency tunings to cover the frequency range $84.2-114.9$ GHz. With this setup, there were a few ranges where there was overlap between independent spectral windows (SPWs). Over this frequency range the ALMA half power beam width (HPBW), or primary beam (PB), ranges between $61''$ and $45''$. Observations in band 6 were performed in a 7-point mosaic, using a hexagonal pattern: the central pointing overlaps the other 6 pointings by about half the ALMA PB, i.e. close to Nyquist sampling. We scanned the band 6 using eight frequency tunings, covering the frequency range $212.0-272.0$ GHz. With this configuration there is no frequency overlap between independent SPWs, and there are no gaps in frequency. The frequency coverage of the individual frequency setups is shown in Fig.~\ref{fig_rms}. Over this frequency range the ALMA PB ranges between $30''$ and $23''$.

\vspace{2mm}

Observations in bands 3 and 6 were taken with ALMA's  compact array configurations, C34-2 and C34-1, respectively. The observations used between 30 and 35 antennas in each bands, resulting in synthesized beam sizes of $3.5''\times2.0''$ and $1.5''\times1.0''$ from the low to high frequency ends of bands 3 and 6, respectively.

Flux calibration was performed on planets or Jupiter's moons, with passband and phase calibration determined from nearby quasars. Calibration and imaging was done using the Common Astronomy Software Application package (CASA). The calibrated visibilities were inverted using the CASA task \verb CLEAN \, using natural weighting to create data cubes at different channel resolutions. In particular, for the line search in the 3-mm and 1-mm bands, we created data cubes at 8.0 MHz channel$^{-1}$ and 31.25 MHz channel$^{-1}$, equivalent to 24 and 37 km s$^{-1}$ per channel, respectively. 

The final rms varies mildly as a function of frequency, but such variation is mostly dominated by atmospheric lines across the 1-mm window and by the loss of sensitivity at the SPW edges. We find an average rms of 0.53 mJy beam$^{-1}$ in the 1-mm cube at 31.25 MHz channel$^{-1}$ resolution (no PB corrected). This is shown in Fig.~\ref{fig_rms}: the increased noise in some parts of the spectrum can be explained by the expected variation of the sky transmission.

\begin{figure}[t]
\includegraphics[scale=0.42]{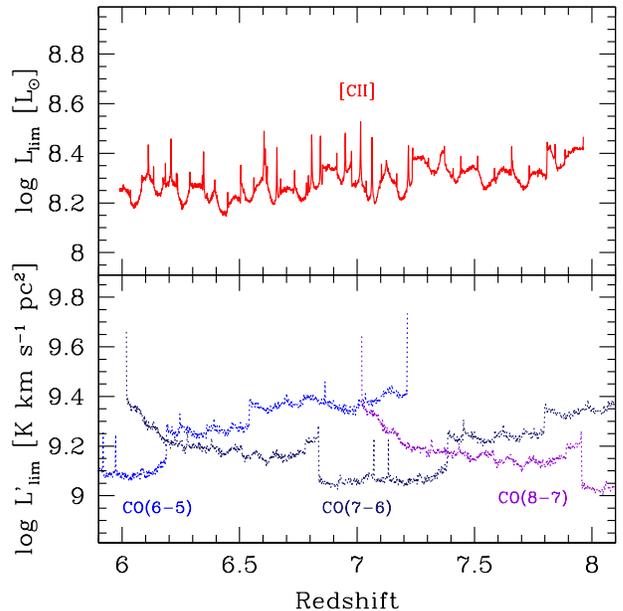}
\caption{Luminosity limit and redshift coverage over $6<z<8$ reached in our 1mm and 3mm scans, for relevant CO transitions and the [CII] line. The (3$\sigma$) limits plotted here are computed assuming point-source emission, and are based on the observed noise per channel, scaled for a line width of 300 km s$^{-1}$. }
\label{fig:lumlimit}
\end{figure}

The noise behaviour across the observed frequencies translates into a detection limit on the [CII] luminosity in the 1-mm scan, as well as luminosity limits for the high-$J$ CO lines covered by our 3-mm scan in the redshift range $6<z<8$. Fig. \ref{fig:lumlimit} show the luminosity limits obtained in our spectral scans. Assuming a typical line width of 300 km s$^{-1}$, we find that the $3\sigma$ detection limit for the [CII] luminosity varies between $L_{\rm [CII]}=(1.6-2.5)\times10^8\ L_\sun$. Over the same redshift range, the 3$\sigma$ detection limit for the CO(6-5), CO(7-6) and CO(8-7) lines also varies with frequency ranging between $L'_{\rm CO}=(1.2-2.5)\times10^9$ (K km s${-1}$ pc$^2$).

\vspace{2mm}

\subsection{Multi-wavelength ancillary data}\label{sec_ancillary}

The {\it Hubble} UDF is the cosmological field with the deepest observations in all important wave-bands, with 18,000 catalogued galaxies \citep{coe06}.  In this study, we use a sample of 58 $z>5.5$ optical dropout galaxies that were located within the 1 arcmin$^2$ region covered by our 1-mm mosaic in the {\it Hubble} UDF. This sample, enabled by the great depth of the UDF in all optical and NIR bands, was compiled from several searches for such objects based on {\it Hubble} Space Telescope ({\it HST}) imaging over the deepest area, the 4.7 arcmin$^2$ eXtremely Deep Field \citep[XDF; ][]{illingworth13}. This includes all available {\it HST} Advanced Camera for Surveys (ACS) optical and Wide Field Camera 3 (WFC3) IR data from the HUDF09, HUDF12 and from the Cosmic Assembly Near-infrared Deep Extragalactic Legacy Survey (CANDELS) programs. For more details about these datasets, see \citet{bouwens14} and references therein. In particular, we use the most complete catalog of galaxies selected to have photometric redshifts in the range $5.5-8.5$ compiled by \citet{schenker13}, \citet{mclure13} and \citet{bouwens15}. None of these objects have optical or NIR spectroscopy or previously confirmed redshift that could allow us to directly look for the [CII] line emission in our data cube. The full list of drop--out galaxies investigated in this paper is given in Table~\ref{table:sources}. 

The {\it HST} observations and dropout catalogs are complemented with recent spectroscopic constraints and redshifts from deep Multi Unit Spectroscopic Explorer (MUSE) integrations taken with the Very Large Telescope (VLT) as part of the Guaranteed Time Observations (GTO) on the {\it Hubble} UDF (Bacon et al., in preparation). These correspond to optical Integral Field Unit (IFU) spectroscopic observations covering the full region targeted by our ASPECS program, comprising the wavelength range between 4650-9200 \AA. As such, this provides effective coverage of the Lyman-$\alpha$ line emission out to redshifts $z<6.6$ down to unprecedented levels. Since our targets are at $6<z<8$, they can be observed by MUSE in Ly-$\alpha$ emission in the wavelength range 8500-9200 \AA. Over the area covered by our ASPECS observations, the MUSE spectra reach an emission line limiting flux at $5\sigma$ of $\sim2\times10^{-19}$ erg sec$^{-1}$ cm$^{2}$ for a point source within a $1"$ aperture and at 8500-9200 \AA (measured where no sky emission lines exist).

\subsection{Photometric redshifts}

Photometric redshifts and redshift probability distributions ($P(z)$) for the sources in this study are computed based using the best-fit chi-square at a given redshift derived from the full set of {\it HST} optical and NIR photometry for a source using the \texttt{EAZY} photometric redshift code \citep{brammer08}. We use the \texttt{EAZY\_v1.0} template set supplemented by the spectral energy distribution (SED) templates from the Galaxy Evolutionary Synthesis Models \citep[\texttt{GALEV;}][]{kotulla09}. A flat redshift prior is assumed. As such, the full optical and NIR SEDs are taken into account in the computation of the most likely redshifts and distributions.

\section{Results and analysis}
\label{sect:results}

Since the depth of our current 1-mm observations is limited, in most cases we need to guide our search for [CII] line emitters by previously selected $z>6$ galaxy candidates. The reasoning is the following: for a relatively bright normal star forming galaxy with SFR$=10\ M_{\sun}$ yr$^{-1}$, we expect an IR luminosity of $\sim1\times10^{11}\ L_{\sun}$. Galaxies with this level of star formation in the local Universe are observed to have [CII] to FIR ratios of $\sim5\times10^{-3}$, and thus under these conditions, for a typical line width of 200 km s$^{-1}$ we expect the strength of the [CII] line to vary between $\sim1-2$ mJy from $z=6$ to 8. At this level, we would only expect to detect [CII] lines from these galaxies at S/N$<6$ at the depth of our observations (0.4 mJy per 75 km s$^{-1}$ channel). Therefore, blind selection of these targets is limited to the most significant candidates only.

Our strategy to search for [CII] line emission in galaxies at 6$<$z$<$8 thus follows two parallel approaches: (1) we blindly select all the positive line peaks with significances above 5.3$\sigma$ based on the fidelity assessment in the 1.2-mm cube (see \S \ref{sec_stats}). We filter this sample by rejecting all the line peaks that are associated within $1''$ with an optical counterpart with a photometric redshift $<5.0$. This leaves us with line peaks that either do not have an optical association or that have a counterpart with $z_{\rm phot}>5.5$. (2) We specifically look for positive line peaks at significances of $>4.5\sigma$ that lie within 1.0'' from the position of optical dropout galaxies with $z_{\rm phot}=5.5-8.5$. As stated above, our observations covered 58 such optical dropout galaxies that were located within the HPBW of our mosaic (see Table~\ref{table:sources}). While our spectroscopic coverage permits us to search for [CII] line emission in the redshift range $z=6-8$, we allow for a extended range in the photometric redshift of the optical counterparts to take into consideration the typical uncertainty of these estimates, $\delta z\sim0.5$. We choose a distance of $1.0''$ to be about half the size of our synthesized beam. 

We find a total of 14 [CII] line candidates, two of which are blindly detected and selected based on their high significance in the data cube. Another 12 candidates are selected based on their association to a nearby optical dropout. Details on these searches are provided below.

\subsection{Line search procedure}

 For the line search, we used a data cube that had been de-convolved using a natural weighting scheme, without primary beam correction using the CASA task \texttt{CLEAN}. This ensures a similar noise behaviour across the image, for all source positions. Primary beam correction is later on taken into account for all subsequent quantitative analysis. In order not to introduce any priors to the inverted cube related to particular sources in the field, we do not apply the {\it clean} algorithm and directly work on the {\it dirty} images.  We do not subtract the continuum, as this requires a specific frequency range to subtract on and this could potentially affect detection of lines in the selected frequency ranges. Instead, after the lines have been searched for, we remove sources coincident with strong continuum or lines sources manually.

We performed a line search using the Astronomical Image Processing System (AIPS) task \texttt{SERCH}. The \texttt{SERCH}  task uses a Gaussian kernel to convolve the data cube along the frequency axis with an expected input line width, and reports all channels and pixels having a signal-to-noise ratio over the specified limit. For our search, we used a set of different Gaussian kernels, from 180 to 500 km s$^{-1}$ in order to ensure that all possible line widths were considered. We searched for all line peaks with signal to noise (S/N) values above 3.5. Here, the S/N is defined as the maximum significance level achieved after convolving over the various Gaussian kernels. This means that the the signal and noise levels are defined after averaging each image plane over several channels. We average over velocity ranges from 180 to 500 km s$^{-1}$, which corresponds to the typical line widths observed in the ISM of star-forming galaxies \citep[e.g.,][]{goto15, aravena16}.

Once all line peaks were identified, we used the IDL routine \texttt{CLUMPFIND} \citep{williams11} to isolate individual line peaks and generate a list of line candidates. A full list of 4200 positive line peaks with S/N$=3.5-10$ was thus obtained in the 1-mm data cube using this procedure. We note that the blind line search procedure described here is independent of that described in \citet[][; Paper~I]{walter16}, although the searches use a similar line detection algorithm. Both searches were performed to provide independent checks on the reliability of our algorithms, and they result in a similar list of blindly selected line candidates. This can be quantified by the very similar levels of {\it fidelity} and {\it completeness} of the 1-mm line candidate samples obtained below compared to that of Walter et al. (2016; see Fig. 4 in Paper~I): in both independent searches we find {\it fidelity} levels of $\sim5-10\%$, $\sim50\%$ and $\sim90-100\%$ at S/N of 4.0, 5.0, and 6.0, respectively. We also find similar {\it completeness} levels in both searches, of $\sim50\%$, $\sim80\%$ and $\sim100\%$ at 1-mm line flux densities $\sim0.5, 1.0$ and 1.5 mJy, respectively. Finally, we note that there is little dependency of the completeness on the line widths, particularly for those between $200-500$ km/s (this is discussed in more detail in Walter et al. 2016). However, both algorithms appear to recover fewer candidates with line widths below $\sim$100 km/s.

\subsection{Purity and completeness}\label{sec_stats}

\vspace{2mm}
\begin{figure}[t]
\includegraphics[scale=0.5]{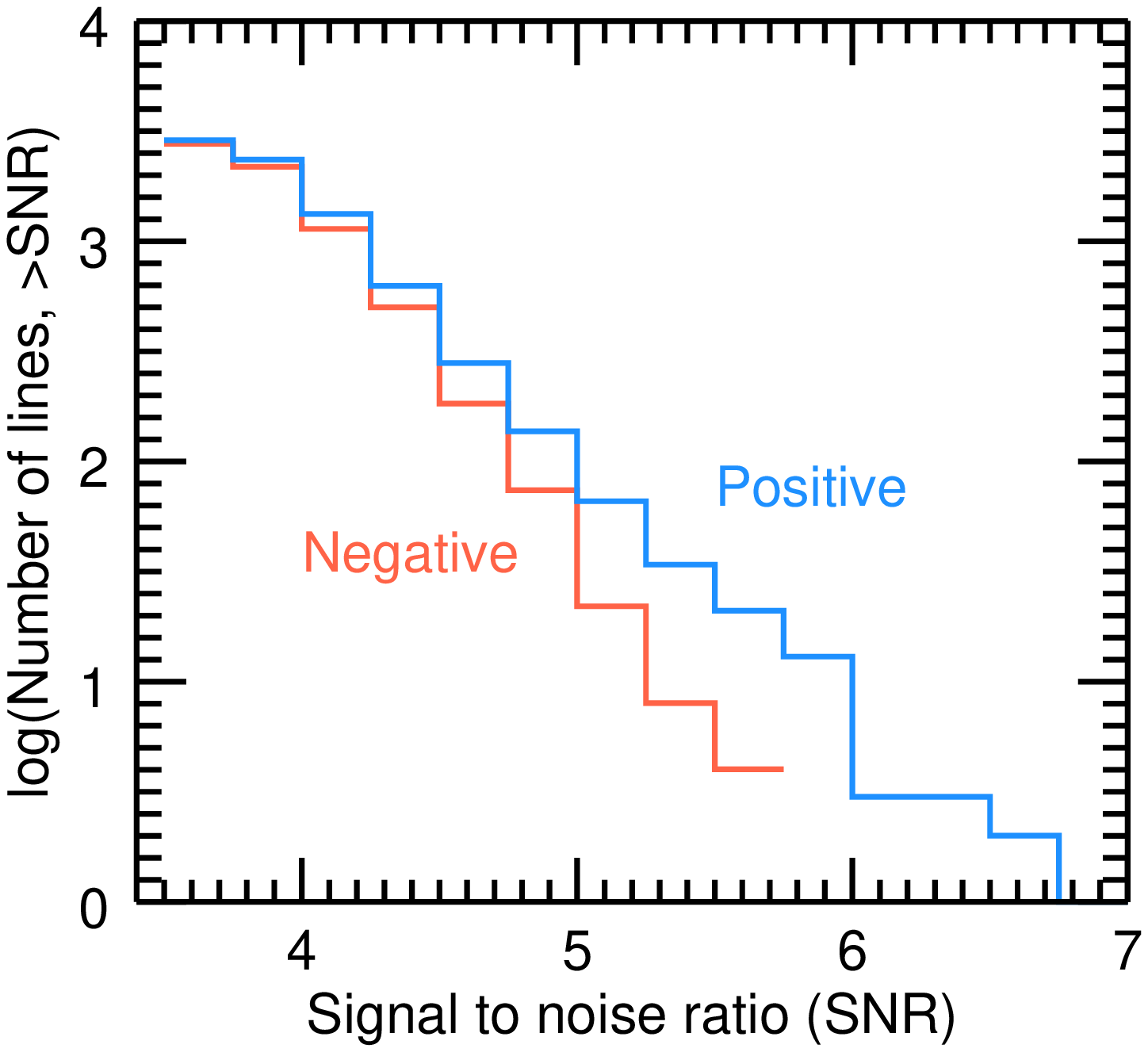}
\includegraphics[scale=0.5]{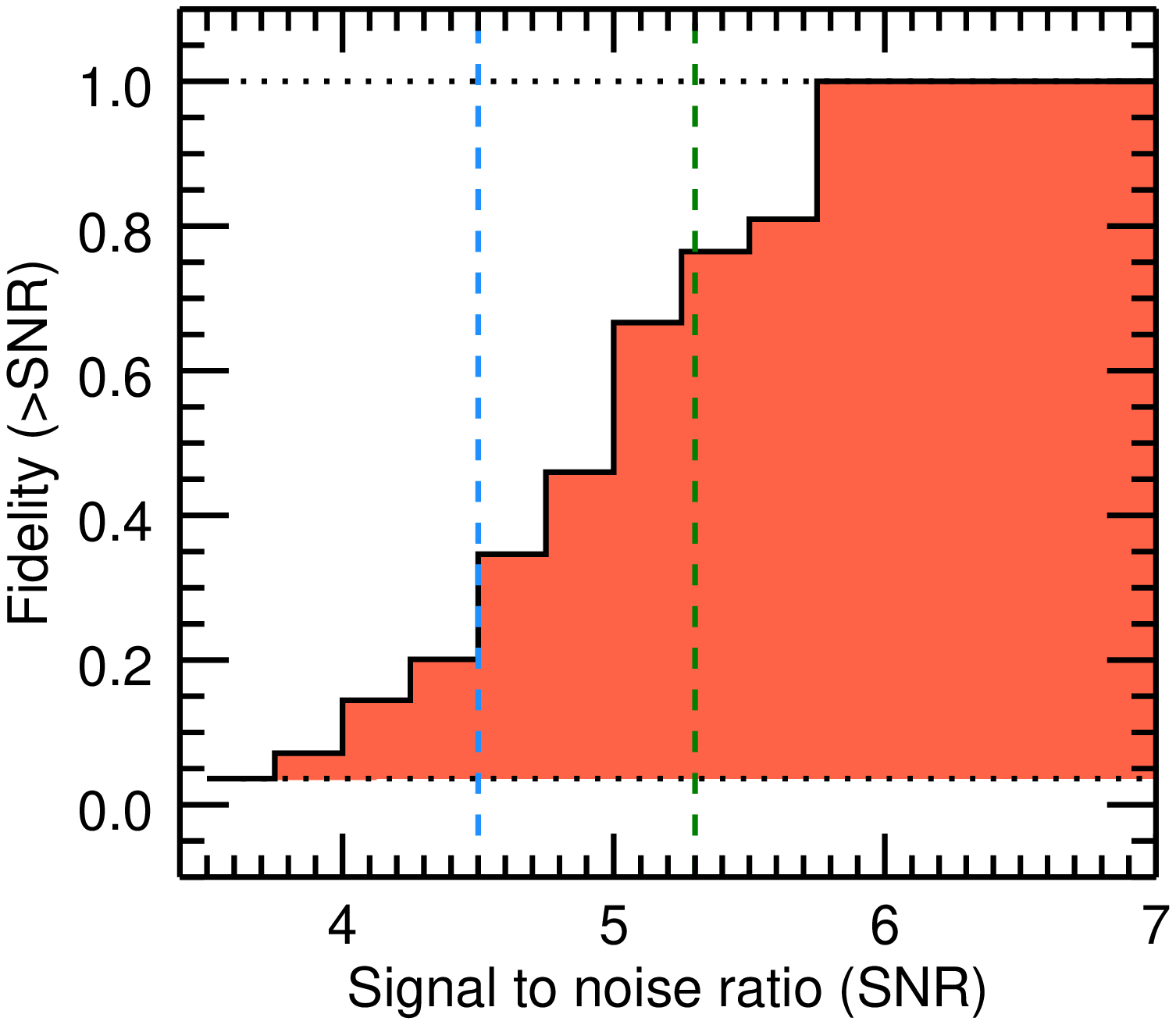}
\caption{{\it (Top:)} Number of positive and negative peaks as a function of signal to noise (S/N) for the [CII] line search performed our ALMA 1mm cube. More positive candidate lines are recovered than negative ones, implying that true astrophysical signal is recovered. {\it (Bottom:)} {\it Fidelity} assessment of the line search in the 1.2-mm cube. The dotted lines represent the level at which the purity becomes null at low S/N, and become unity at $S/N>6$. The dashed lines represent the threshold at which we choose to consider positive line peaks as candidates for [CII] line emission: The green line represents the threshold used for {\it blind} detections that lacked $z>5.5$ counterparts (fidelity level $>70\%$), and the blue line corresponds to the threshold used for line identifications with optical counterparts at $z>5.5$ (purity level $>40\%$).}\label{fig:purity}
\end{figure}

\vspace{2mm}
\begin{figure}[t]
\includegraphics[scale=0.5]{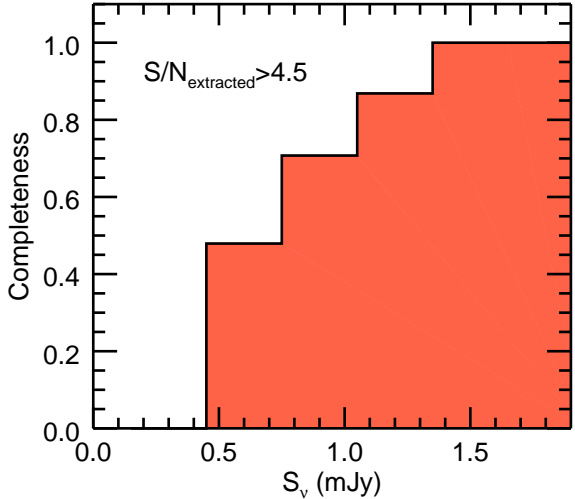}
\caption{Completeness of the [CII] line search for the ALMA 1mm cube as a function of the measured line flux density at the peak position. Since the noise level varies along the frequency axis, we only compare the completeness level for artificial input sources where our extracted S/N level is above our adopted threshold of S/N$>$4.5.\label{fig:completeness}}
\end{figure}

To establish the significance level at which we can rely on a given line peak selected by our search, we need to measure the {\it fidelity} of our line catalog (also termed as ``purity''). We quantify the {\it fidelity} level of our sample based on the number of negative peaks in our 1-mm cube, using the same line search procedure. The idea is that negative line emitters are unphysical, i.e. they are a good way to quantify the contamination of the list of positive line candidates. We define the fidelity $P$ as (see also Paper~I):

\begin{equation}
P (>S/N)= 1 - \frac{N_{\rm negative}}{N_{\rm positive}},
\end{equation}

where $N_{\rm positive}$ and $N_{\rm negative}$ are the number of positive and negative line candidates found above a given S/N threshold. The top panel in Fig.~\ref{fig:purity} shows the number of positive and negative line peaks above a certain S/N found by our line search algorithm. We find 2875 positive versus 2771 negative line peaks identified within the HPBW or PB of our data, or an excess of 104 positive line peaks. The bottom panel of Fig.~\ref{fig:purity}  also shows the fidelity as a function of the derived S/N. At S/N$=5.8$, we reach 100\% {\it fidelity} in our selection, while at S/N$=4.9$ we find 50\% {\it fidelity}. We note that this {\it fidelity} level is almost identical to that found in an independent blind line search of CO line emitters for the full 1\,mm datacube presented in Walter et al. (2016; Paper~I). An important conclusion is that the sample properties are not statistically affected by the different line searching methods used in both studies. Based on this, we set a threshold at the 70\% fidelity level or S/N$=5.3$ for considering a {\it blind} candidate line detection (without optical counterparts), and at 40\% {\it fidelity} level or S/N$=4.5$ for the line candidates associated with optical dropouts at $z>5.5$.

\begin{table}
\caption{Optical dropout galaxies with $z_{\rm phot}=5.5-8.5$ covered in our ALMA 1mm mosaic at the HPBW. Sources are selected from the catalogs of \citet{bouwens15}, \citet{schenker13} and \citet{mclure13}. Columns: (1) Adopted source name; (2), (3) Right ascension and declination; (4) Apparent magnitude in the {\it HST} F160W band; (5) Photometric redshift computed with \texttt{EAZY}.} \label{table:sources}
\begin{tabular}{ccccc}
\hline
 Source name & RA & Dec & m$_{\rm F160W}$ & z$_{\rm phot}$ \\
   & (J2000) & (J2000) & AB &  \\
 (1)  & (2) & (3) & (4) & (5) \\
\hline\hline
ID02 &  3:32:39.39 & $-$27:46:11.1 &   29.6 &   5.69 \\ 
ID03 &  3:32:38.58 & $-$27:46:52.1 &   28.9 &   6.10 \\ 
ID05 &  3:32:39.17 & $-$27:46:48.8 &   30.8 &   5.96 \\ 
ID07 &  3:32:36.75 & $-$27:46:44.7 &   28.5 &   6.24 \\ 
ID08 &  3:32:38.38 & $-$27:46:43.7 &   29.5 &   6.17 \\ 
ID09 &  3:32:38.80 & $-$27:46:34.3 &   29.7 &   5.89 \\ 
ID10 &  3:32:39.79 & $-$27:46:33.7 &   28.2 &   6.03 \\ 
ID11 &  3:32:37.70 & $-$27:46:32.7 &   29.3 &   6.32 \\ 
ID13 &  3:32:37.68 & $-$27:46:21.5 &   28.8 &   6.03 \\ 
ID14 &  3:32:38.45 & $-$27:46:19.8 &   29.1 &   6.17 \\ 
ID15 &  3:32:39.99 & $-$27:46:19.5 &   29.2 &   6.17 \\ 
ID16 &  3:32:39.84 & $-$27:46:18.9 &   27.2 &   6.10 \\ 
ID17 &  3:32:38.26 & $-$27:46:18.4 &   28.2 &   6.10 \\ 
ID18 &  3:32:37.83 & $-$27:46:18.0 &   28.9 &   5.96 \\ 
ID19 &  3:32:38.72 & $-$27:46:17.7 &   29.5 &   6.24 \\ 
ID20 &  3:32:38.64 & $-$27:46:16.4 &   28.5 &   6.32 \\ 
ID21 &  3:32:38.54 & $-$27:46:17.5 &   27.9 &   5.82 \\ 
ID22 &  3:32:38.57 & $-$27:46:15.0 &   29.9 &   6.10 \\ 
ID23 &  3:32:39.08 & $-$27:46:09.1 &   29.8 &   6.46 \\ 
ID24 &  3:32:38.33 & $-$27:46:09.7 &   29.2 &   5.76 \\ 
ID25 &  3:32:37.95 & $-$27:46:02.0 &   29.6 &   6.24 \\ 
ID26 &  3:32:36.49 & $-$27:46:41.7 &   25.5 &   6.03 \\ 
ID27 &  3:32:37.46 & $-$27:46:32.7 &   26.3 &   6.54 \\ 
ID28 &  3:32:37.23 & $-$27:46:31.4 &   29.5 &   5.69 \\ 
ID29 &  3:32:37.51 & $-$27:46:01.0 &   29.4 &   5.69 \\ 
ID30 &  3:32:36.97 & $-$27:45:57.6 &   27.3 &   6.03 \\ 
ID31 &  3:32:38.28 & $-$27:46:17.2 &   26.1 &   6.10 \\ 
ID32 &  3:32:37.40 & $-$27:46:04.5 &   26.7 &   6.10 \\ 
ID33 &  3:32:36.97 & $-$27:45:57.5 &   27.3 &   6.10 \\ 
ID36 &  3:32:39.72 & $-$27:46:21.3 &   28.5 &   7.08 \\ 
ID37 &  3:32:40.18 & $-$27:46:19.0 &   29.5 &   6.84 \\ 
ID38 &  3:32:38.17 & $-$27:46:17.3 &   30.5 &   6.10 \\ 
ID39 &  3:32:39.13 & $-$27:46:16.2 &   29.6 &   6.54 \\ 
ID40 &  3:32:38.35 & $-$27:46:11.8 &   28.3 &   7.00 \\ 
ID41 &  3:32:38.14 & $-$27:46:04.8 &   29.3 &   6.24 \\ 
ID42 &  3:32:37.54 & $-$27:46:01.8 &   29.2 &   6.92 \\ 
ID43 &  3:32:37.31 & $-$27:46:42.0 &   29.8 &   6.39 \\ 
ID44 &  3:32:37.35 & $-$27:46:24.5 &   28.9 &   6.32 \\ 
ID45 &  3:32:39.20 & $-$27:46:32.2 &   29.0 &   8.29 \\ 
ID46 &  3:32:37.63 & $-$27:46:01.4 &   28.2 &   8.11 \\ 
ID47 &  3:32:37.79 & $-$27:46:00.1 &   28.2 &   8.02 \\ 
ID49 &  3:32:36.60 & $-$27:46:22.1 &   30.3 &   7.66 \\ 
ID51 &  3:32:37.44 & $-$27:46:51.3 &   28.5 &   6.80 \\ 
ID52 &  3:32:36.91 & $-$27:46:51.7 &   30.2 &   6.80 \\ 
ID53 &  3:32:39.22 & $-$27:46:14.9 &   30.1 &   6.80 \\ 
ID54 &  3:32:38.65 & $-$27:46:04.1 &   29.4 &   6.60 \\ 
ID55 &  3:32:39.00 & $-$27:46:48.2 &   28.2 &   6.60 \\ 
ID57 &  3:32:37.29 & $-$27:46:17.5 &   28.5 &   6.90 \\ 
ID58 &  3:32:37.09 & $-$27:46:44.1 &   29.8 &   7.20 \\ 
ID59 &  3:32:39.31 & $-$27:46:18.1 &   29.2 &   7.50 \\ 
\hline
\end{tabular}
\end{table}

We check the {\it completeness} level of our sample by inserting 200 fake sources into the 1-mm data cube. We run our line search algorithm on this cube and compare the number of recovered sources with the input catalog. We use the same procedure as in Walter et al. (2016; Paper~I) to insert sources, and we refer the reader to this paper for details. In short, fake emission line sources are inserted at random positions and frequencies in the cube, assuming a single 3-dimensional Gaussian profile with varying line widths and input fluxes. Since the noise level of our cube is affected by different atmospheric features and SPW edges along the frequency axis, the significances of each line detection will vary. We thus compute the completeness level only for those line peaks which are extracted with a S/N level above 4.5, matching the cut in significance level we have set. Figure \ref{fig:completeness} shows the completeness level obtained in this way, as a function of flux density computed at the position of the line peak. For sources with S/N$>4.5$, our sample reaches 100\% {\it completeness} at line peak fluxes above 1.5\,mJy, while we reach down to 50\% {\it completeness} at very faint flux levels of 0.7\,mJy. 

\vspace{2mm}
\begin{figure*}[t]
\centering
\includegraphics[scale=0.4]{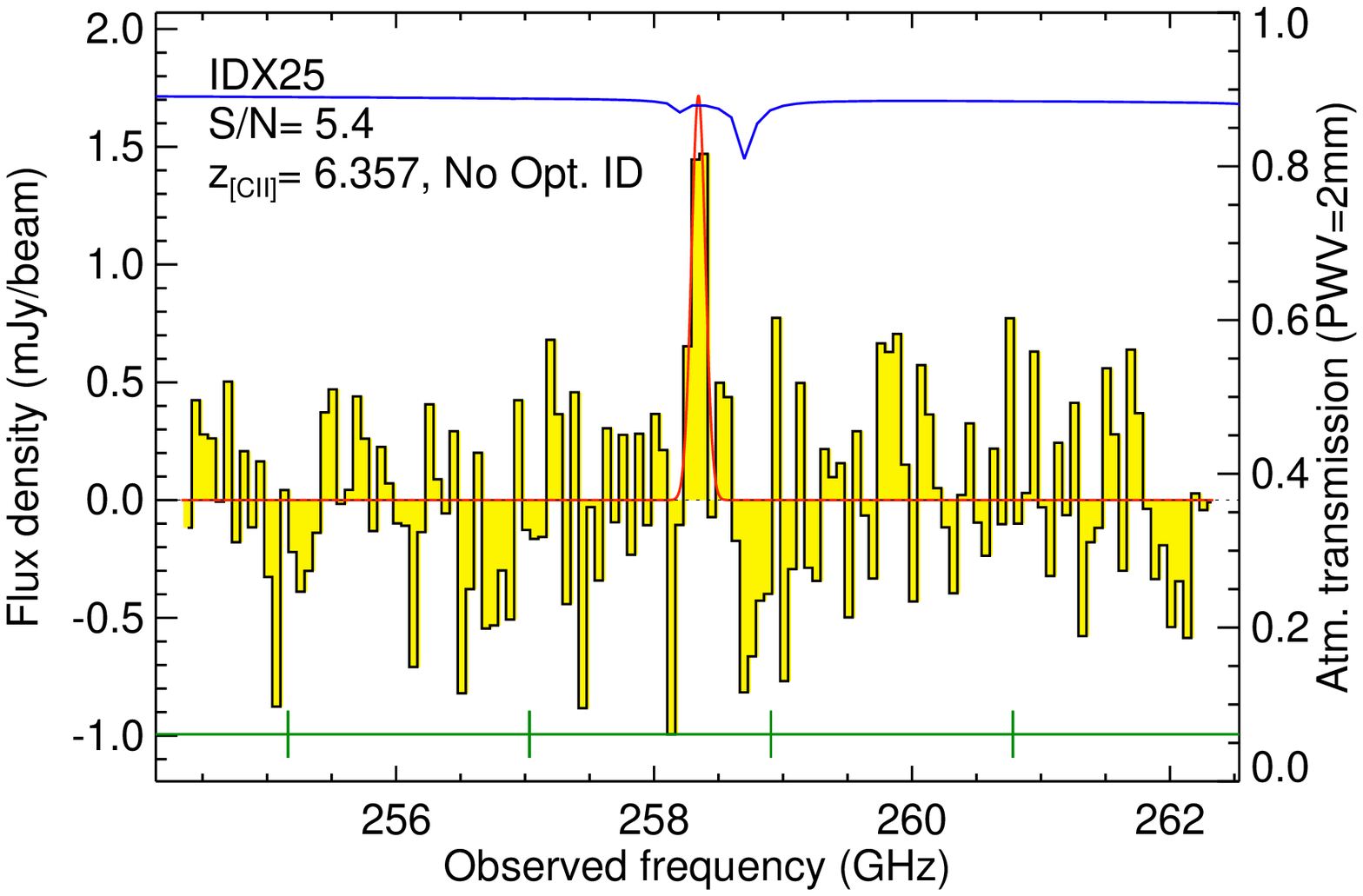}
\includegraphics[scale=0.4]{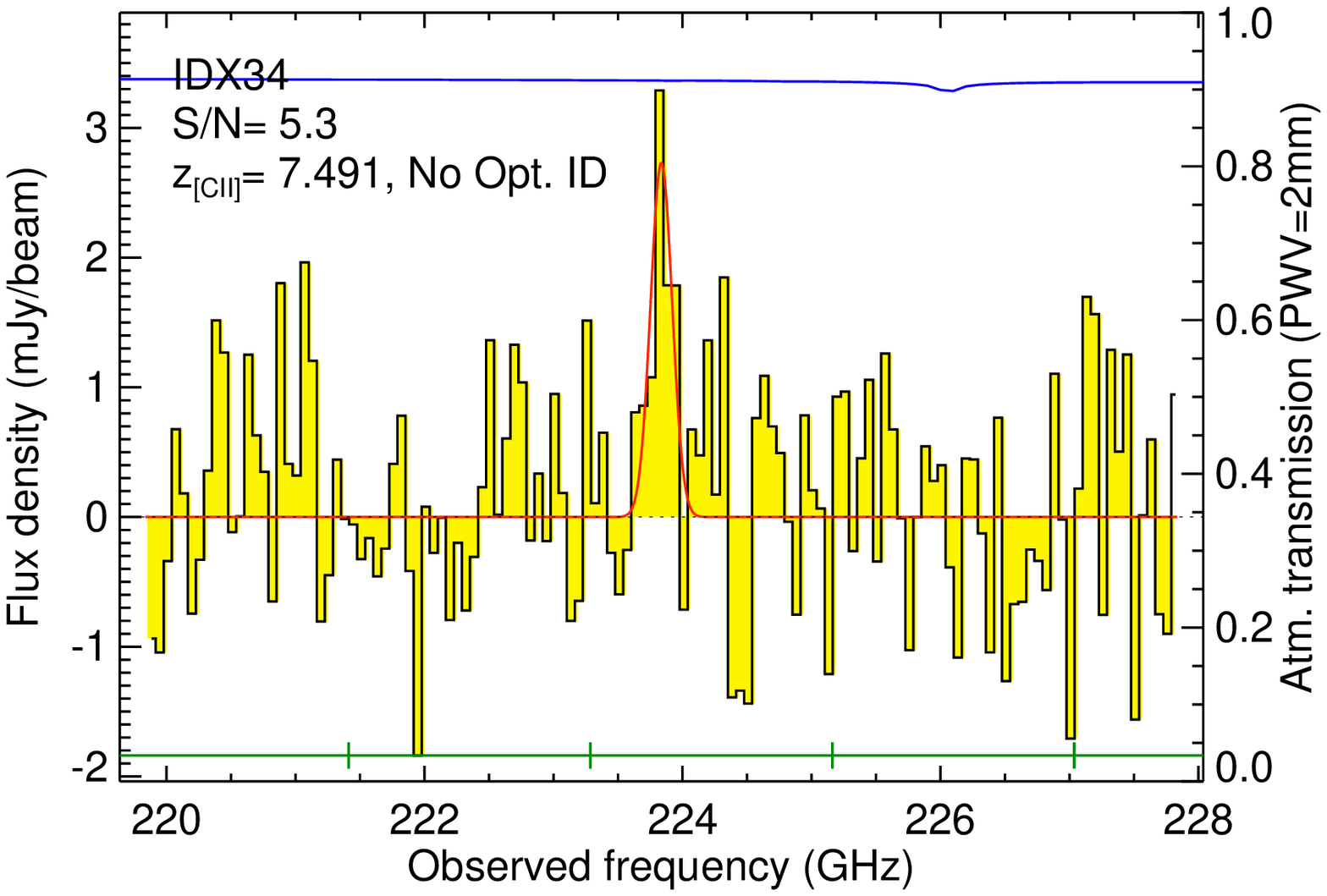}
\caption{ALMA band-6 spectra of the two blind [CII] line emitter candidates at $z>6$ (PB-corrected), selected to have significances above 5.3$\sigma$ (70\% fidelity). In both cases, there is no obvious optical counterpart within $1''$. Both spectra are shown in 62.5 MHz channels (or $\sim75$ km s$^{-1}$ at 250 GHz). A frequency range of 8 GHz is shown, centered around the identified candidate line. The spectra have been computed at the position of the peak S/N as computed by our line search algorithm. The red curve shows a Gaussian fit to the candidate line. The blue solid line shows the atmospheric transmission for a precipitable water vapour (PWV) of 2.0mm, typical for 1-mm observations. The green lines represent the location of each SPW and its edges. \label{fig:blind_spectra}}
\end{figure*}

\vspace{2mm}
\begin{figure*}[t]
\centering
\includegraphics[scale=1.5]{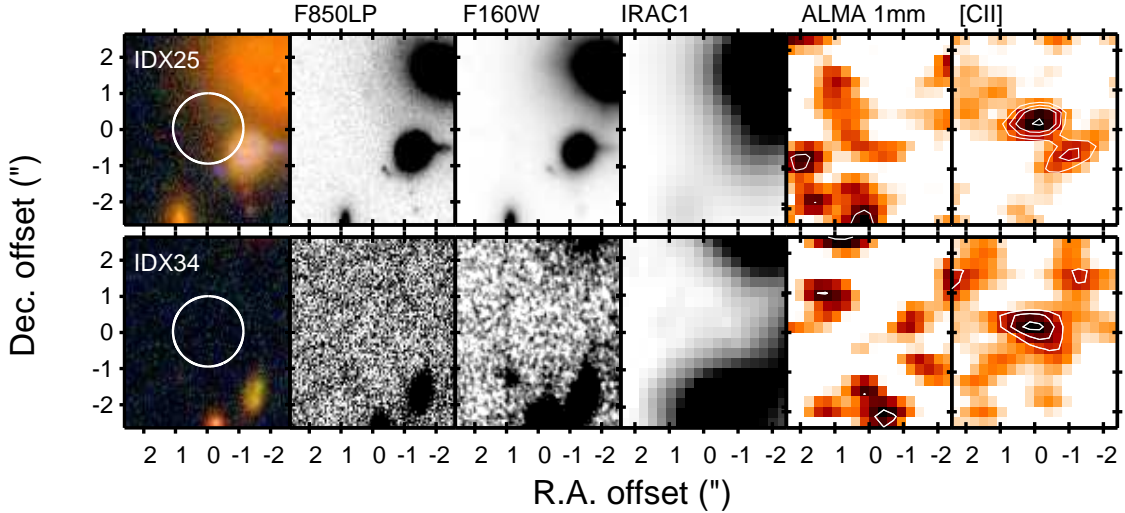}
\caption{Image postage stamps for the two {\it blind} [CII] line candidates in the ALMA UDF survey ($5''\times5''$ in size). From left to right, we show cutouts of a HST F435W/F850LP/F105W color composite image, F850LP, F160W, IRAC channel 1 (3.6$\mu$m) and the ALMA 1.2-mm continuum and the collapsed [CII] map. Images are centered at the candidate [CII] line position. The white circle represents the $1''$ search radius used in the optical counterpart identification. The white contour levels in the right panel for the ALMA 1.2-mm continuum and [CII] line maps are shown at 2, 3, 4 and 5$\sigma$.\label{fig:blind_stamps}}
\end{figure*}

\subsection{Blind [CII] line candidates}

\label{sec:blind}

We look for possible [CII] line candidates by directly searching for significant positive line peaks in our 1.2-mm data cube. We restrict this search by setting a limit on the {\it fidelity} level down to $P=$70\%, which is equivalent to a line peak S/N$>5.3$. Since detection of a single line in the 1-mm cube could correspond to CO line emission at lower redshift, we further restrict the sample to those objects that either do not have an obvious optical counterpart in the HUDF catalogs or that have a counterpart with a photometric redshift $z>5.5$. The reason to select this {\it fidelity} level, compared to the 60\% level set in Paper~I for detections, is to statistically increase the chances that the identified peak is real as we are explicitly looking for objects that lack of or have a faint optical counterpart.  

Using this procedure, we found two positive line peaks with S/N$=5.3-5.4$ in the 1-mm data cube in which no optical counterpart was identified. Figure \ref{fig:blind_spectra} shows the millimeter spectra for these line candidates. If we associate these candidate line peaks to the [CII] emission, they would correspond to redshifts of $\sim6.36$ and 7.50 for sources IDX25 and IDX34, respectively. 

For each spectrum we fit a single Gaussian to the candidate line emission, thereby deriving an integrated intensity ($I_{\rm [CII]}$) and line full width at half maximum (FWHM). We checked for extended emission by collapsing each candidate emission line along the frequency axis, and producing a integrated (moment~0) map. This collapsed map was then fitted with a two-dimensional Gaussian. In all cases, the emission was found unresolved and the integrated line intensity obtained in the collapsed map was found to be consistent with the value derived from the fitted spectrum. The integrated intensity as well as the resulting [CII] luminosities are given in Table~\ref{table:properties}.

In Fig.~\ref{fig:blind_spectra}, the location of each frequency setup (SPW) is represented by green bars. Since our frequency setup does not provide frequency overlap between adjacent SPWs, the noise is expected to slightly increase at the SPW edges. This could potentially lead to spurious line peaks along the frequency axis. However, inspection of Fig.~\ref{fig:blind_spectra} shows that all the [CII] line candidates presented here are located comfortably away (at least $\sim100$ MHz; or $>120$ km s$^{-1}$) from a SPW edge. Furthermore, our line search algorithm, by construction, would discard such line peaks to begin with, as it first convolves the cube along the frequency axis, and thus effectively searches for significant line peaks in the average images centered at a particular channel.

Fig.~\ref{fig:blind_spectra} also shows the atmospheric transmission as a function of frequency for PWV$=2.0$ (blue curve). Source IDX25 is the only one from these candidates that lies right on top of an atmospheric absorption line, suggesting that this line peak might be produced by this atmospheric feature. However, an atmospheric line should translate in an increase of the overall noise level at a particular frequency and thus not necessarily act to mimic a fake source both in space and frequency. As we will see later on, there is tentative evidence for the parallel detection of CO(6--5) line emission from this source in the 3-mm data cube. For the other two line candidates, there is no further evidence to confirm their emission, and only future spectroscopy with ALMA would be able to confirm their reality. 

Figure \ref{fig:blind_stamps} shows multi-wavelength postage stamps centered at the location of the candidate line emission. In both cases, there is no obvious optical counterpart. IDX25 falls close to a system of nearby galaxies suggesting the background line emission, if real, could be gravitationally lensed. While there is no quantitative supporting evidence for this scenario, the {\it HST} images (see color composite) reveal several blueish sources along the line of sight of the southern bright source. These blue sources do not coincide with the candidate line emission, however this is expected if the source emission is coming from different regions in the source plane. The only way to verify these blind detections is with deep ALMA follow up.

\subsection{Optically selected [CII] line candidates}

We searched for additional fainter [CII] candidates, by cross matching our catalog of line peaks at S/N$>4.5$ with the sample of optical dropout galaxies at $z_{\rm phot}=5.5-8.5$ discussed in \S \ref{sec_ancillary}. In this case, we select all line peaks at S/N$>4.5$ that lie within $1''$ of the optical position, i.e. less than one synthesized ALMA beam in the 1\,mm band. In the redshift range $z=5.5-8.5$, $1''$ corresponds to $4.7-8.1$ kpc. The targeted significance level translates into a fidelity of 40\% for line detection. Statistically, this implies that roughly 60\% of our sample of line candidates correspond to spurious detections. Conversely, the positional prior implies that even if the {\it fidelity} level is relatively low at the chosen S/N threshold, the coincidence of a line candidate with an optical dropout galaxy will increase the chances that it corresponds to a real [CII] detection. 

\begin{figure*}[t]
\centering
\includegraphics[scale=0.33]{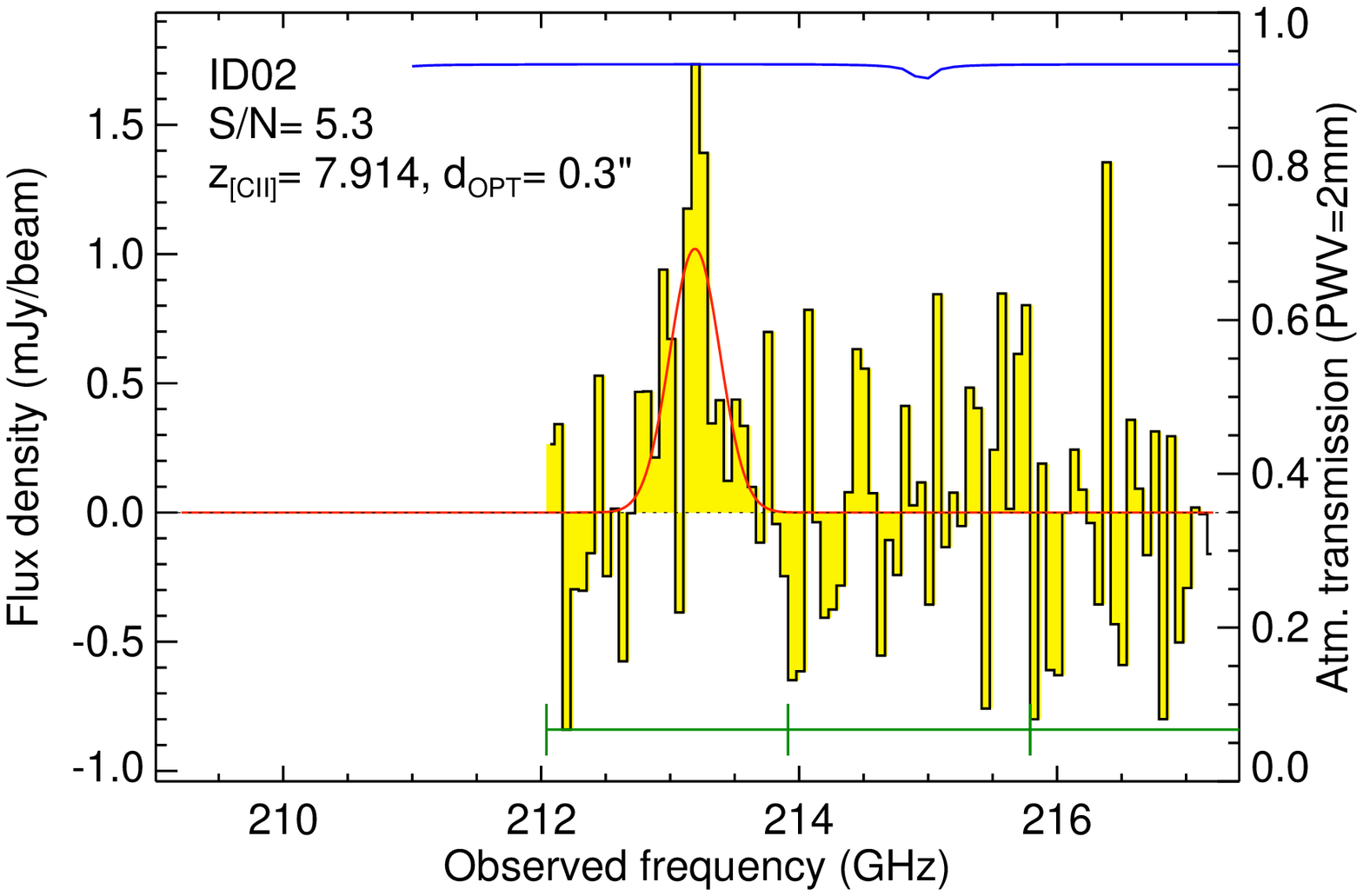}
\includegraphics[scale=0.33]{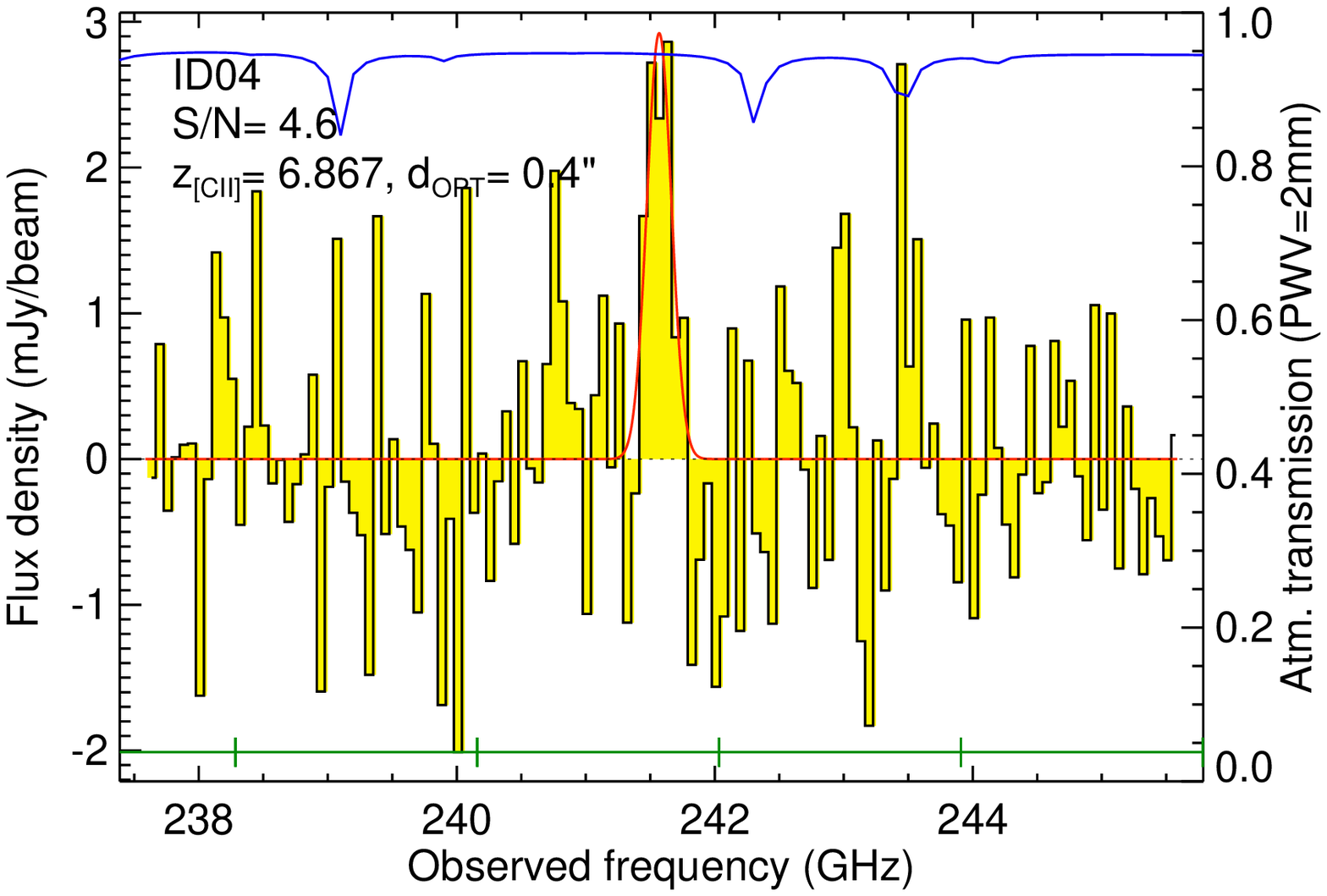}
\includegraphics[scale=0.33]{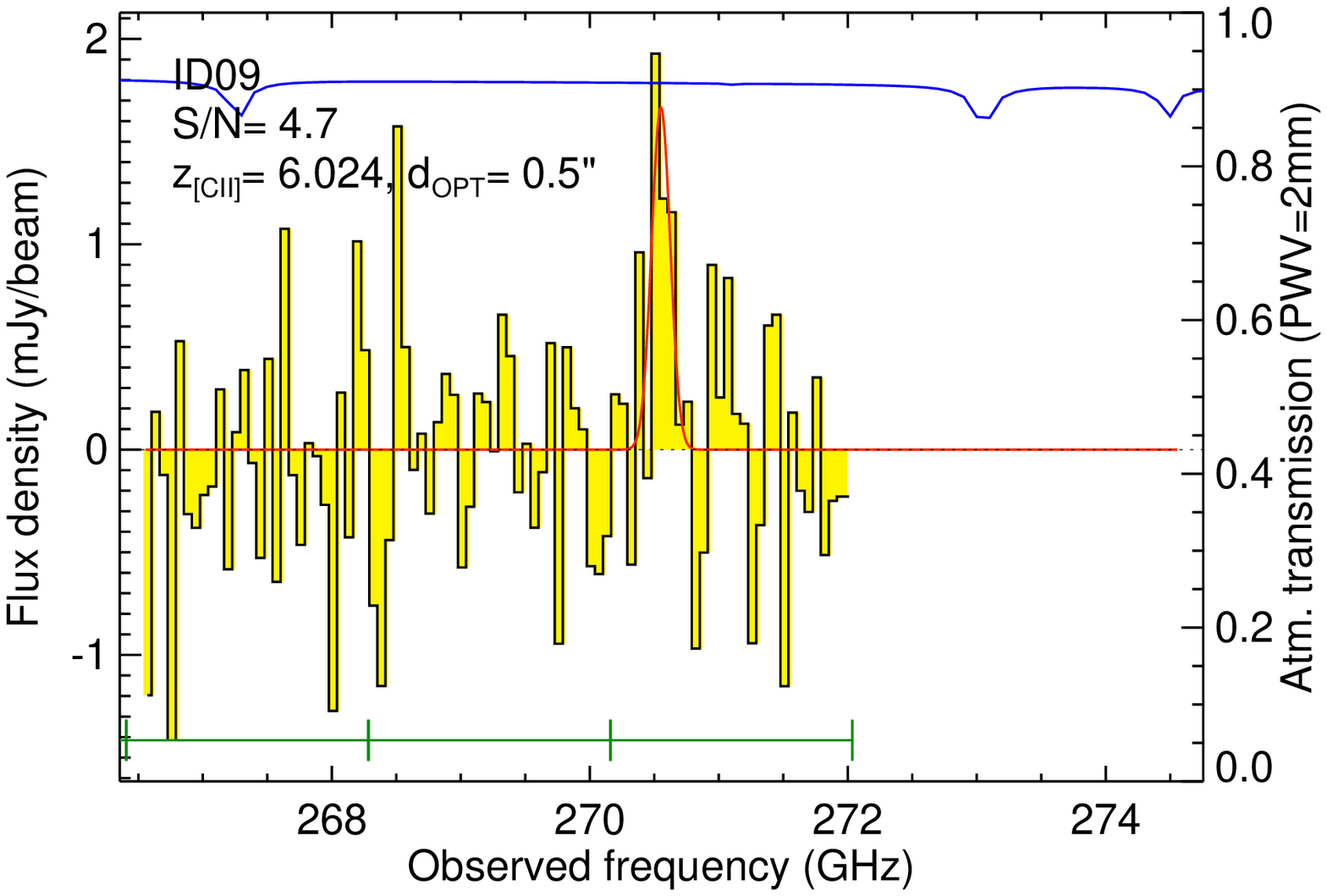}\\
\includegraphics[scale=0.33]{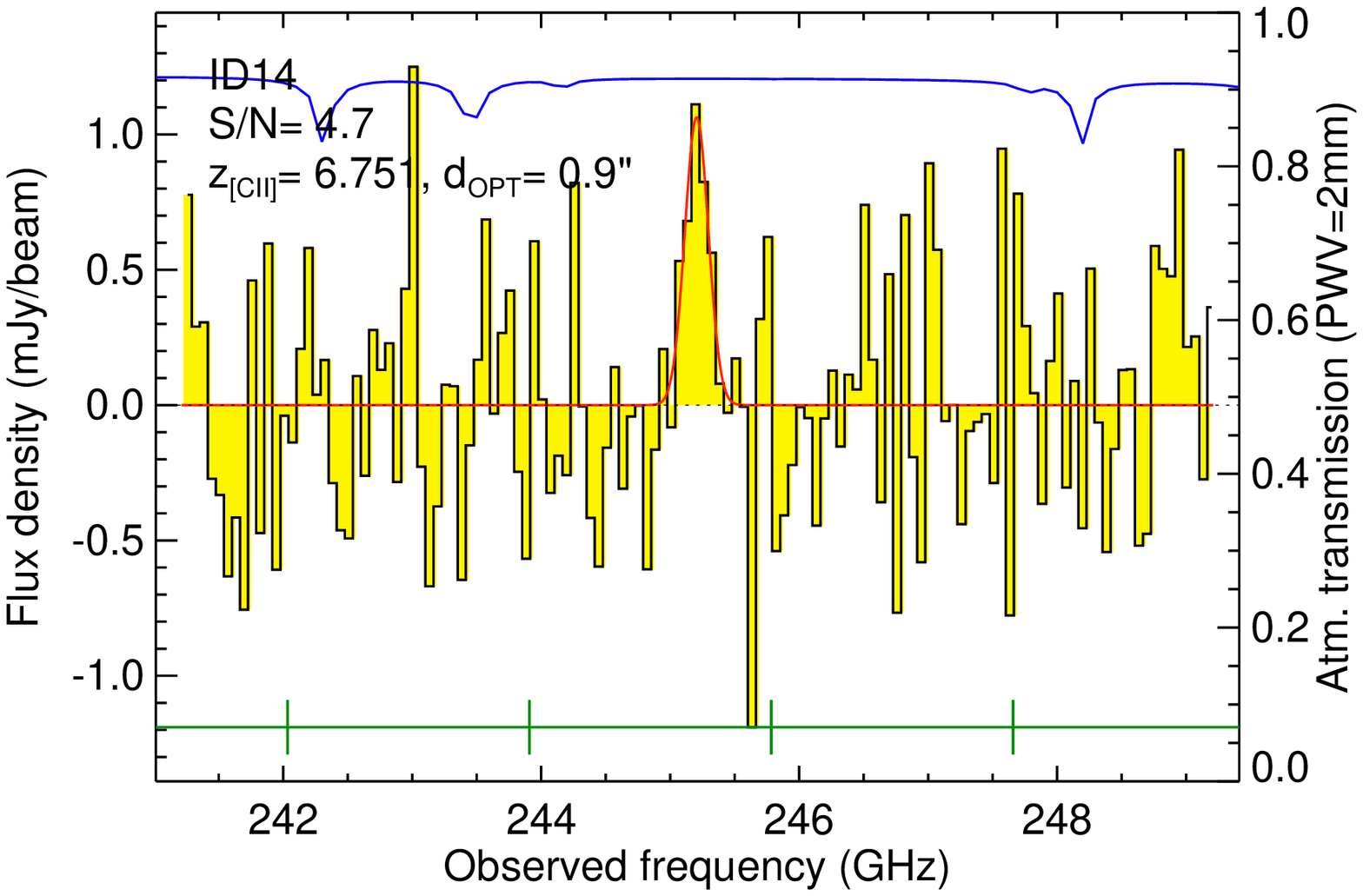}
\includegraphics[scale=0.33]{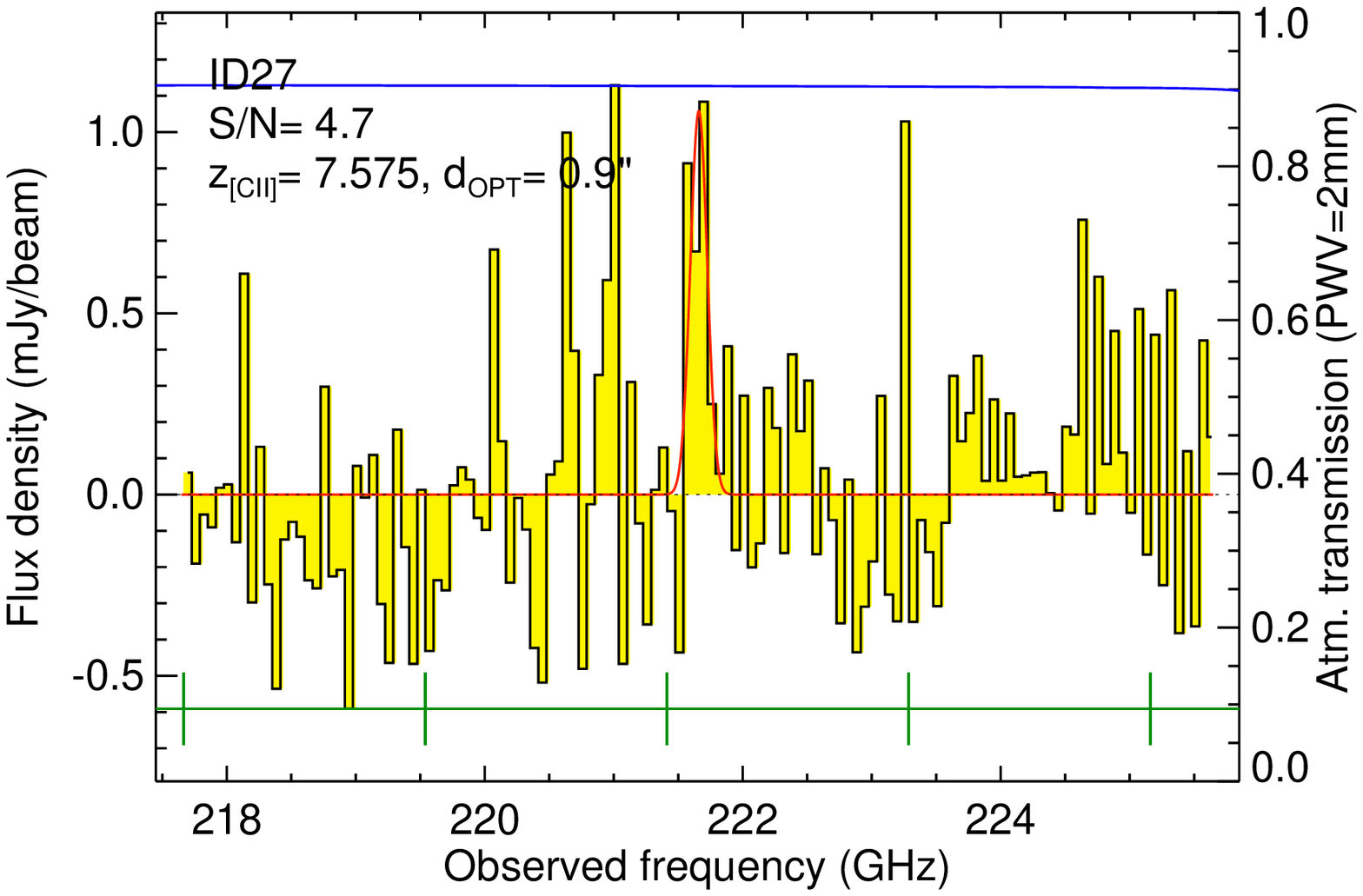}
\includegraphics[scale=0.33]{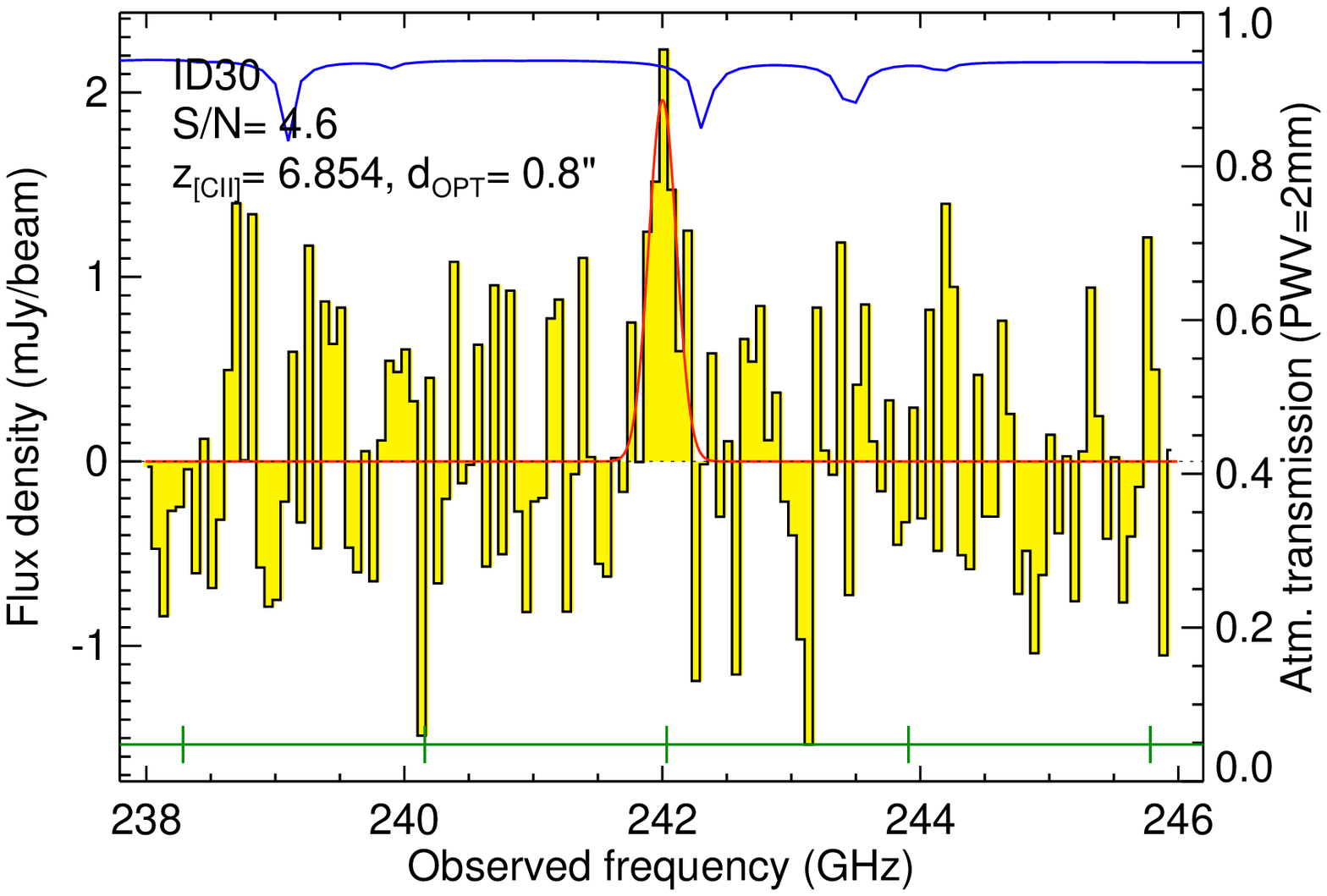}\\
\includegraphics[scale=0.33]{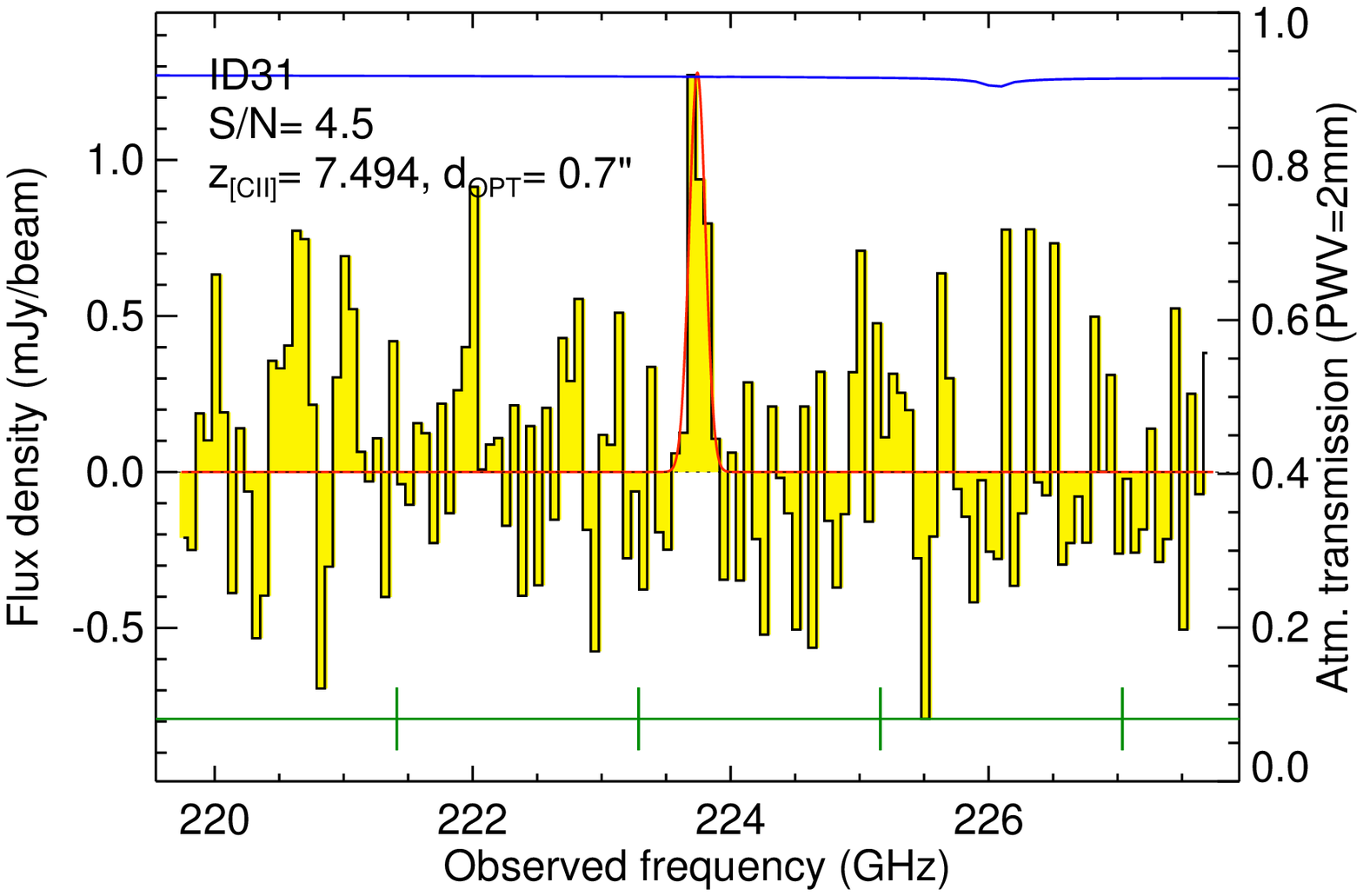}
\includegraphics[scale=0.33]{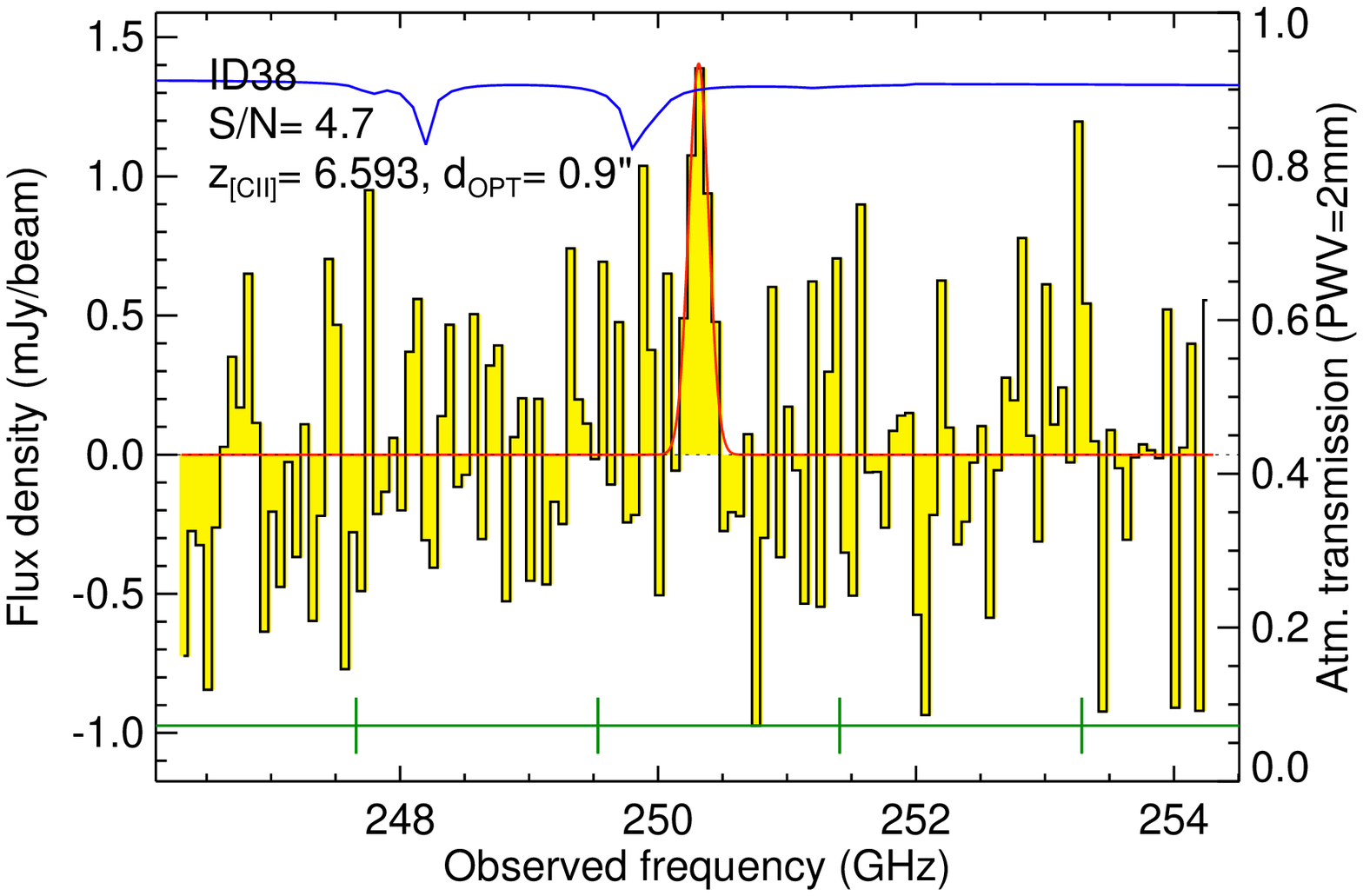}
\includegraphics[scale=0.33]{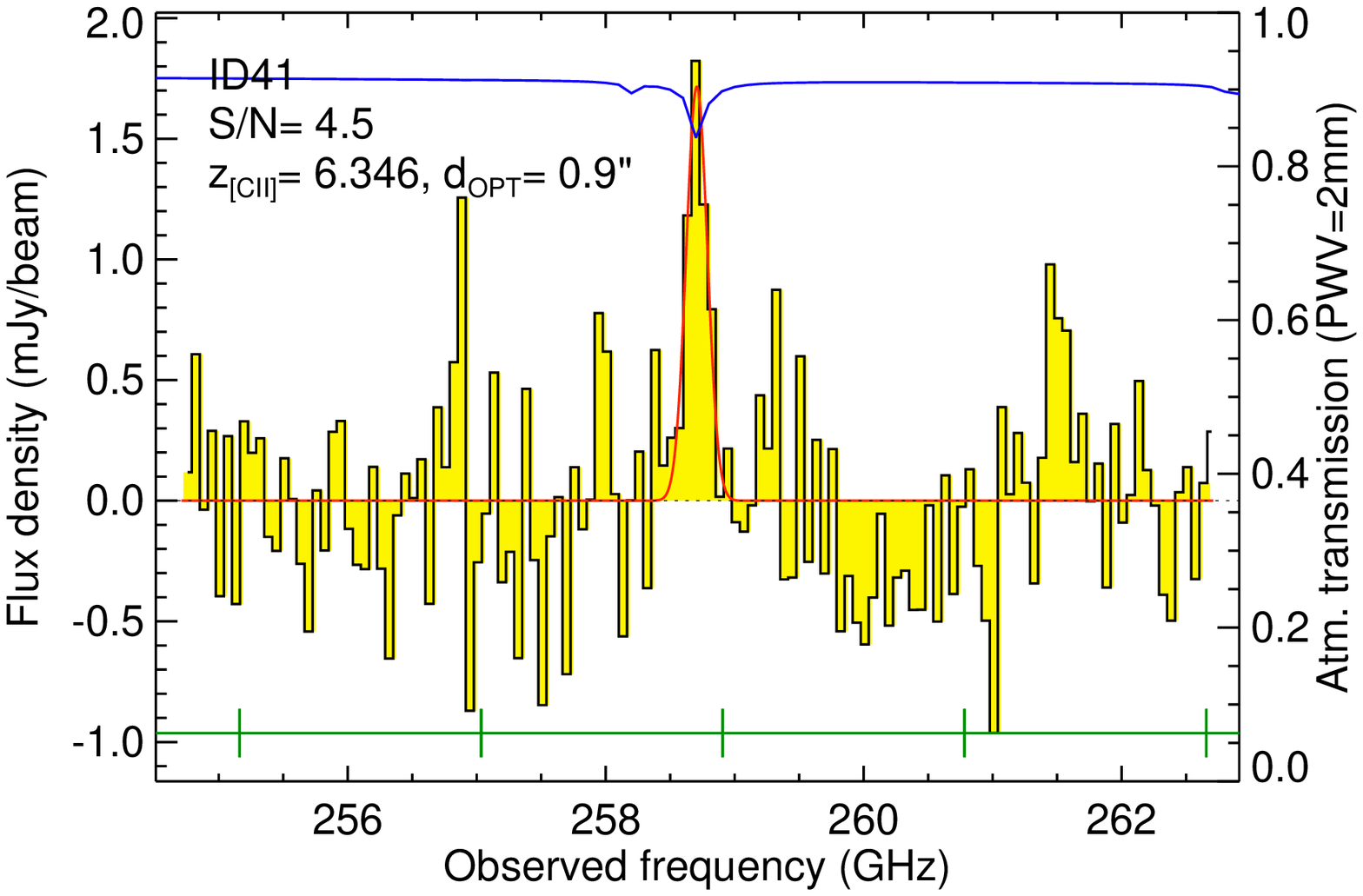}\\
\includegraphics[scale=0.33]{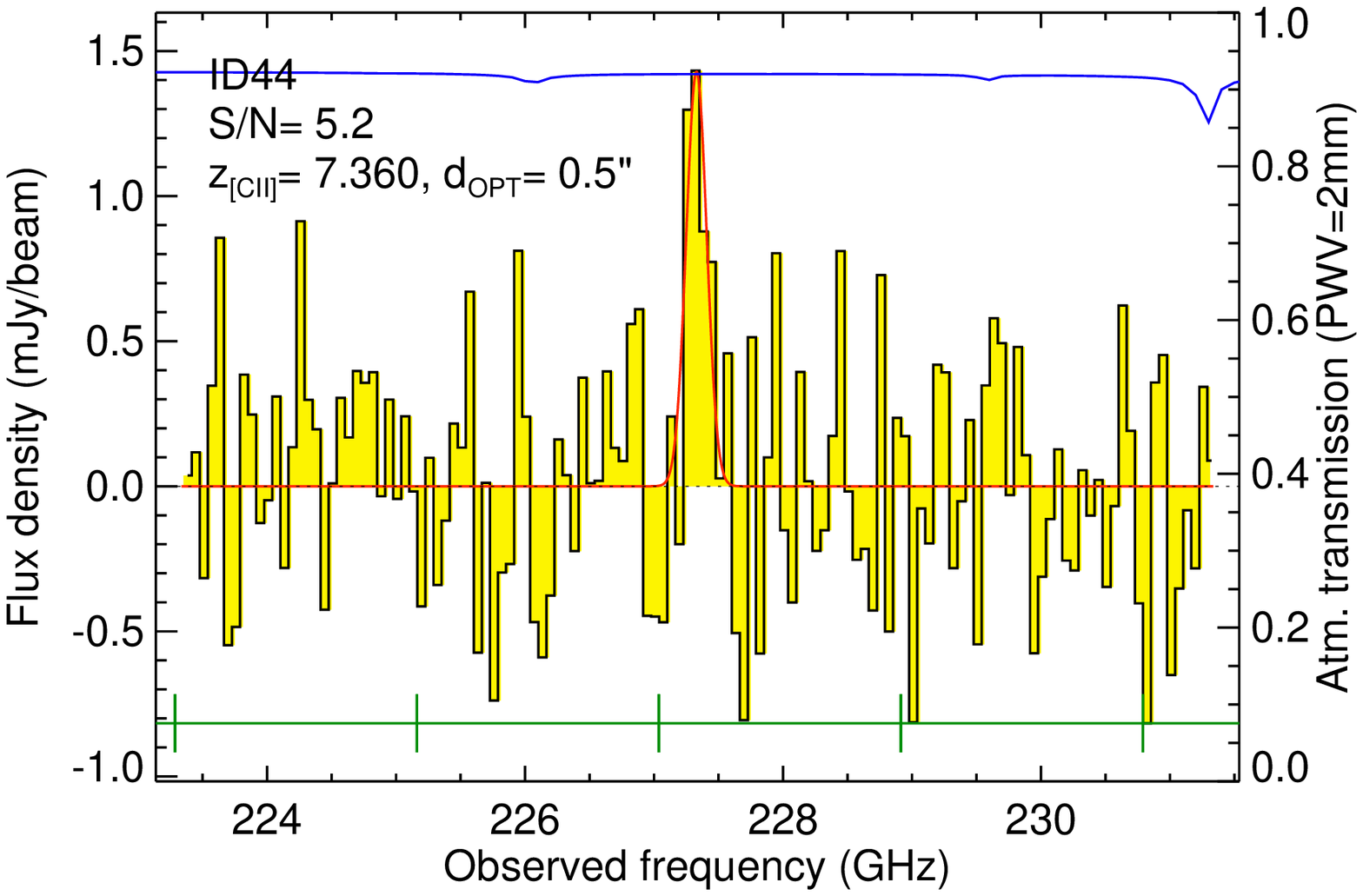}
\includegraphics[scale=0.33]{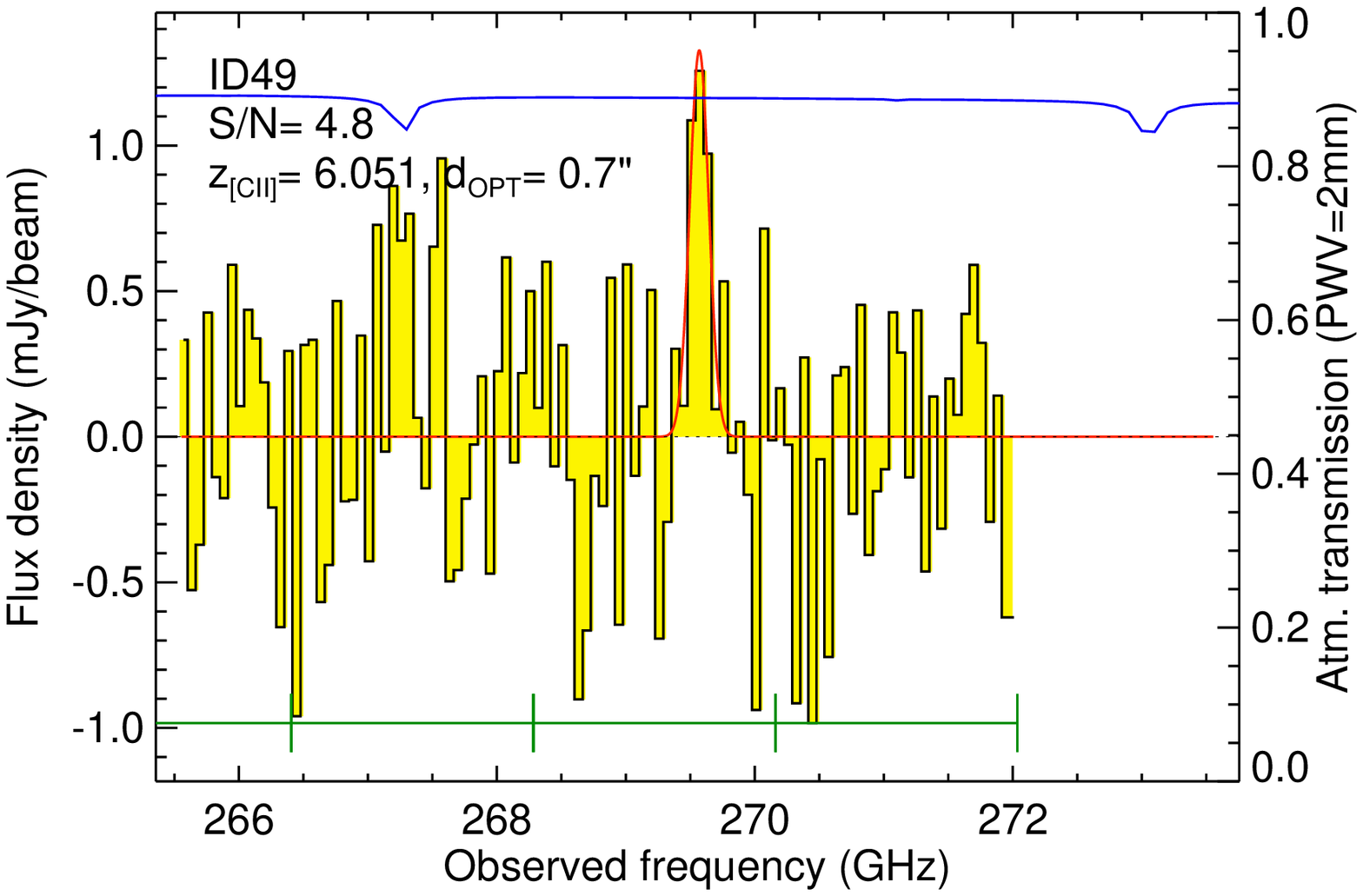}
\includegraphics[scale=0.33]{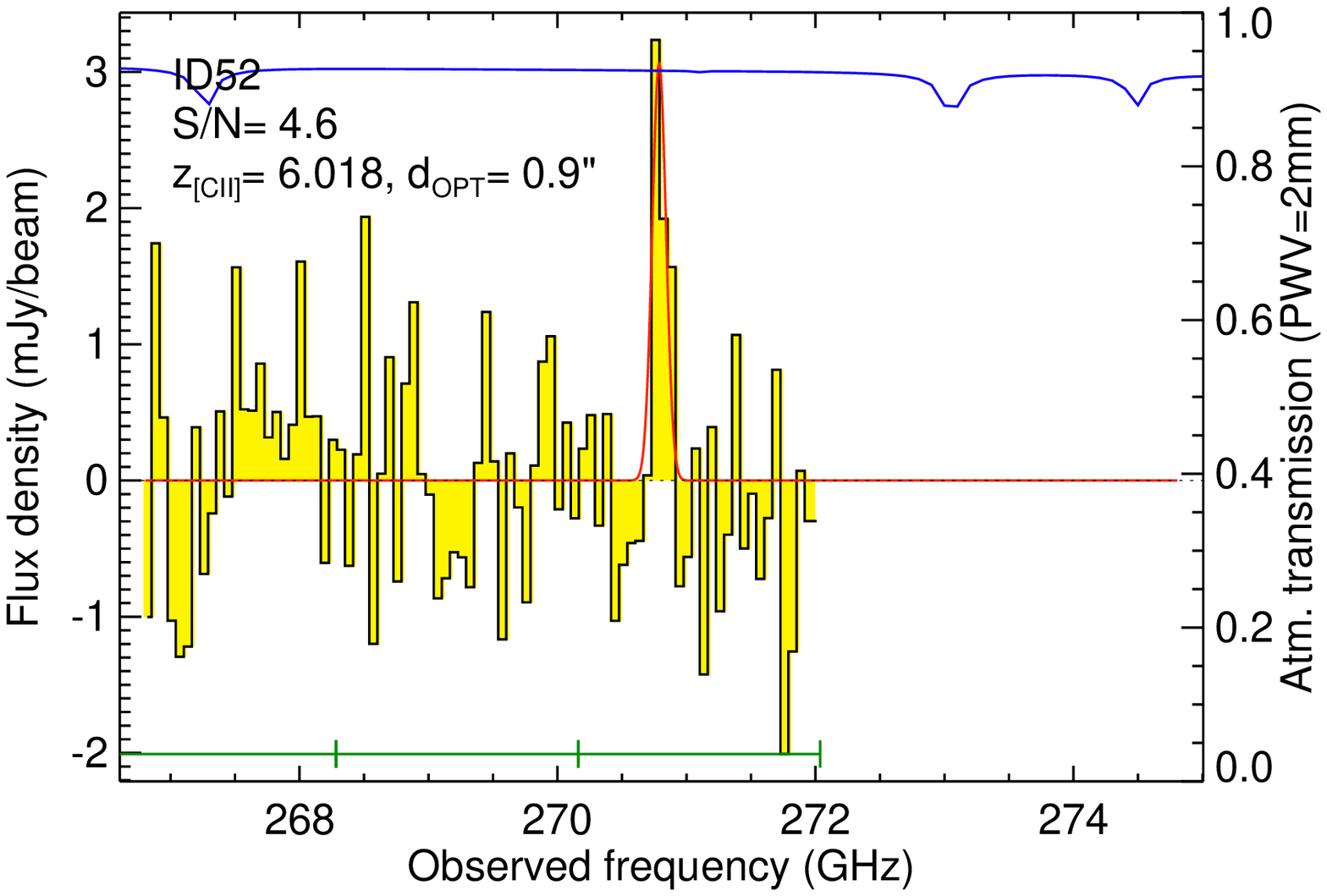}
\caption{ALMA band-6 spectra of the [CII] line emitter candidates at $z>6$ (PB-corrected), selected to have significances above 4.5$\sigma$ (50\% fidelity) and to coincide within $1.0''$ with an optical dropout galaxy with $z_{\rm phot}=5.5-8.5$. All spectra are shown in 62.5 MHz channels (or $\sim75$ km s$^{-1}$ at 250 GHz). A frequency range of 8 GHz is shown, centered around the identified candidate line. The spectra have been computed at the position of the peak S/N as computed by our line search algorithm. The red curve shows a Gaussian fit to the candidate line. The blue solid line shows the atmospheric transmission for a precipitable water vapour (PWV) of 2.0mm, typical for 1-mm observations. The green lines represent the location of each SPW and its edges.\label{fig:lines}}
\end{figure*}

In Fig.~\ref{fig:lines} we present the spectra of the 12 identified [CII] line candidates matched to optical dropout positions. As in the previous section, for each spectrum we fit Gaussian profiles to the candidate line emission. In all cases, the collapsed line map source emission was found to be unresolved and the integrated line intensity obtained in the collapsed map was found to be consistent with the value derived from the fitted spectrum. The integrated intensity as well as the resulting [CII] luminosities are given in Table~\ref{table:properties}.

Figure~\ref{fig:lines} also shows the location of each frequency setup (SPW) and the atmospheric transmission as a function of frequency. In two cases, ID30 and ID41, the candidate line peak is located at the same frequency than a SPW edge. For ID41, this also coincides with the location of an atmospheric line, strongly suggesting that this line peak is produced by a noise artefact. 

\begin{figure*}[t]
\centering
\includegraphics[scale=1.5]{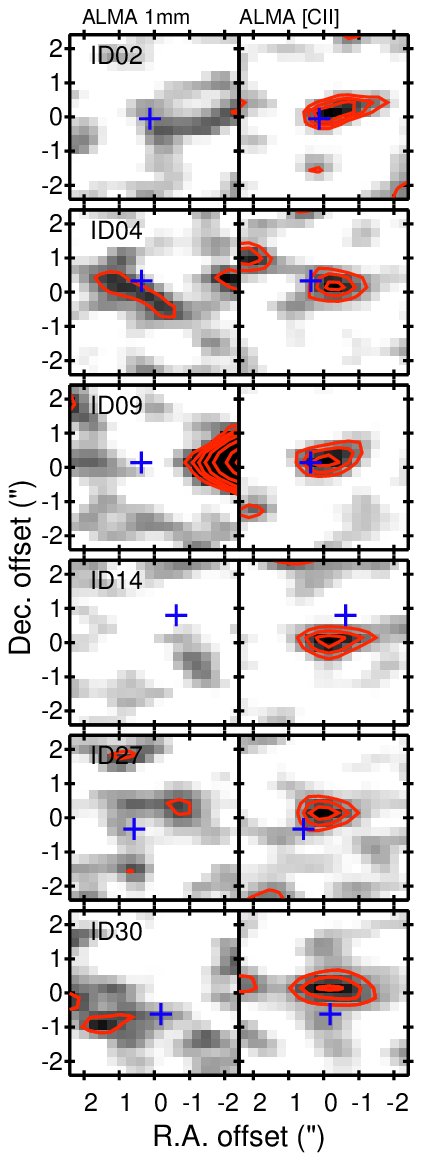}
\hspace{3mm}
\includegraphics[scale=1.5]{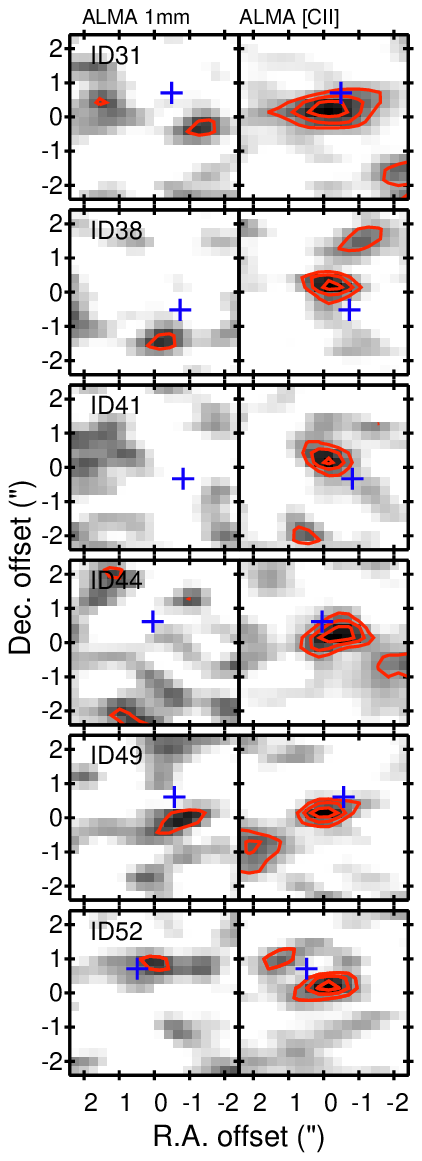}
\caption{Postage stamps for the [CII] line candidates in the ALMA UDF survey ($5''\times5''$ in size) that are associated with an $z>6$ optical dropout galaxy within $1''$. From each object, shown are the ALMA 1.2-mm continuum image and the candidate [CII] line emission map, averaged over the line FWHM. The red contours represent the emission at 2, 3, 4, 5 and 6$\sigma$ levels. The blue cross represents the location of the associated optical dropout galaxy from the catalogs of \citep{bouwens15}.\label{fig:stamps}}
\end{figure*}

In Fig.~\ref{fig:stamps} we present the ALMA 1.2-mm continuum and line maps of our optically selected [CII] line candidates centered at the location of the identified line emission. All of the line candidates are recovered in the collapsed line maps, indicating that the features seen in the spectra are not caused by increased noise at a particular frequency. This is expected since our line search algorithm does average over frequency to isolate individual peaks. By selection, the candidate [CII] line emission is located within $1''$ from the optical high redshift counterpart, however, in the cases of ID02, ID04, ID09, ID31 and ID44, the coincidence is remarkably closer (within $0.5''$). 

It is difficult to argue that closer associations are more likely. While in a statistical sense this is correct, it is always possible that the offset is physical as it has been argued in previous studies \citep{capak15, maiolino15}. None of the sources are detected in continuum emission, yet a few cases show positive, low-significance continuum signal at the location of the candidate [CII] emission (ID04, ID27, ID31).

\vspace{2mm}

\vspace{2mm}

\subsection{Redshift probability distributions}

An important piece of information comes from the comparison between the optical/NIR photometric redshift and their probability distributions with the candidate [CII] redshifts. As mentioned above, the optical galaxies in our study have been selected to have photometric redshifts in the range 5.5 -- 8.5. Although these sources are faint, even for the deep HST images, the photometric redshifts can still be constrained to within $\Delta z=0.5$ based on the redshifted Lyman break. 

\begin{figure}[t]
\centering
\includegraphics[scale=1.]{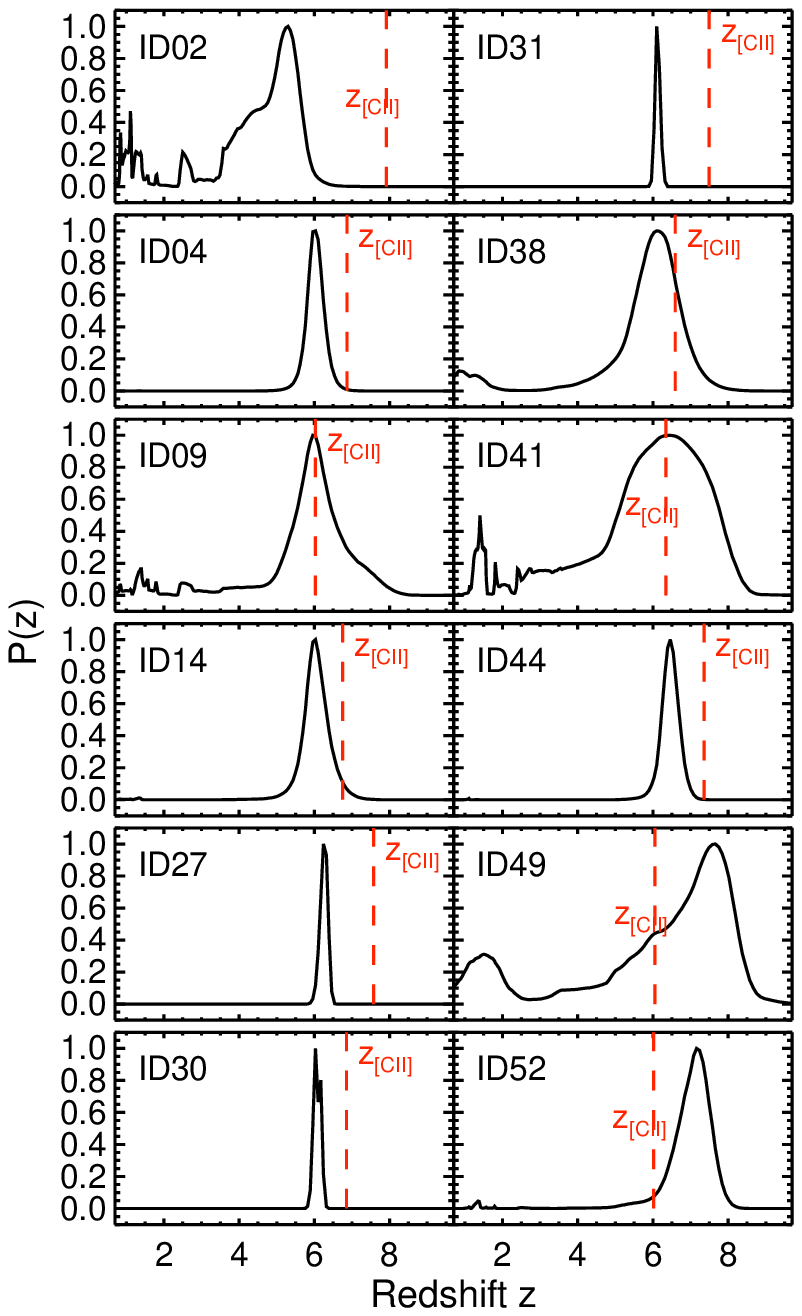}
\caption{Redshift probability distributions, $P(z)$, of the optical sources associated with the [CII] line candidates. By construction, these sources have been selected to have $z_{\rm phot}=5.5-8.5$. The red vertical line represents the redshift obtained by assuming that the line identified in the 1.2-mm cube corresponds to [CII] line emission.}\label{fig:pdz}
\end{figure}

Figure \ref{fig:pdz} shows the redshift probability distributions for the optical associations to the [CII] line candidates. The red line represents the [CII]-based redshift estimate. In all cases, the peak in the probability distribution is produced at $z>5$. In four cases, there is a non negligible chance that it could correspond to a lower redshift galaxy at $z\sim2$ (sources ID02, ID38, ID41 and ID49). In most cases, there is a large disagreement between the redshift implied by the identified millimeter line and the the bulk of the redshift probability distributions. For sources ID02, ID04, ID27, ID30, ID31, ID44 and ID52 this disagreement is above the $95\%$ confidence level ($2\sigma$). For the cases of ID14, ID38, ID41 and ID49, there is better agreement, however in the latter three cases this is likely driven by the poorer constraints on the photometric redshift estimate (i.e. broader probability distributions).

\vspace{2mm}

\subsection{Other redshift possibilities}

Even though we have selected millimeter spectra from locations that are consistent with high--redshift galaxies -- i.e., undetected in the optical or with an optical dropout counterpart -- it is likely that the majority of our candidate [CII] lines are either spurious (based on {\it fidelity}) or have a different identification than [CII]. The latter case could have two reasons: there could be a dust--obscured galaxy along the line of sight that is not visible by HST \citep[e.g. the case of HDF\,850.1, ][]{walter12}, or the high--redshift identification is not correct; e.g., an optical dropout candidate, selected based on the identification of the redshifted 1216\AA \ Lyman break, could also correspond to a lower redshift passive galaxy in which instead the redshifted 4000\AA \ Balmer break has been identified. We have thus used two important sources of information to reject other redshift possibilities and assess the reality of our [CII] line candidates, as described below. 

\vspace{2mm}
\begin{figure}[t]
\centering
\includegraphics[scale=0.5]{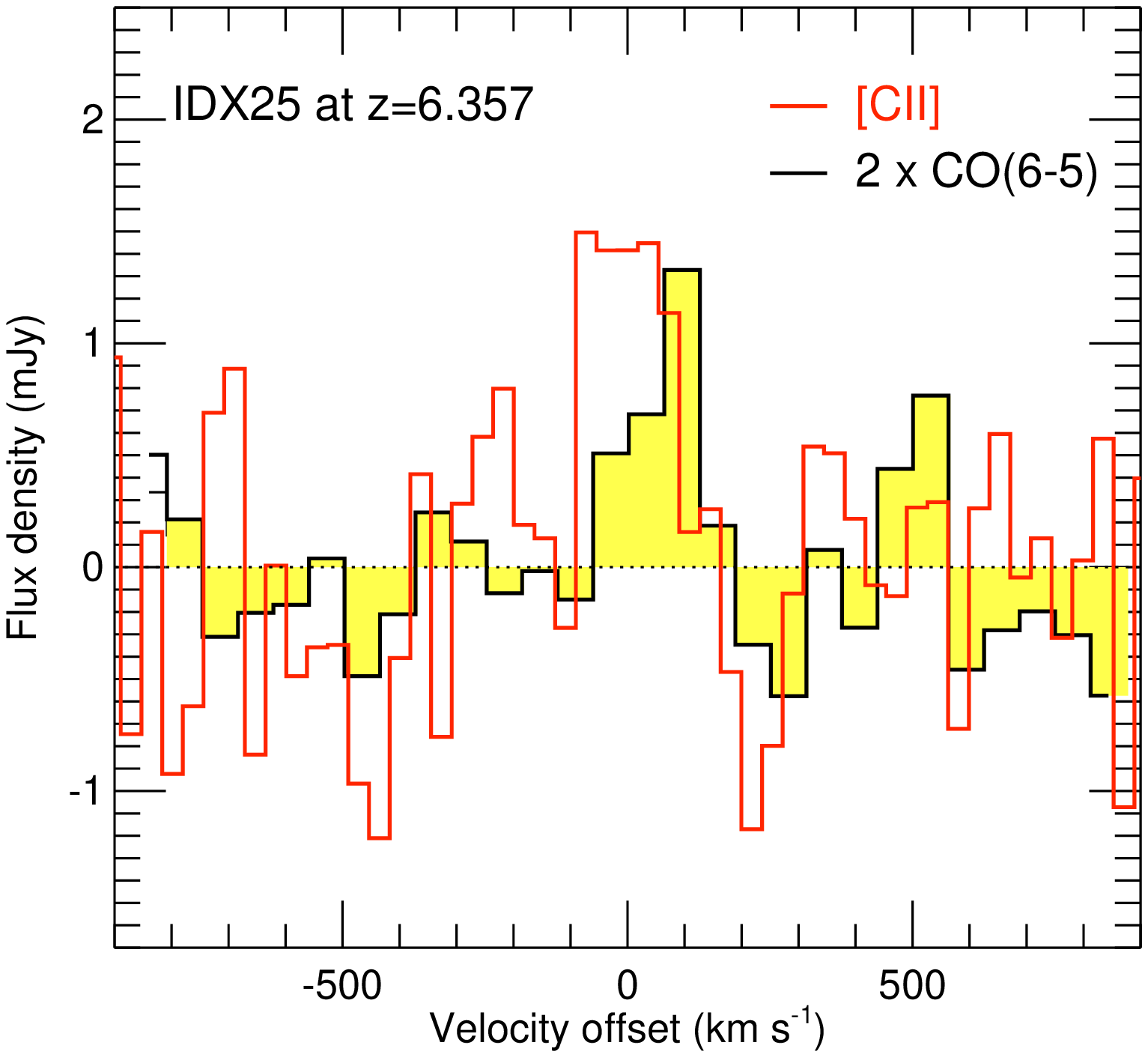}
\caption{ALMA spectra of two candidate millimeter emission lines towards the {\it blind} [CII] candidate IDX25. If we associate the 1-mm line peak to [CII] line emission at $z=6.357$, shown by the red histogram, then we can identify the positive spectral feature observed in the 3-mm spectrum with CO(6-5) line emission, which is shown scaled by a factor of 2 by the black solid line and filled histogram. No other combination of ISM line provides a viable redshift solution.}\label{fig:2lines}
\end{figure}

\subsubsection{MUSE complementary information}

We took advantage of the deep MUSE optical spectroscopy available in the region covered by our 1-mm observations (Bacon et al. in preparation) to assess the reality and redshift of our [CII] candidates. The wavelength coverage of MUSE allows us to cover the Ly-$\alpha$ line at redshifts below $6.6$ thus enabling us us to confirm any [CII] candidate in the range $z=6.0-6.6$. However, we note that it is not expected that a [CII] line emitter would necessarily have Ly-$\alpha$ line emission. Hence, the reality of a given [CII] candidate will not necessarily be subject to the existence of the Ly-$\alpha$ line and an optical redshift confirmation. Conversely,  it is possible to use the MUSE spectroscopic information to check whether the optical dropout corresponds to a lower redshift interloper.

We searched in the MUSE cube for possible optical emission lines at the location of the {\it blind} [CII] line candidates and at the position of the optical counterparts of the lower significance [CII] candidates. We find that no obvious optical lines showed up in the spectra extracted at the locations of the {\it blind} [CII] candidates (sources IDX25 and IDX34), and thus we did not obtain further information on the reality of these objects. Similarly, the MUSE data does not provide good redshift constraints for any of the optical dropout galaxies associated to the [CII] line candidates, based on the spectra extracted at these locations, except in the case of ID52 where MUSE confirms $z=2.379$. This implies that the line peak identified in the 1-mm cube is spurious since no ISM line (e.g. CO, [CI]) at the frequency of the peak matches the MUSE optical redshift.

Finally, in order to place limits on the [CII] line emission from other sources that might be below our detection threshold, we checked the MUSE cube at the location of any of the previously known optical dropouts within the region covered by our 1-mm observations. However, none of them could be confidently confirmed in the redshift range covered by our 1-mm observations for redshifted [CII] emission.

\subsubsection{ALMA 3-mm scan constraints}

For any redshifted FIR emission line detected in our 1-mm spectral scan, the 3-mm spectra provides complementary information on its reality and redshift, as for most redshift possibilities there will be a second line covered (albeit potentially at very low luminosity). Therefore, we can check if our candidate line detection in the 1-mm cube corresponds to a different line than [CII], or if we can indeed confirm the candidate [CII] line emission at high redshift. In particular, for the [CII] line emission at $6<z<8$, we will also cover either of the CO(6-5), CO(7-6) or CO(8-7) emission lines in our 3-mm survey. However, these lines are expected to be significantly fainter than [CII] in the normal star forming galaxies pursued in this study. Therefore, if no other lines are detected in the 3-mm cube, we conclude that the [CII] line identification remains the most plausible option.

We inspected the 3-mm data cube at the location of each of our candidate [CII] line emitters. Figures \ref{fig:zopt1} to \ref{fig:zopt5} (see Appendix) show the spectra of our emission line candidates placed at the different redshift options if the detected 1-mm line were to be identified with a different ISM line. In all plots, we show the spectra in the same scale, normalized to the luminosity $L'/\Delta v$ in units of 10$^7$ K pc$^2$. In most cases, the noise in the 3-mm spectra makes it hard to identify a secondary emission line. 

Only for the case of the {\it blind} [CII] line candidate IDX25, the 3-mm spectra shows a tentative second line corresponding to the CO(6-5) emission line. This would confirm the [CII] line emission, placing this blind line detection at $z=6.3568$. No other redshift possibility matches the unique combination of emission line peaks in the 1-mm and 3-mm data cubes. At the significance of the candidate [CII] detection ($\sim5.4\sigma$), the fidelity level reaches $\sim80\%$. However, as we pointed out in \S \ref{sec:blind}, the 1-mm line falls right at the location of two atmospheric features making this detection uncertain despite its high significance.

\subsection{Continuum emission: Individual sources and stack}

\begin{figure}
\centering
\includegraphics[scale=0.27]{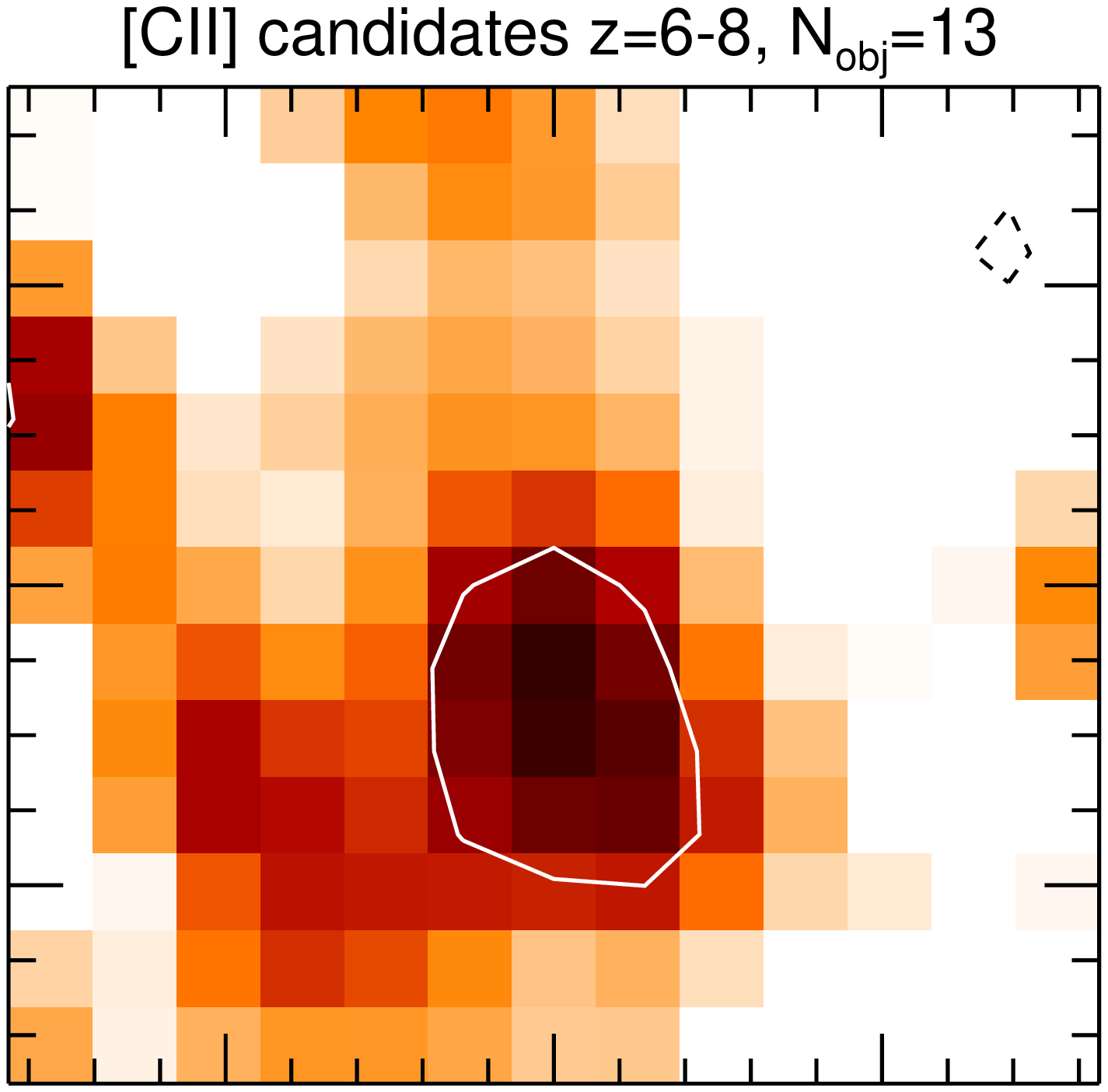}
\includegraphics[scale=0.27]{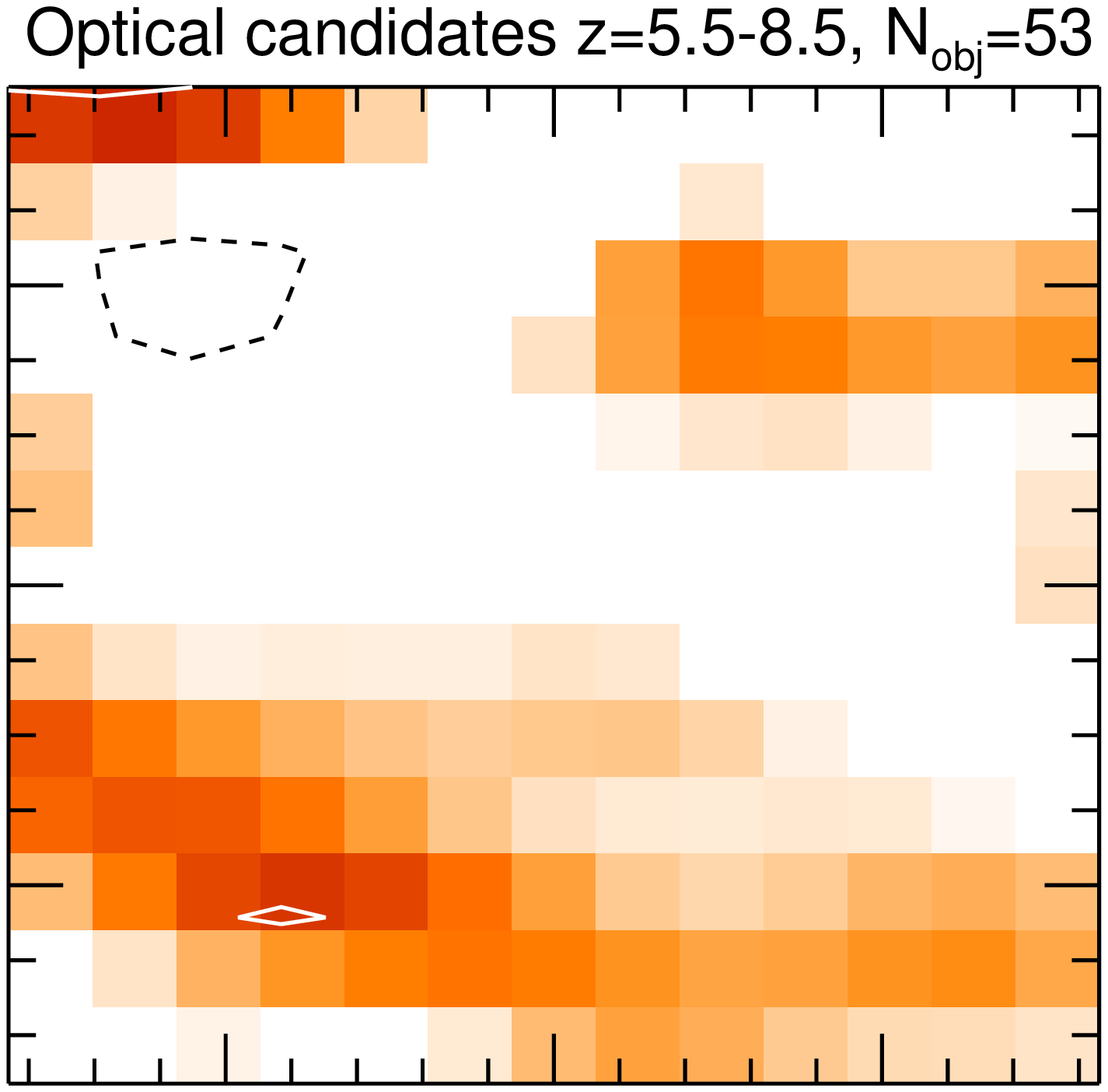}
\caption{({\it Left:}) Average 1.2-mm continuum emission obtained by stacking at the location of all the [CII] line candidates. ({\it Right:}) Average 1.2-mm continuum emission at the location of all the optical galaxies with photometric redshifts $z=5.5-8.5$ considered in this work. The images reach an rms levels of 5.0 and 2.0$\mu$Jy in each case, respectively. Contours levels are given at $-2$ and $2\sigma$. Positive and negative signal are shown in white solid  and black dashed contours, respectively. Stacked continuum emission from the [CII] candidates is tentatively detected with a 1.2-mm flux density of $14\pm5\,\mu$Jy. }\label{fig:stacking}
\end{figure}

The large frequency coverage of our ALMA observations in the 1-mm band resulted in a very deep collapsed continuum image reaching down to 12.7 $\mu$Jy in the central regions of the mosaic \citep[see ][; Paper~II]{aravena16b}. Despite the great depth of the continuum map, none of our sources show obvious continuum emission at the $3\sigma$ level, even though three sources are associated with $\sim2-3\sigma$ blobs in the 1.2-mm map (ID04, ID27 and ID31; see Table 2). Our 1.2-mm continuum map reaches a noise level of $\sim13-35$ $\mu$Jy over the area where our [CII] candidates are located (within PB$=0.5$). This implies an upper limit of $\sim40-100\mu$Jy ($3\sigma$) for the 1.2-mm continuum emission of individual objects (see Table 2). 

To reach deeper continuum levels, we measure the average emission by stacking the 1-mm continuum map at the location of all the optical dropouts covered by our 1.2-mm mosaic and also at the location of our [CII] line candidates. In the first case, we use the optical positions to guide the stacking, while in the second case we stack at the location of the [CII] peaks. Using the positions of the optical associations in the second case, when available, do not alter the results. For the stacking, we extract image cutouts centered at the location of our sources. From this, we compute a weighted average of the individual cutout images, using the sensitivity pattern of the mosaic as weight. In order to avoid contamination from bright sources, we explicitly avoid sources that are within $5''$ from the brightest 6 continuum sources detected in our 1.2-mm mosaic. 

The outcome of the stacking is shown in Fig. \ref{fig:stacking}. The stack of the [CII] line candidates yields a tentative detection at the $\sim3\sigma$ level with a measured peak flux density of $14\pm5\,\mu$Jy. The uncertainty here merely represents the noise in the stacked image. No detection is found in the stack of the optical dropout galaxies yielding a $3\sigma$ upper limit of 6$\mu$Jy at 1.2-mm. This might not be too surprising given that we expect about 40\% of the sample to be composed of real sources based on our purity estimates. However, the detection in the continuum does not provide information of the redshift of these sources or the reality of the [CII] line emission, since their continuum emission could be produced by sources at lower redshifts.

We use these continuum measurements to place a constraint on the far IR luminosity and SFR of our sources. We adopt modified black body dust model of the form $S_\nu\propto(1+z)D_{\rm L}^{-2} M_{\rm d} \kappa_{\nu} [B_\nu(T_{\rm d}) - B_\nu(T_{\rm BG})]$ that becomes optically thick at a wavelength of 100$\mu$m. Here,  $S_{\nu}$ represents the observed flux density, $z$ is the source redshift, $D_{\rm L}$ is the luminosity distance, $B_{\nu}$ is the Planck function and $T_{\rm BG}$ is the cosmic microwave background temperature at the source redshift. $M_{\rm d}$ and $T_{\rm d}$ are the dust mass and temperature, respectively. We consider a dust absorption coefficient of the form $\kappa_\nu=\kappa_0(\nu/\nu_0)^\beta$,  where we adopt a dust emissivity index $\beta=1.8$ \citep{kovacs06,magdis11} and $\kappa_0=0.4$ cm$^2$ gr$^{-1}$ at 250 GHz \citep{krugel94}. We assume a typical dust temperature $T_{\rm d}=50$ K at $z\sim6-8$, motivated by recent predictions for the evolution of the dust temperature as $T_{\rm d}=35\times((1+z)/2.5)^{0.32}$ K \citep[see ][; but also Bethermin et al. 2015]{bouwens16}.

For the redshift range $z=6-8$, the average 1.2-mm flux measurement on the [CII] candidate sample translates into a FIR luminosity ($40-500~\mu$m) $L_{\rm FIR}=(2.7\pm0.9)\times10^{10}\ L_\sun$. This luminosity value is almost insensitive to small variations on the emissivity index, with $\beta\sim1.5-1.8$, or variations in redshift as it varies little in the range $z=6-8$. For a \citet{chabrier03} initial mass function, this translates into an average IR-derived SFR of $3\pm1$ $M_\sun$ yr$^{-1}$. However, according to \citet{dacunha15}, the detectability of the dust continuum against the CMB would decrease drastically at lower dust temperatures. Assuming lower dust temperatures in our calculations, $T_{\rm d}=25$ K, yield FIR luminosity and SFR$_{\rm IR}$ estimates a factor $(2.5-3.0)$ lower. This suggests that the non-detection of most of the [CII] line candidates could be due to low intrinsic dust temperatures, as well as low SFR$_{\rm IR}$.

This tentative detection thus raises the question about the sources that should have been detected in 1.2-mm continuum emission based on their UV-based SFRs. This is the case for ID27 and ID31, which have SFR$_{\rm UV}\sim10$ and 12 $M_\sun$ yr$^{-1}$. In both cases, however, there is positive continuum peak within $1''$ from the [CII] candidate position. At the location of the 1.2-mm continuum peak position, we measure $S_{\rm 1.2mm}=34\pm12\,\mu$Jy ($2.8\sigma$) and $30\pm13\,\mu$Jy ($2.3\sigma$) for ID27 and ID31, respectively. Using the models presented above {\bf ($T_{\rm d}=50$ K)}, these fluxes yield IR-derived SFRs of $\sim8$ and $7$ $M_\sun$ yr$^{-1}$, respectively, comparable to the UV-derived SFR values.

The average IR-derived SFR estimate obtained for the [CII] candidate sample is comparable to its average UV--derived SFR of 2.5 $M_\sun$ yr$^{-1}$. On average for this sample, we find an IR excess IRX=SFR$_{\rm IR}$/SFR$_{\rm UV}\sim1$, suggesting that $\sim50\%$ of the emission related to star formation in these sources is obscured. For the two individual sources with marginal 1.2-mm continuum detections, ID27 and ID31, this ratio ranges between $\sim0.6-0.8$. Our measurements for the CII candidate sample are consistent with the results reported in \citet{bouwens16} using larger samples of star forming galaxies at $z=4-10$ \citep[see also][]{capak15}. Note that this also implies SFR$_{\rm tot}=$ SFR$_{\rm UV+IR}\sim2\times$SFR$_{\rm UV}$, and thus to recover the total SFR value for our galaxies, we need to apply an average correction of $\sim2$ to the UV-derived SFR (see \S 4.1).

\begin{table*}
\centering
\caption{Properties of the $z>6$ [CII] line candidates. Columns: (1) Source name; (2), (3) Right ascension and declination of the candidate [CII] line emitter; (4) Distance between the location of the [CII] emission and the nearest optical dropout galaxy at $z=5.5-8.5$; (5) Frequency of the candidate line emission; (6) Signal to noise ratio of the candidate line emission; (7) Spectroscopic redshift based on the assumption the identified line is [CII]; (8) Photometric redshift; (9) Integrated line flux; (10) [CII] line luminosity; (11) UV-derived SFR; (12) Flux density at 1.2-mm. Limits are given at the $3\sigma$ level. }
\begin{tabular}{cccccccccccc}
\hline\hline
 Source name & RA$_{\rm [CII]}$ & Dec$_{\rm [CII]}$ & Dist$_{\rm opt}$ & $\nu$ & SNR & z$_{\rm [CII]}$ & $z_{\rm phot}$ & $I_{\rm line}$ & $L_{\rm [CII]}$ & SFR$_{\rm UV}$ & $S_{\rm 1.2mm}$ \\
   & (J2000) & (J2000) & ($"$) & (GHz) &  &  &  & (Jy km s$^{-1}$) & ($10^8\ L_\sun$) & ($M_\sun$ yr$^{-1}$) & ($\mu$Jy) \\
 (1)  & (2) & (3) & (4) & (5)  & (6) & (7) & (8) & (9)  & (10) & (11) & (12) \\
\hline
IDX25 &  3:32:37.36 & -27:46:10.0 & \ldots &  258.337 &    5.4 &   6.357 &  \ldots & $ 0.25\pm 0.06$ & $ 2.6\pm 0.6$ & \ldots & $<40$ \\
IDX34 &  3:32:35.75 & -27:46:36.7 & \ldots &  223.837 &    5.3 &   7.491 &  \ldots & $ 0.72\pm 0.08$ & $ 9.3\pm 1.0$ & \ldots & $<100$ \\ 
\hline
ID02 &  3:32:39.37 & -27:46:11.2 &    0.3 &  213.212 &    5.3 &   7.914 & 5.7 & $ 0.65\pm 0.13$ & $ 9.2\pm 1.8$ &  0.6 & $<55$ \\ 
ID04$^\dagger$ &  3:32:39.53 & -27:46:49.6 &    0.6 &  241.587 &    4.6 &   6.867 & 6.1 & $ 0.81\pm 0.09$ & $ 9.2\pm 1.1$ &  0.4 & $82\pm33$  \\ 
ID09 &  3:32:38.76 & -27:46:34.6 &    0.6 &  270.556 &    4.7 &   6.024 & 5.9 & $ 0.31\pm 0.07$ & $ 3.0\pm 0.7$ &  0.3 & $<42$   \\ 
ID14 &  3:32:38.49 & -27:46:20.8 &    0.9 &  245.212 &    4.7 &   6.751 & 6.2 & $ 0.28\pm 0.06$ & $ 3.1\pm 0.7$ &  0.7 & $<42$   \\ 
ID27 &  3:32:37.40 & -27:46:32.5 &    0.9 &  221.650 &    4.7 &   7.575 & 6.5 & $ 0.22\pm 0.05$ & $ 2.9\pm 0.7$ & 10.5 & $34\pm12 $  \\ 
ID30 &  3:32:36.97 & -27:45:57.1 &    0.8 &  241.993 &    4.6 &   6.854 & 6.0 & $ 0.61\pm 0.09$ & $ 7.0\pm 1.1$ &  4.0 & $<65$ \\ 
ID31 &  3:32:38.31 & -27:46:18.1 &    0.7 &  223.743 &    4.5 &   7.494 & 6.1 & $ 0.25\pm 0.06$ & $ 3.3\pm 0.8$ & 12.4 & $30\pm13 $ \\ 
ID38 &  3:32:38.22 & -27:46:16.9 &    0.9 &  250.306 &    4.7 &   6.593 & 6.1 & $ 0.31\pm 0.08$ & $ 3.3\pm 0.8$ &  0.2 & $<42 $  \\ 
ID41 &  3:32:38.19 & -27:46:04.6 &    0.9 &  258.712 &    4.5 &   6.346 & 6.2 & $ 0.38\pm 0.09$ & $ 3.9\pm 0.9$ &  0.4 & $<44$  \\ 
ID44 &  3:32:37.34 & -27:46:25.3 &    0.5 &  227.337 &    5.2 &   7.360 & 6.3 & $ 0.35\pm 0.09$ & $ 4.4\pm 1.1$ &  1.2 & $<40$  \\ 
ID49 &  3:32:36.63 & -27:46:22.9 &    0.7 &  269.556 &    4.8 &   6.051 & 7.7 & $ 0.25\pm 0.06$ & $ 2.4\pm 0.6$ &  0.1 & $<40 $ \\ 
ID52 &  3:32:36.86 & -27:46:52.6 &    0.9 &  270.806 &    4.6 &   6.018 & 6.8 & $ 0.43\pm 0.06$ & $ 4.0\pm 0.6$ &  0.1 & $<70 $ \\ 
\hline\hline
\end{tabular}
\\
\noindent $^\dagger$ This source lies very close to the edge of the mosaic, being likely that the continuum peak measured at the $2.5\sigma$ level is spurious.
\label{table:properties}
\end{table*}

\section{Analysis and discussion}\label{sect:discussion}

\subsection{$L_{\rm [CII]}$ vs SFR}

Based on observations of local galaxies, it has recently been suggested that a scaling relation between the SFR and the [CII] luminosity of galaxies of different types exist \citep{delooze11, delooze14, sargsyan12, cormier15, vallini15, olsen15,herreracamus15}. \citet{delooze14} argued that galaxies with lower metallicities and higher dust temperatures, as expected for high-redshift galaxies, would tend to have larger scatter in this relationship, with possibly increasing [CII] luminosities for a given SFR value. However, recent [CII] line observations of a sample of dwarf galaxies show that low metallicity galaxies do not have systematically higher [CII] to IR luminosity ratios \citep{cormier15}. It is thus interesting to examine the location of our [CII] line candidates in the SFR vs.\ [CII] luminosity plot (SFR--$L_{\rm [CII]}$).

\begin{figure}[t]
\centering
\includegraphics[scale=0.4]{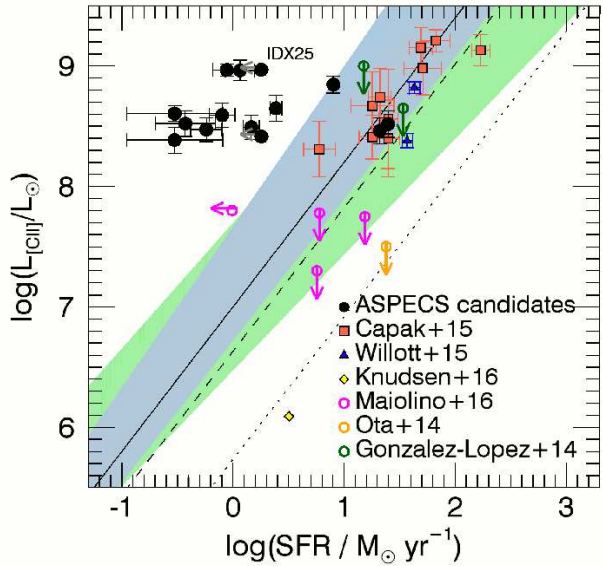}
\caption{$L_{\rm [CII]}$ vs.\ SFR relationship for our [CII] line candidates (black circles) compared to high-redshift galaxies from the literature . The green shaded area represents the relation (including dispersion) for local star forming galaxies and starbursts found by \citet{delooze14}. The blue shaded area shows the relation found for local low metallicity dwarfs and irregular galaxies \citep[also by][]{delooze14}. The solid, dashed and dotted liness represent the models obtained by \citet{vallini15} for galaxies with 1, 0.2 and 0.05 solar metallicity, respectively. The `horizontal' location of our line candidates in this plot is due to our sensitivity limit. We note that the statistical expectation is that 60\% of our line candidates are not real. \label{fig:sfr_cii}} 
\end{figure}

Figure~\ref{fig:sfr_cii} presents the [CII] luminosity and SFR for our [CII] line candidates in the UDF, compared to previous [CII] line detections of non--quasar galaxies at $z>5$ \citep{gonzalezlopez14,ota14,capak15,willott15,maiolino15,knudsen16}. In this plot, we also compare our observations with the relationships found for local galaxies with different environments and metallicities \citep[e.g.,][]{delooze14} and simulations of high-redshift galaxies \citep{vallini15}. Most of these studies, including the recent [CII] detections at $z>5$ quote total SFRs, including both obscured and unobscured star formation components. Since the IR--derived SFR appears to be comparable to the UV-derived SFR in our [CII] line candidate sample, SFR$_{\rm IR}$/SFR$_{\rm UV}\sim1$ or SFR$_{\rm tot}\sim 2\times$SFR$_{\rm UV}$ (see \S 3.7), we simply apply a factor of 2 correction to each individual UV--derived SFR value in our sample. This effectively translates into a shift of 0.3 dex to higher SFRs in Fig.~\ref{fig:sfr_cii}.

Two important things can be extracted from Fig.~\ref{fig:sfr_cii}. First, most of the high--redshift galaxies with previous [CII] detections seem to agree, at least to first order, with the local calibrations for SFR--$L_{\rm [CII]}$. Secondly, only three of our candidates are located in this region of the plot, with the rest having too low SFRs. These sources would need to have obscured SFRs a factor $>10$ larger than their SFR$_{\rm UV}$ to be placed in the local SFR--[CII] relationship. However, if these sources had SFR$_{\rm IR}\sim5-10$ M$_\sun$ yr$^{-1}$ -- implying they are highly obscured as most of their emission would be produced in the IR -- they would still be at the limit of what can be individually detected in our 1.2-mm continuum map, assuming $T_{\rm d}\sim50$ K, similar to what is observed in ID27 and ID31 discussed in \S 3.7.

It would therefore be tempting to simply discard the [CII] line candidates that have SFR$_{\rm UV}<5-10\ M_\sun$ yr$^{-1}$. However, the possibility that low SFR galaxies could indeed have significant [CII] line emission has recently been suggested by the ALMA detection of [CII] line emission in the vicinity of the normal star forming galaxy BDF-3299 at $z=7.1$, shown as an SFR upper limit in Fig. \ref{fig:sfr_cii} \citep{maiolino16}.

Furthermore, and along the same line, most of the galaxies so far detected in [CII] emission at $z>5$ have been pre-selected based on their SFR, a strong Ly-$\alpha$ line or high-equivalent width, with typical SFR values above 25 $M_\sun$ yr$^{-1}$. However, our sample is mostly composed of galaxies with SFR$<10$ $M_\sun$ yr$^{-1}$ selected only based on the existence of a Lyman break (Sec.~\ref{sec_ancillary}). Thus, it is possible that the physical properties of the different samples could differ significantly (e.g. metallicities). In any case, we do not claim that all of the lower--SFR [CII] candidates are real. In fact, based on statistical grounds, we expect $\sim$60\% of them to be not real lines, as discussed in Sec.~\ref{sec_stats}.

\subsection{Constraints on luminosity function}

The number of [CII] line candidates in our ALMA survey of the UDF allow us to investigate the [CII] luminosity function in the redshift range $6<z<8$. Since all of our sources are just candidates, all the numbers derived here in the following effectively correspond to upper limits until we are able to confirm at least one of of these objects. In the future this can be done by either deeper integrations to confirm the candidate lines and/or by securing their redshift by detecting a second emission line.

We compute the number counts in two ways. The first assumes that only the brightest [CII] line candidate (IDX04) is real, and the second uses all the line candidates correcting for purity and completeness at the significance level of the sample at $S/N=4.5$, using the values computed in \S \ref{sec_stats}. We thus measure the [CII] luminosity function by counting the number of [CII] line candidates within the cosmic comoving volume spanned between $z=6$ to $z=8$. For the $\sim$1\,arcmin$^2$ area of the sky covered by our survey this volume corresponds to 4678 Mpc$^3$. 

\begin{figure}[t]
\centering
\includegraphics[scale=0.4]{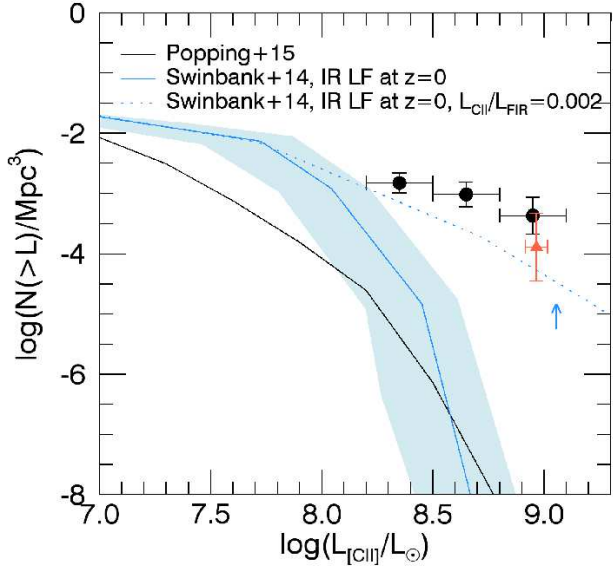}
\caption{Observational constraints on the number density of [CII] line emitters at $6<z<8$ ([CII] luminosity function) from ASPECS. Since all of our targets are only candidates so far, we are only able to put upper limits to the bright end of the [CII] luminosity function. The red triangle represents the number density assuming that only the brightest candidate is real. The black circles assumes the unlikely scenario that all candidates are real. The blue upward arrow is the measurement by \citet{swinbank14} at z=4.4. The blue solid line and shaded area corresponds to the local [CII] luminosity function derived by \citet{swinbank14} based on the local IR luminosity function and the variation of the [CII]/IR line ratio with IR luminosity. The blue dotted curve represents the [CII] luminosity function derived from local IR luminosity function but assuming a constant (average) [CII]/IR luminosity ratio of 0.002 \citep[following ][]{swinbank14}. The black solid line represents the recent predictions by \citet{popping16}.}\label{fig:lumfunc}
\end{figure}

Figure \ref{fig:lumfunc} shows the observed [CII] luminosity function at $6<z<8$ in the UDF. We compare our measurements with the recent constraints obtained by \citet{swinbank14} based on the serendipitous detection of two [CII] line emitters at $z=4.4$. We also compare our measurements with the expected [CII] luminosity function for local galaxies \citep{swinbank14}. The latter uses the local IR luminosity function \citep{sanders03} and extrapolates it using either the apparent variation of the [CII]/IR luminosity ratio with IR luminosity \citep{brauher08}, or a fixed ratio of [CII]/IR$=0.002$. We also compare to recent predictions for the [CII] luminosity function at $z=6-8$ by \citet{popping16}. 

Our observations show a disagreement of more than an order of magnitude on the number counts with respect to recent models \citep[e.g.][]{popping16}. They tend to agree with the IR-derived local [CII] luminosity function for a fixed [CII]/IR ratio as well as a rough agreement with \citet{swinbank14} measurements based on the two bright submillimeter galaxies (lower limit). If at least one of our [CII] line candidates is confirmed, this would point to a less abrupt fall on the [CII] number counts at high redshift than predicted by current models.

\section{Conclusions}\label{sect:conclusion}

We have used our molecular ALMA survey of the {\it Hubble} UDF to search for [CII] line candidates in the redshift range $6.0<z<8.0$. We do this by blindly searching for significant line peaks in the 1-mm data cube, and by specifically looking around 58 known dropout galaxies with photometric redshifts that are consistent with the redshift range covered by [CII] ($5.5<z<8.5$). Our survey field within the UDF was chosen such that the number of known drop--out sources was maximised. The spectroscopic survey reaches an approximately uniform depth of L$_{\rm [CII]}\sim$(1.6-2.5)$\times$10$^{8}$\,L$_{\odot}$ at an angular resolution of $\sim$\,1$"$, well matched to the expected size of the galaxies at that redshift.

We discuss our statistical tools to search for line emission in the vicinity of the drop--out galaxies. These include assessing the {\it fidelity} (fraction of positive vs.\ negative emission line candidates) as well as the {\it completeness} (our ability to recover artificial objects) of the line search. We find that there are more positive line candidates than (physically implausible) negative ones, assuring us that we are recovering actual signal in our spectroscopic survey at the positions of interest. We end up with 14 [CII] line candidates above a S/N cutoff of S/N$>$4.5. All of the candidate lines are spatially unresolved, with implied radii of $<0.5"$ ($<$3\,kpc). Most of them are located away from the edges of individual spectral windows; the latter could lead to spurious sources due to increased noise. 

None of the candidates are detected in the 1\,mm continuum, which is consistent with the recent finding by \citet{capak15} that [CII] lines are more easily detected in z$\sim$5 Lyman--break galaxies than the underlying dust continuum. When stacking all high--redshift drop--out galaxies, we derive a 3$\sigma$ upper limit for the continuum emission of 6\,$\mu$Jy\,beam$^{-1}$. When stacking only at the location of the [CII] line candidates, we find a tentative detection of the dust continuum with a flux of $14\pm5\ \mu$Jy. This implies a dust--obscured star formation rate of 3\,M$_\odot$\,yr$^{-1}$.

We compare the [CII] luminosities of our line candidates with the UV--based star formation rates and compare these with relations that have recently been discussed in the literature. These include galaxies and starbursts, as well as lower--metallicity dwarf/irregular galaxies. We find that the three highest--SFR objects have candidate [CII] lines with luminosities that are consistent with the low--redshift $L_{\rm [CII]}$ vs.\ UV-derived SFR relation. The other candidates have significantly higher [CII] luminosities. Given our {\it fidelity} analysis, we expect that $60\%$ of these sources to be spurious. A possible conclusion would be that some of the sources have elevated [CII] fluxes compared to expectations based on individual SFRs. Similar such claims have recently been reported in the literature \citep{maiolino16}. Future, deeper observations with ALMA will shed light on this issue. We note that confirming the objects of interest will request significantly less time with ALMA than the original survey, as no mosaic, nor frequency scans, would be required. Approved, deeper ALMA cycle~3 data of the same field will also improve the reliability of some of the candidates presented here.

Based on the available information, we put first constraints on the [CII] luminosity function at $z\sim6-8$. Even if only one of our line candidates was real (a scenario greatly favoured by our statistical analysis) we find a source density that is consistent with the value derived by \citet{swinbank14} based on blindly detected [CII] emitters at higher luminosities at z$\sim$4.4. However these numbers are in conflict with a local (z=0) [CII] luminosity function derived by \citet{swinbank14} assuming a varying [CII]/IR luminosity ratio with IR luminosity. They are consistent though with a [CII] luminosity function that assumes a constant [CII]/IR$=0.002$ luminosity ratio \citep{swinbank14}. On the other hand, the high--redshift constraints so far appear to give significantly higher number densities than the recent models by \citet{popping16}.

The observations presented here demonstrate that even in ALMA early science, [CII] luminosities can be reached that enable the studies of some of the faintest HST drop--out galaxies at $6.0<z<8.0$. Future, deeper and wider surveys with ALMA will be needed to improve the significance of the detections, and to improve the overall number statistics. With the fully completed ALMA now available, these goals now appear to be within reach. The full UDF appears to be the best field choice for such a survey, as the highest--redshift galaxy population is best--characterised in this field. In the near future, optical/NIR spectroscopy from JWST Guaranteed Time efforts will also provide accurate redshifts for the highest redshift galaxies in the UDF that would eventually enable pushing [CII] line studies to unprecedented depths through stacking.

\bibliographystyle{apj}
\bibliography{alma_udf_cii_MA_v2}

\begin{thebibliography}{}
\expandafter\ifx\csname natexlab\endcsname\relax\def\natexlab#1{#1}\fi

\bibitem[{{Aravena} {et~al.}(2016{\natexlab{a}}){Aravena}, {Spilker},
  {Bethermin}, {Bothwell}, {Chapman}, {de Breuck}, {Furstenau},
  {G{\'o}nzalez-L{\'o}pez}, {Greve}, {Litke}, {Ma}, {Malkan}, {Marrone},
  {Murphy}, {Stark}, {Strandet}, {Vieira}, {Weiss}, {Welikala}, {Wong}, \&
  {Collier}}]{aravena16}
{Aravena}, M., {Spilker}, J.~S., {Bethermin}, M., {et~al.} 2016{\natexlab{a}},
  \mnras, 457, 4406

\bibitem[{{Aravena} {et~al.}(2016{\natexlab{b}}){Aravena}, {Decarli}, {Walter},
  {Da Cunha}, {Bauer}, {Carilli}, {Daddi}, {Elbaz}, {Ivison}, {Riechers},
  {Smail}, {Swinbank}, {Weiss}, {Anguita}, {Assef}, {Bell}, {Bertoldi},
  {Bacon}, {Bouwens}, {Cortes}, {Cox}, {G{\'o}nzalez-L{\'o}pez}, {Hodge},
  {Ibar}, {Inami}, {Infante}, {Karim}, {Le F{\`e}vre}, {Magnelli}, {Ota},
  {Popping}, {Sheth}, {van der Werf}, \& {Wagg}}]{aravena16b}
{Aravena}, M., {Decarli}, R., {Walter}, F., {et~al.} 2016{\natexlab{b}}, ArXiv
  e-prints, arXiv:1607.06769

\bibitem[{{Beckwith} {et~al.}(2006){Beckwith}, {Stiavelli}, {Koekemoer},
  {Caldwell}, {Ferguson}, {Hook}, {Lucas}, {Bergeron}, {Corbin}, {Jogee},
  {Panagia}, {Robberto}, {Royle}, {Somerville}, \& {Sosey}}]{beckwith06}
{Beckwith}, S.~V.~W., {Stiavelli}, M., {Koekemoer}, A.~M., {et~al.} 2006, \aj,
  132, 1729

\bibitem[{{Bouwens} {et~al.}(2016){Bouwens}, {Aravena}, {Decarli}, {Walter},
  {da Cunha}, {Labbe}, {Bauer}, {Bertoldi}, {Carilli}, {Chapman}, {Daddi},
  {Hodge}, {Ivison}, {Karim}, {Le Fevre}, {Magnelli}, {Ota}, {Riechers},
  {Smail}, {van der Werf}, {Weiss}, {Cox}, {Elbaz}, {Gonzalez-Lopez},
  {Infante}, {Oesch}, {Wagg}, \& {Wilkins}}]{bouwens16}
{Bouwens}, R., {Aravena}, M., {Decarli}, R., {et~al.} 2016, ArXiv e-prints,
  arXiv:1606.05280

\bibitem[{{Bouwens} {et~al.}(2014){Bouwens}, {Bradley}, {Zitrin}, {Coe},
  {Franx}, {Zheng}, {Smit}, {Host}, {Postman}, {Moustakas}, {Labb{\'e}},
  {Carrasco}, {Molino}, {Donahue}, {Kelson}, {Meneghetti}, {Ben{\'{\i}}tez},
  {Lemze}, {Umetsu}, {Broadhurst}, {Moustakas}, {Rosati}, {Jouvel},
  {Bartelmann}, {Ford}, {Graves}, {Grillo}, {Infante}, {Jimenez-Teja}, {Lahav},
  {Maoz}, {Medezinski}, {Melchior}, {Merten}, {Nonino}, {Ogaz}, \&
  {Seitz}}]{bouwens14}
{Bouwens}, R.~J., {Bradley}, L., {Zitrin}, A., {et~al.} 2014, \apj, 795, 126

\bibitem[{{Bouwens} {et~al.}(2015){Bouwens}, {Illingworth}, {Oesch}, {Trenti},
  {Labb{\'e}}, {Bradley}, {Carollo}, {van Dokkum}, {Gonzalez}, {Holwerda},
  {Franx}, {Spitler}, {Smit}, \& {Magee}}]{bouwens15}
{Bouwens}, R.~J., {Illingworth}, G.~D., {Oesch}, P.~A., {et~al.} 2015, \apj,
  803, 34

\bibitem[{{Brammer} {et~al.}(2008){Brammer}, {van Dokkum}, \&
  {Coppi}}]{brammer08}
{Brammer}, G.~B., {van Dokkum}, P.~G., \& {Coppi}, P. 2008, \apj, 686, 1503

\bibitem[{{Brauher} {et~al.}(2008){Brauher}, {Dale}, \& {Helou}}]{brauher08}
{Brauher}, J.~R., {Dale}, D.~A., \& {Helou}, G. 2008, \apjs, 178, 280

\bibitem[{{Capak} {et~al.}(2015){Capak}, {Carilli}, {Jones}, {Casey},
  {Riechers}, {Sheth}, {Carollo}, {Ilbert}, {Karim}, {Lefevre}, {Lilly},
  {Scoville}, {Smolcic}, \& {Yan}}]{capak15}
{Capak}, P.~L., {Carilli}, C., {Jones}, G., {et~al.} 2015, \nat, 522, 455

\bibitem[{{Carilli} \& {Walter}(2013)}]{carilli13}
{Carilli}, C.~L., \& {Walter}, F. 2013, \araa, 51, 105

\bibitem[{{Chabrier}(2003)}]{chabrier03}
{Chabrier}, G. 2003, \pasp, 115, 763

\bibitem[{{Coe} {et~al.}(2006){Coe}, {Ben{\'{\i}}tez}, {S{\'a}nchez}, {Jee},
  {Bouwens}, \& {Ford}}]{coe06}
{Coe}, D., {Ben{\'{\i}}tez}, N., {S{\'a}nchez}, S.~F., {et~al.} 2006, \aj, 132,
  926

\bibitem[{{Cormier} {et~al.}(2015){Cormier}, {Madden}, {Lebouteiller}, {Abel},
  {Hony}, {Galliano}, {R{\'e}my-Ruyer}, {Bigiel}, {Baes}, {Boselli},
  {Chevance}, {Cooray}, {De Looze}, {Doublier}, {Galametz}, {Hughes},
  {Karczewski}, {Lee}, {Lu}, \& {Spinoglio}}]{cormier15}
{Cormier}, D., {Madden}, S.~C., {Lebouteiller}, V., {et~al.} 2015, \aap, 578,
  A53

\bibitem[{{da Cunha} {et~al.}(2013){da Cunha}, {Groves}, {Walter}, {Decarli},
  {Weiss}, {Bertoldi}, {Carilli}, {Daddi}, {Elbaz}, {Ivison}, {Maiolino},
  {Riechers}, {Rix}, {Sargent}, \& {Smail}}]{dacunha13}
{da Cunha}, E., {Groves}, B., {Walter}, F., {et~al.} 2013, \apj, 766, 13

\bibitem[{{da Cunha} {et~al.}(2015){da Cunha}, {Walter}, {Smail}, {Swinbank},
  {Simpson}, {Decarli}, {Hodge}, {Weiss}, {van der Werf}, {Bertoldi},
  {Chapman}, {Cox}, {Danielson}, {Dannerbauer}, {Greve}, {Ivison}, {Karim}, \&
  {Thomson}}]{dacunha15}
{da Cunha}, E., {Walter}, F., {Smail}, I.~R., {et~al.} 2015, \apj, 806, 110

\bibitem[{{De Looze} {et~al.}(2011){De Looze}, {Baes}, {Bendo}, {Cortese}, \&
  {Fritz}}]{delooze11}
{De Looze}, I., {Baes}, M., {Bendo}, G.~J., {Cortese}, L., \& {Fritz}, J. 2011,
  \mnras, 416, 2712

\bibitem[{{De Looze} {et~al.}(2014){De Looze}, {Cormier}, {Lebouteiller},
  {Madden}, {Baes}, {Bendo}, {Boquien}, {Boselli}, {Clements}, {Cortese},
  {Cooray}, {Galametz}, {Galliano}, {Gracia-Carpio}, {Isaak}, {Karczewski},
  {Parkin}, {Pellegrini}, {Remy-Ruyer}, {Spinoglio}, {Smith}, {Sturm}, \&
  {Wilson}}]{delooze14}
{De Looze}, I., {Cormier}, D., {Lebouteiller}, V., {et~al.} 2014, ArXiv
  e-prints, arXiv:1402.4075

\bibitem[{{Gonz{\'a}lez-L{\'o}pez} {et~al.}(2014){Gonz{\'a}lez-L{\'o}pez},
  {Riechers}, {Decarli}, {Walter}, {Vallini}, {Neri}, {Bertoldi}, {Bolatto},
  {Carilli}, {Cox}, {da Cunha}, {Ferrara}, {Gallerani}, \&
  {Infante}}]{gonzalezlopez14}
{Gonz{\'a}lez-L{\'o}pez}, J., {Riechers}, D.~A., {Decarli}, R., {et~al.} 2014,
  \apj, 784, 99

\bibitem[{{Goto} \& {Toft}(2015)}]{goto15}
{Goto}, T., \& {Toft}, S. 2015, \aap, 579, A17

\bibitem[{{Gullberg} {et~al.}(2015){Gullberg}, {De Breuck}, {Vieira},
  {Wei{\ss}}, {Aguirre}, {Aravena}, {B{\'e}thermin}, {Bradford}, {Bothwell},
  {Carlstrom}, {Chapman}, {Fassnacht}, {Gonzalez}, {Greve}, {Hezaveh},
  {Holzapfel}, {Husband}, {Ma}, {Malkan}, {Marrone}, {Menten}, {Murphy},
  {Reichardt}, {Spilker}, {Stark}, {Strandet}, \& {Welikala}}]{gullberg15}
{Gullberg}, B., {De Breuck}, C., {Vieira}, J.~D., {et~al.} 2015, \mnras, 449,
  2883

\bibitem[{{Herrera-Camus} {et~al.}(2015){Herrera-Camus}, {Bolatto}, {Wolfire},
  {Smith}, {Croxall}, {Kennicutt}, {Calzetti}, {Helou}, {Walter}, {Leroy},
  {Draine}, {Brandl}, {Armus}, {Sandstrom}, {Dale}, {Aniano}, {Meidt},
  {Boquien}, {Hunt}, {Galametz}, {Tabatabaei}, {Murphy}, {Appleton}, {Roussel},
  {Engelbracht}, \& {Beirao}}]{herreracamus15}
{Herrera-Camus}, R., {Bolatto}, A.~D., {Wolfire}, M.~G., {et~al.} 2015, \apj,
  800, 1

\bibitem[{{Illingworth} {et~al.}(2013){Illingworth}, {Magee}, {Oesch},
  {Bouwens}, {Labb{\'e}}, {Stiavelli}, {van Dokkum}, {Franx}, {Trenti},
  {Carollo}, \& {Gonzalez}}]{illingworth13}
{Illingworth}, G.~D., {Magee}, D., {Oesch}, P.~A., {et~al.} 2013, \apjs, 209, 6

\bibitem[{{Knudsen} {et~al.}(2016){Knudsen}, {Richard}, {Kneib}, {Jauzac},
  {Clement}, {Drouart}, {Egami}, \& {Lindroos}}]{knudsen16}
{Knudsen}, K.~K., {Richard}, J., {Kneib}, J.-P., {et~al.} 2016, ArXiv e-prints,
  arXiv:1603.02277

\bibitem[{{Kotulla} {et~al.}(2009){Kotulla}, {Fritze}, {Weilbacher}, \&
  {Anders}}]{kotulla09}
{Kotulla}, R., {Fritze}, U., {Weilbacher}, P., \& {Anders}, P. 2009, \mnras,
  396, 462

\bibitem[{{Kov{\'a}cs} {et~al.}(2006){Kov{\'a}cs}, {Chapman}, {Dowell},
  {Blain}, {Ivison}, {Smail}, \& {Phillips}}]{kovacs06}
{Kov{\'a}cs}, A., {Chapman}, S.~C., {Dowell}, C.~D., {et~al.} 2006, \apj, 650,
  592

\bibitem[{{Kruegel} \& {Siebenmorgen}(1994)}]{krugel94}
{Kruegel}, E., \& {Siebenmorgen}, R. 1994, \aap, 288, 929

\bibitem[{{Magdis} {et~al.}(2011){Magdis}, {Daddi}, {Elbaz}, {Sargent},
  {Dickinson}, {Dannerbauer}, {Aussel}, {Walter}, {Hwang}, {Charmandaris},
  {Hodge}, {Riechers}, {Rigopoulou}, {Carilli}, {Pannella}, {Mullaney},
  {Leiton}, \& {Scott}}]{magdis11}
{Magdis}, G.~E., {Daddi}, E., {Elbaz}, D., {et~al.} 2011, \apjl, 740, L15

\bibitem[{{Maiolino} {et~al.}(2005){Maiolino}, {Cox}, {Caselli}, {Beelen},
  {Bertoldi}, {Carilli}, {Kaufman}, {Menten}, {Nagao}, {Omont}, {Wei{\ss}},
  {Walmsley}, \& {Walter}}]{maiolino05}
{Maiolino}, R., {Cox}, P., {Caselli}, P., {et~al.} 2005, \aap, 440, L51

\bibitem[{{Maiolino} {et~al.}(2015{\natexlab{a}}){Maiolino}, {Carniani},
  {Fontana}, {Vallini}, {Pentericci}, {Ferrara}, {Vanzella}, {Grazian},
  {Gallerani}, {Castellano}, {Cristiani}, {Brammer}, {Santini}, {Wagg}, \&
  {Williams}}]{maiolino16}
{Maiolino}, R., {Carniani}, S., {Fontana}, A., {et~al.} 2015{\natexlab{a}},
  \mnras, 452, 54

\bibitem[{{Maiolino} {et~al.}(2015{\natexlab{b}}){Maiolino}, {Carniani},
  {Fontana}, {Vallini}, {Pentericci}, {Ferrara}, {Vanzella}, {Grazian},
  {Gallerani}, {Castellano}, {Cristiani}, {Brammer}, {Santini}, {Wagg}, \&
  {Williams}}]{maiolino15}
---. 2015{\natexlab{b}}, \mnras, 452, 54

\bibitem[{{McLure} {et~al.}(2011){McLure}, {Dunlop}, {de Ravel}, {Cirasuolo},
  {Ellis}, {Schenker}, {Robertson}, {Koekemoer}, {Stark}, \&
  {Bowler}}]{mclure11}
{McLure}, R.~J., {Dunlop}, J.~S., {de Ravel}, L., {et~al.} 2011, \mnras, 418,
  2074

\bibitem[{{McLure} {et~al.}(2013){McLure}, {Dunlop}, {Bowler}, {Curtis-Lake},
  {Schenker}, {Ellis}, {Robertson}, {Koekemoer}, {Rogers}, {Ono}, {Ouchi},
  {Charlot}, {Wild}, {Stark}, {Furlanetto}, {Cirasuolo}, \&
  {Targett}}]{mclure13}
{McLure}, R.~J., {Dunlop}, J.~S., {Bowler}, R.~A.~A., {et~al.} 2013, \mnras,
  432, 2696

\bibitem[{{Oesch} {et~al.}(2015){Oesch}, {van Dokkum}, {Illingworth},
  {Bouwens}, {Momcheva}, {Holden}, {Roberts-Borsani}, {Smit}, {Franx},
  {Labb{\'e}}, {Gonz{\'a}lez}, \& {Magee}}]{oesch15}
{Oesch}, P.~A., {van Dokkum}, P.~G., {Illingworth}, G.~D., {et~al.} 2015,
  \apjl, 804, L30

\bibitem[{{Olsen} {et~al.}(2015){Olsen}, {Greve}, {Narayanan}, {Thompson},
  {Toft}, \& {Brinch}}]{olsen15}
{Olsen}, K.~P., {Greve}, T.~R., {Narayanan}, D., {et~al.} 2015, \apj, 814, 76

\bibitem[{{Ota} {et~al.}(2014){Ota}, {Walter}, {Ohta}, {Hatsukade}, {Carilli},
  {da Cunha}, {Gonz{\'a}lez-L{\'o}pez}, {Decarli}, {Hodge}, {Nagai}, {Egami},
  {Jiang}, {Iye}, {Kashikawa}, {Riechers}, {Bertoldi}, {Cox}, {Neri}, \&
  {Weiss}}]{ota14}
{Ota}, K., {Walter}, F., {Ohta}, K., {et~al.} 2014, \apj, 792, 34

\bibitem[{{Pentericci} {et~al.}(2014){Pentericci}, {Vanzella}, {Fontana},
  {Castellano}, {Treu}, {Mesinger}, {Dijkstra}, {Grazian}, {Brada{\v c}},
  {Conselice}, {Cristiani}, {Dunlop}, {Galametz}, {Giavalisco}, {Giallongo},
  {Koekemoer}, {McLure}, {Maiolino}, {Paris}, \& {Santini}}]{pentericci14}
{Pentericci}, L., {Vanzella}, E., {Fontana}, A., {et~al.} 2014, \apj, 793, 113

\bibitem[{{Popping} {et~al.}(2016){Popping}, {van Kampen}, {Decarli}, {Spaans},
  {Somerville}, \& {Trager}}]{popping16}
{Popping}, G., {van Kampen}, E., {Decarli}, R., {et~al.} 2016, ArXiv e-prints,
  arXiv:1602.02761

\bibitem[{{Riechers} {et~al.}(2013){Riechers}, {Bradford}, {Clements},
  {Dowell}, {P{\'e}rez-Fournon}, {Ivison}, {Bridge}, {Conley}, {Fu}, {Vieira},
  {Wardlow}, {Calanog}, {Cooray}, {Hurley}, {Neri}, {Kamenetzky}, {Aguirre},
  {Altieri}, {Arumugam}, {Benford}, {B{\'e}thermin}, {Bock}, {Burgarella},
  {Cabrera-Lavers}, {Chapman}, {Cox}, {Dunlop}, {Earle}, {Farrah}, {Ferrero},
  {Franceschini}, {Gavazzi}, {Glenn}, {Solares}, {Gurwell}, {Halpern},
  {Hatziminaoglou}, {Hyde}, {Ibar}, {Kov{\'a}cs}, {Krips}, {Lupu}, {Maloney},
  {Martinez-Navajas}, {Matsuhara}, {Murphy}, {Naylor}, {Nguyen}, {Oliver},
  {Omont}, {Page}, {Petitpas}, {Rangwala}, {Roseboom}, {Scott}, {Smith},
  {Staguhn}, {Streblyanska}, {Thomson}, {Valtchanov}, {Viero}, {Wang},
  {Zemcov}, \& {Zmuidzinas}}]{riechers13}
{Riechers}, D.~A., {Bradford}, C.~M., {Clements}, D.~L., {et~al.} 2013, \nat,
  496, 329

\bibitem[{{Robertson} {et~al.}(2010){Robertson}, {Ellis}, {Dunlop}, {McLure},
  \& {Stark}}]{robertson10}
{Robertson}, B.~E., {Ellis}, R.~S., {Dunlop}, J.~S., {McLure}, R.~J., \&
  {Stark}, D.~P. 2010, \nat, 468, 49

\bibitem[{{Sanders} {et~al.}(2003){Sanders}, {Mazzarella}, {Kim}, {Surace}, \&
  {Soifer}}]{sanders03}
{Sanders}, D.~B., {Mazzarella}, J.~M., {Kim}, D.-C., {Surace}, J.~A., \&
  {Soifer}, B.~T. 2003, \aj, 126, 1607

\bibitem[{{Sargsyan} {et~al.}(2012){Sargsyan}, {Lebouteiller}, {Weedman},
  {Spoon}, {Bernard-Salas}, {Engels}, {Stacey}, {Houck}, {Barry}, {Miles}, \&
  {Samsonyan}}]{sargsyan12}
{Sargsyan}, L., {Lebouteiller}, V., {Weedman}, D., {et~al.} 2012, \apj, 755,
  171

\bibitem[{{Schenker} {et~al.}(2012){Schenker}, {Stark}, {Ellis}, {Robertson},
  {Dunlop}, {McLure}, {Kneib}, \& {Richard}}]{schenker12}
{Schenker}, M.~A., {Stark}, D.~P., {Ellis}, R.~S., {et~al.} 2012, \apj, 744,
  179

\bibitem[{{Schenker} {et~al.}(2013){Schenker}, {Robertson}, {Ellis}, {Ono},
  {McLure}, {Dunlop}, {Koekemoer}, {Bowler}, {Ouchi}, {Curtis-Lake}, {Rogers},
  {Schneider}, {Charlot}, {Stark}, {Furlanetto}, \& {Cirasuolo}}]{schenker13}
{Schenker}, M.~A., {Robertson}, B.~E., {Ellis}, R.~S., {et~al.} 2013, \apj,
  768, 196

\bibitem[{{Stacey} {et~al.}(1991){Stacey}, {Geis}, {Genzel}, {Lugten},
  {Poglitsch}, {Sternberg}, \& {Townes}}]{stacey91}
{Stacey}, G.~J., {Geis}, N., {Genzel}, R., {et~al.} 1991, \apj, 373, 423

\bibitem[{{Swinbank} {et~al.}(2014){Swinbank}, {Simpson}, {Smail}, {Harrison},
  {Hodge}, {Karim}, {Walter}, {Alexander}, {Brandt}, {de Breuck}, {da Cunha},
  {Chapman}, {Coppin}, {Danielson}, {Dannerbauer}, {Decarli}, {Greve},
  {Ivison}, {Knudsen}, {Lagos}, {Schinnerer}, {Thomson}, {Wardlow}, {Wei{\ss}},
  \& {van der Werf}}]{swinbank14}
{Swinbank}, A.~M., {Simpson}, J.~M., {Smail}, I., {et~al.} 2014, \mnras, 438,
  1267

\bibitem[{{Treu} {et~al.}(2013){Treu}, {Schmidt}, {Trenti}, {Bradley}, \&
  {Stiavelli}}]{treu13}
{Treu}, T., {Schmidt}, K.~B., {Trenti}, M., {Bradley}, L.~D., \& {Stiavelli},
  M. 2013, \apjl, 775, L29

\bibitem[{{Vallini} {et~al.}(2015){Vallini}, {Gallerani}, {Ferrara},
  {Pallottini}, \& {Yue}}]{vallini15}
{Vallini}, L., {Gallerani}, S., {Ferrara}, A., {Pallottini}, A., \& {Yue}, B.
  2015, \apj, 813, 36

\bibitem[{{Venemans} {et~al.}(2016){Venemans}, {Walter}, {Zschaechner},
  {Decarli}, {De Rosa}, {Findlay}, {McMahon}, \& {Sutherland}}]{venemans16}
{Venemans}, B.~P., {Walter}, F., {Zschaechner}, L., {et~al.} 2016, \apj, 816,
  37

\bibitem[{{Venemans} {et~al.}(2012){Venemans}, {McMahon}, {Walter}, {Decarli},
  {Cox}, {Neri}, {Hewett}, {Mortlock}, {Simpson}, \& {Warren}}]{venemans12}
{Venemans}, B.~P., {McMahon}, R.~G., {Walter}, F., {et~al.} 2012, \apjl, 751,
  L25

\bibitem[{{Walter} {et~al.}(2009){Walter}, {Riechers}, {Cox}, {Neri},
  {Carilli}, {Bertoldi}, {Weiss}, \& {Maiolino}}]{walter09}
{Walter}, F., {Riechers}, D., {Cox}, P., {et~al.} 2009, \nat, 457, 699

\bibitem[{{Walter} {et~al.}(2012){Walter}, {Decarli}, {Carilli}, {Bertoldi},
  {Cox}, {da Cunha}, {Daddi}, {Dickinson}, {Downes}, {Elbaz}, {Ellis}, {Hodge},
  {Neri}, {Riechers}, {Weiss}, {Bell}, {Dannerbauer}, {Krips}, {Krumholz},
  {Lentati}, {Maiolino}, {Menten}, {Rix}, {Robertson}, {Spinrad}, {Stark}, \&
  {Stern}}]{walter12}
{Walter}, F., {Decarli}, R., {Carilli}, C., {et~al.} 2012, \nat, 486, 233

\bibitem[{{Walter} {et~al.}(2016){Walter}, {Decarli}, {Aravena}, {Carilli},
  {Bouwens}, {da Cunha}, {Daddi}, {Ivison}, {Riechers}, {Smail}, {Swinbank},
  {Weiss}, {Anguita}, {Assef}, {Bacon}, {Bauer}, {Bell}, {Bertoldi}, {Chapman},
  {Colina}, {Cortes}, {Cox}, {Dickinson}, {Elbaz}, {G{\'o}nzalez-L{\'o}pez},
  {Ibar}, {Inami}, {Infante}, {Hodge}, {Karim}, {Le Fevre}, {Magnelli}, {Neri},
  {Oesch}, {Ota}, {Popping}, {Rix}, {Sargent}, {Sheth}, {van der Wel}, {van der
  Werf}, \& {Wagg}}]{walter16}
{Walter}, F., {Decarli}, R., {Aravena}, M., {et~al.} 2016, ArXiv e-prints,
  arXiv:1607.06768

\bibitem[{{Wang} {et~al.}(2013){Wang}, {Wagg}, {Carilli}, {Walter}, {Lentati},
  {Fan}, {Riechers}, {Bertoldi}, {Narayanan}, {Strauss}, {Cox}, {Omont},
  {Menten}, {Knudsen}, {Neri}, \& {Jiang}}]{wang13}
{Wang}, R., {Wagg}, J., {Carilli}, C.~L., {et~al.} 2013, \apj, 773, 44

\bibitem[{{Williams} {et~al.}(2011){Williams}, {de Geus}, \&
  {Blitz}}]{williams11}
{Williams}, J.~P., {de Geus}, E.~J., \& {Blitz}, L. 2011, {Clumpfind:
  Determining Structure in Molecular Clouds}, Astrophysics Source Code Library,
  ascl:1107.014

\bibitem[{{Willott} {et~al.}(2015){Willott}, {Carilli}, {Wagg}, \&
  {Wang}}]{willott15}
{Willott}, C.~J., {Carilli}, C.~L., {Wagg}, J., \& {Wang}, R. 2015, \apj, 807,
  180

\end{thebibliography}

\acknowledgements

We thank the anonymous referee for her/his positive feedback and useful comments. M.A.~acknowledges partial support from FONDECYT through grant 1140099. FW, IRS, and RJI acknowledge support through ERC grants COSMIC--DAWN, DUSTYGAL, and COSMICISM, respectively. FEB and LI acknowledge Conicyt grants Basal-CATA PFB--06/2007 and Anilo ACT1417. FEB also acknowledge support from FONDECYT Regular 1141218 (FEB), and the Ministry of Economy, Development, and Tourism's Millennium Science Initiative through grant IC120009, awarded to The Millennium Institute of Astrophysics, MAS. EdC gratefully acknowledges the Australian Research Council as the recipient of a Future Fellowship (project FT150100079). DR acknowledges support from the National Science Foundation under grant number AST-1614213 to Cornell University. IRS also acknowledges support from STFC (ST/L00075X/1) and a Royal Society / Wolfson Merit award. Support for RD and BM was provided by the DFG priority program 1573 `The physics of the interstellar medium'.  AK and FB acknowledge support by the Collaborative Research Council 956, sub-project A1, funded by the Deutsche Forschungsgemeinschaft (DFG). PI acknowledges Conicyt grants Basal-CATA PFB--06/2007 and Anillo ACT1417. RJA was supported by FONDECYT grant number 1151408. This paper makes use of the following ALMA data: ADS/JAO.ALMA\#2013.1.00146.S and ADS/JAO.ALMA\#2013.1.00718.S. ALMA is a partnership of ESO (representing its member states), NSF (USA) and NINS (Japan), together with NRC (Canada), NSC and ASIAA (Taiwan), and KASI (Republic of Korea), in cooperation with the Republic of Chile. The Joint ALMA Observatory is operated by ESO, AUI/NRAO and NAOJ. The 3mm-part of ALMA project had been supported by the German ARC.

\clearpage

\appendix

\section{Assessment of redshift possibilities based on the 3-mm scan}

\begin{figure*}[ht]
\centering
\includegraphics[scale=0.65]{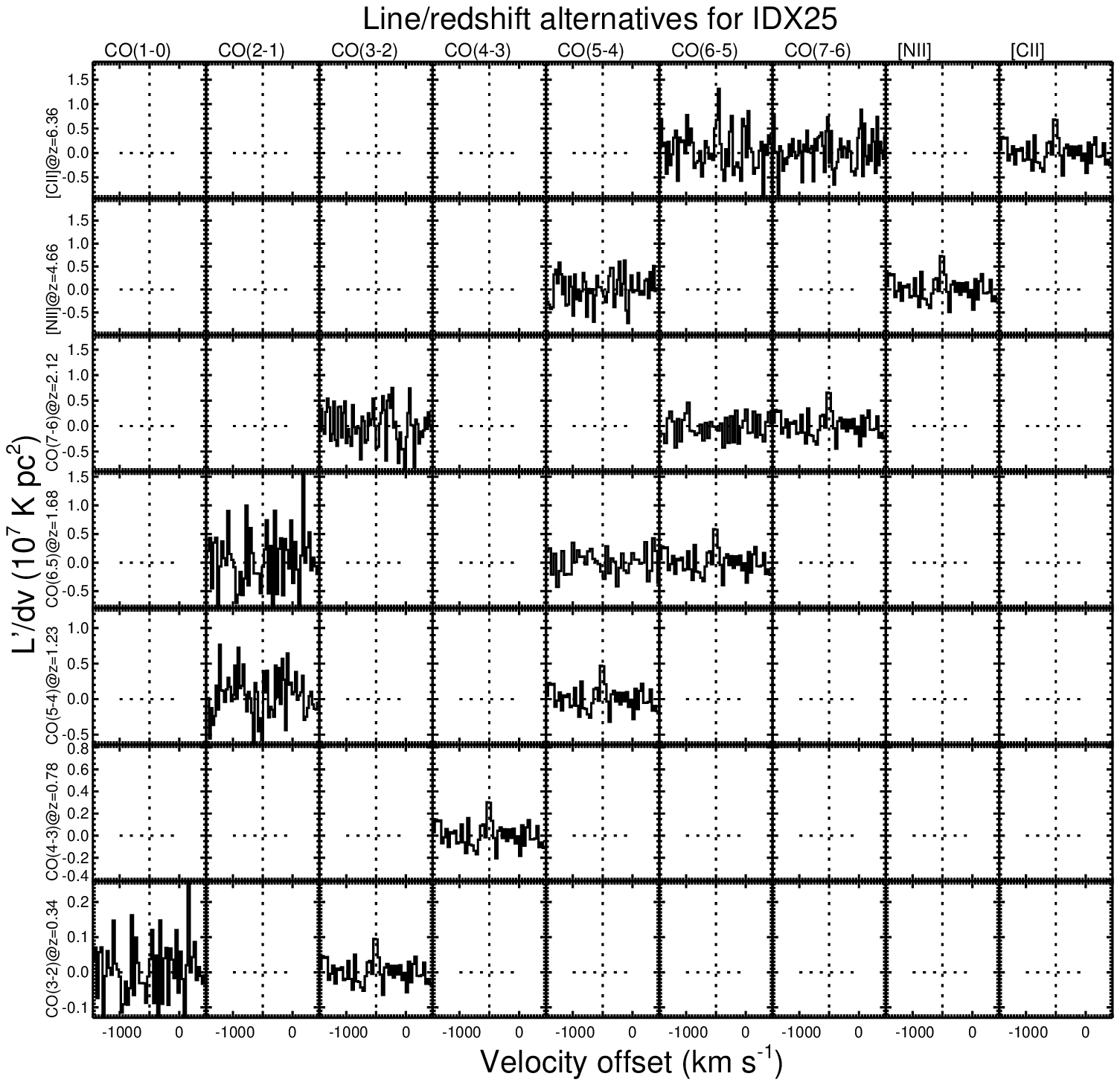}
\includegraphics[scale=0.65]{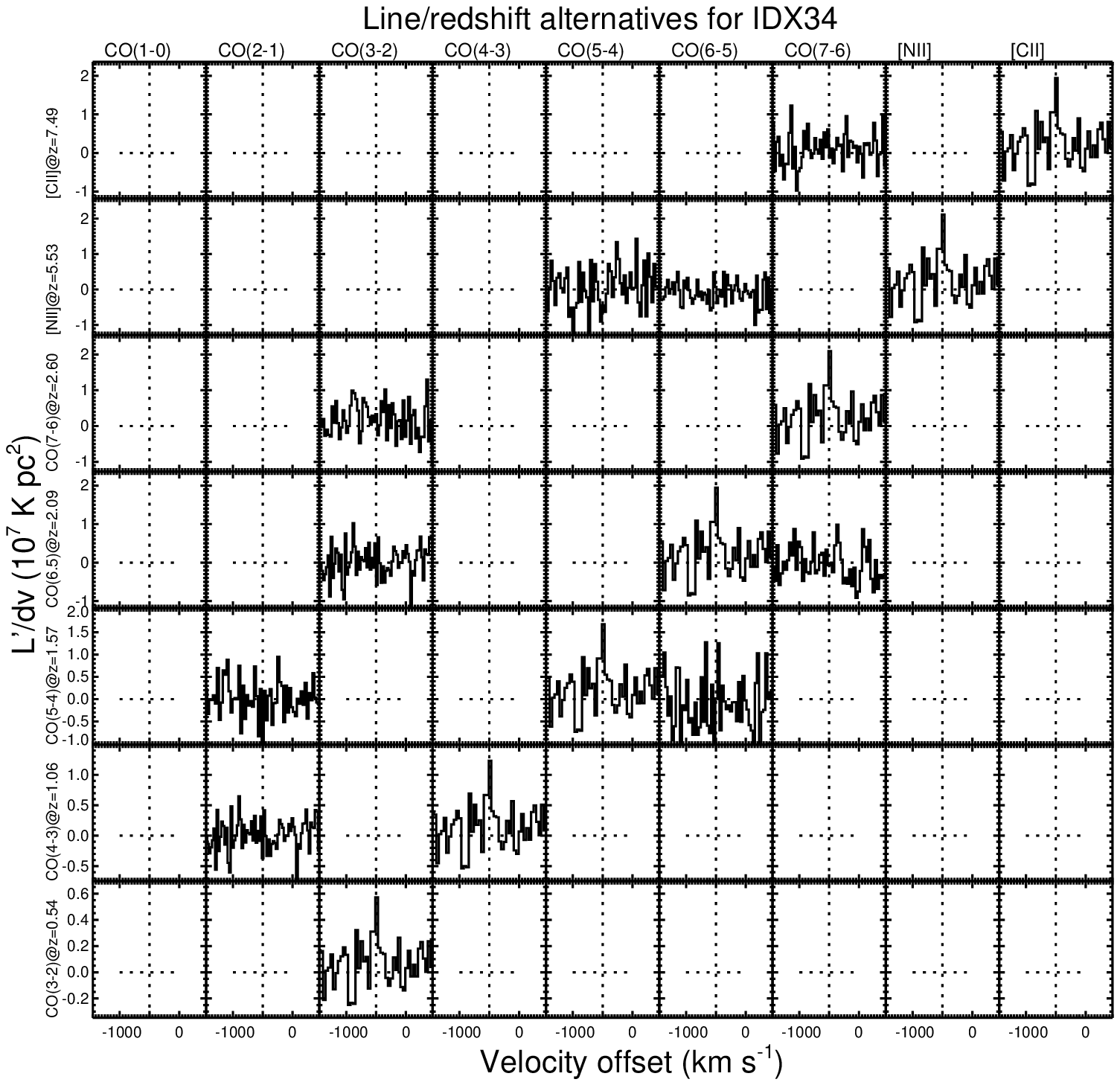}
\caption{Line identification for the {\it blind} [CII] line candidates with no optical association, IDX25 and IDX34. Each row represents the identification of an emission line and redshift. For each row, the different cells show the spectra around other lines covered by our 3-mm and 1-mm scans at the identified redshift. As expected for distant $z>6$ galaxies, the only line identified corresponds to [CII], and other lines as CO(6-5), CO(7-6) or CO(8-7) are too faint to be detected by our 3-mm scan.\label{fig:zopt1}}
\end{figure*}

\begin{figure*}[ht]
\centering
\includegraphics[scale=0.65]{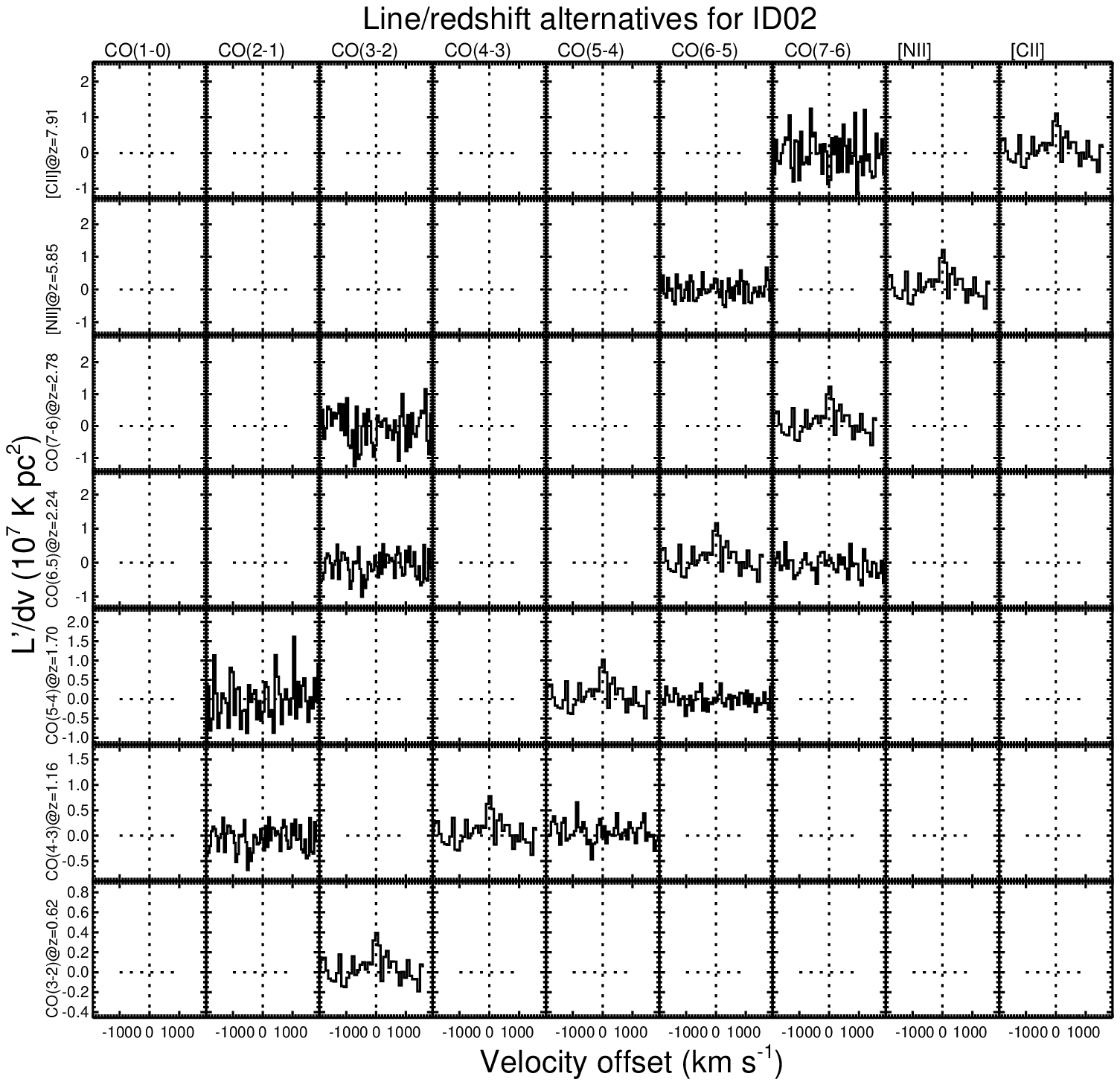}
\includegraphics[scale=0.65]{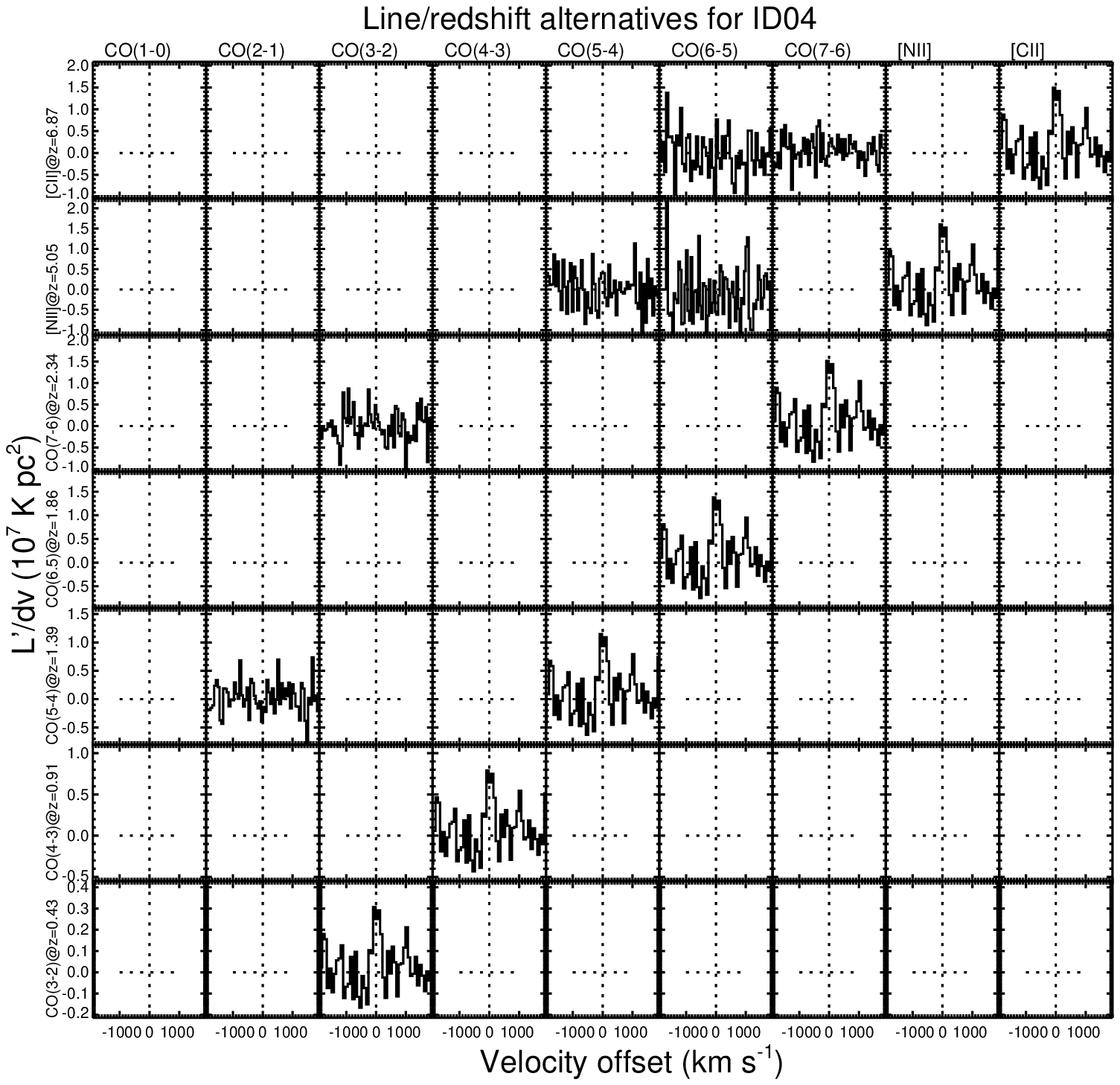}
\caption{Line identification for the optical dropout galaxies associated with a 1-mm line candidate, ID02 and ID04. Each row represents the identification of an emission line and redshift. For each row, the different cells show the spectra around other lines covered by our 3-mm and 1-mm scans at the identified redshift. As expected for distant $z>6$ galaxies, the only line identified corresponds to [CII], and other lines as CO(6-5), CO(7-6) or CO(8-7) are too faint to be detected by our 3-mm scan.\label{fig:zopt2}}
\end{figure*}

\begin{figure*}[ht]
\centering
\includegraphics[scale=0.65]{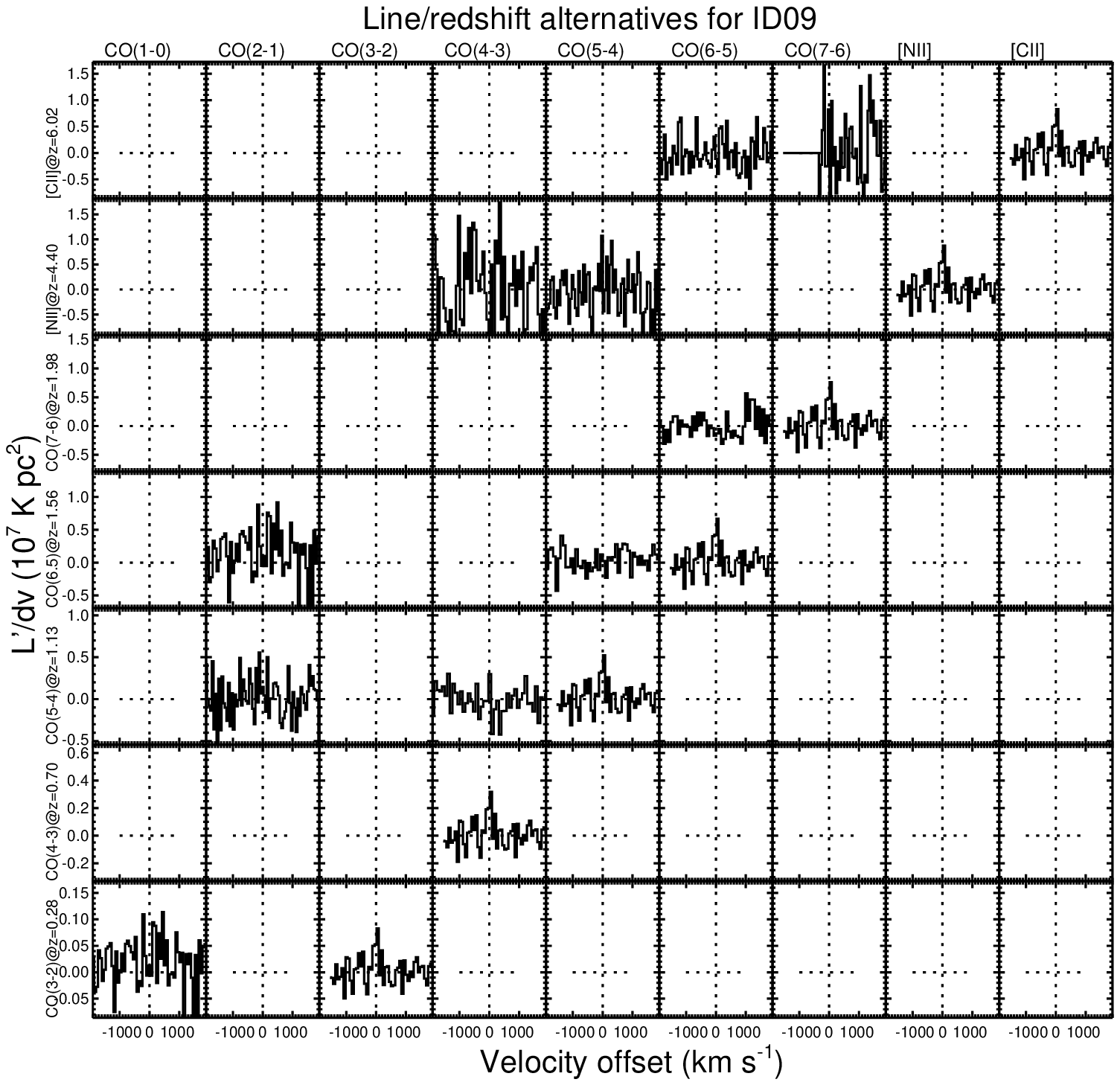}
\includegraphics[scale=0.65]{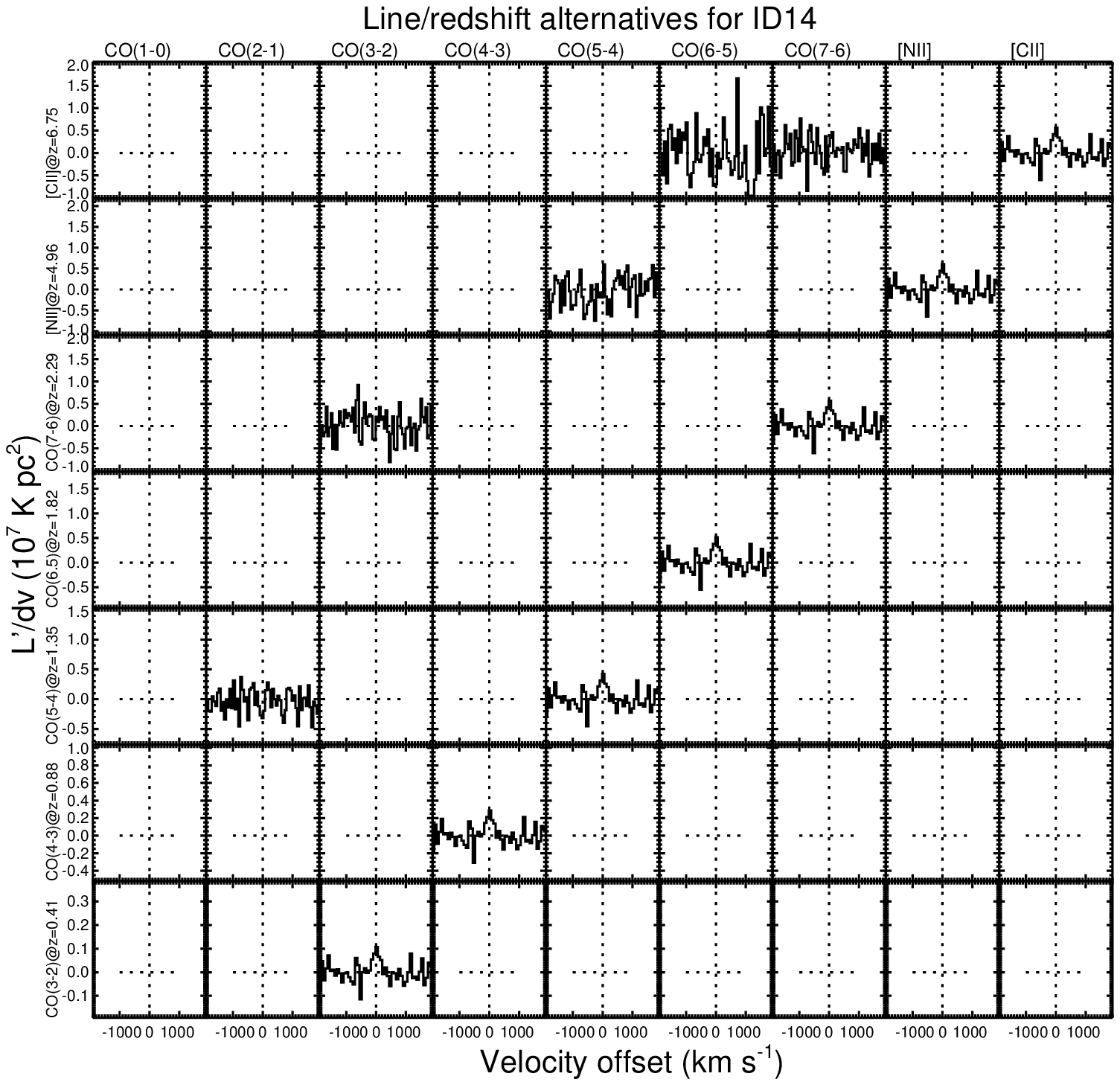}
\caption{Line identification for the optical dropout galaxies associated with a 1-mm line candidate, ID09 and ID14. Each row represents the identification of an emission line and redshift. For each row, the different cells show the spectra around other lines covered by our 3-mm and 1-mm scans at the identified redshift. As expected for distant $z>6$ galaxies, the only line identified corresponds to [CII], and other lines as CO(6-5), CO(7-6) or CO(8-7) are too faint to be detected by our 3-mm scan.\label{fig:zopt3}}
\end{figure*}

\begin{figure*}[ht]
\centering
\includegraphics[scale=0.65]{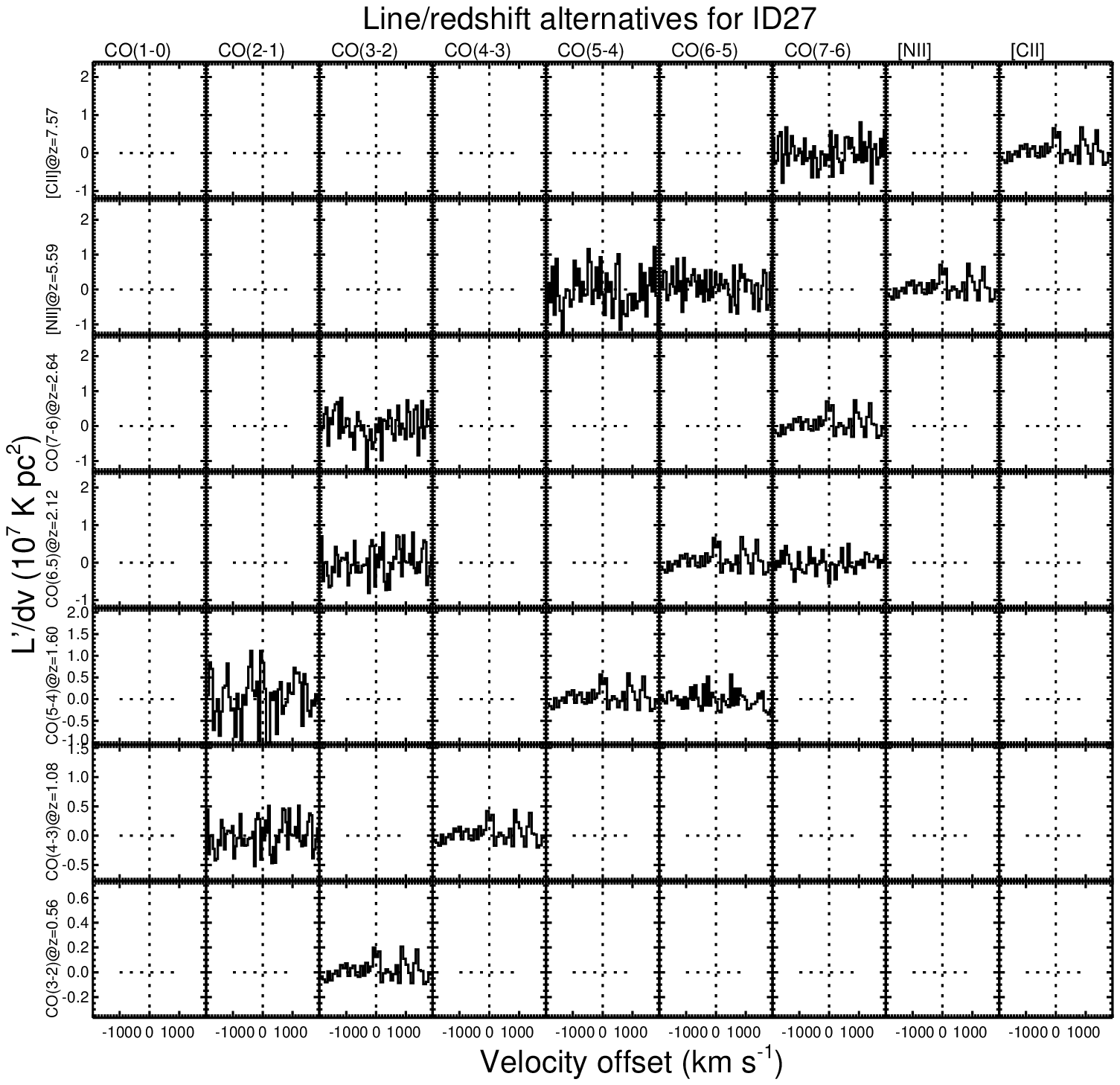}
\includegraphics[scale=0.65]{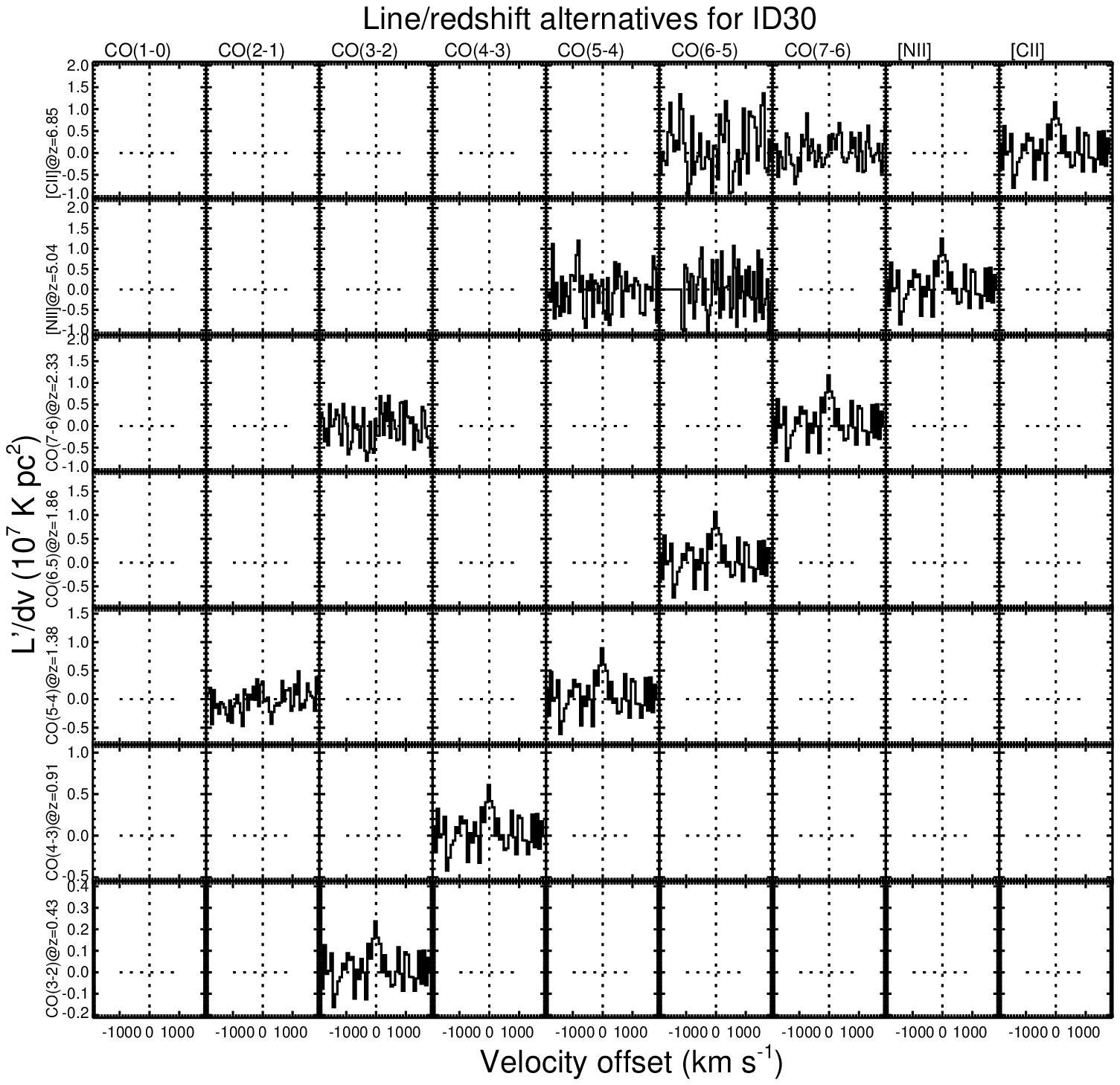}
\caption{Line identification for the optical dropout galaxies associated with a 1-mm line candidate, ID27 and ID30. Each row represents the identification of an emission line and redshift. For each row, the different cells show the spectra around other lines covered by our 3-mm and 1-mm scans at the identified redshift. As expected for distant $z>6$ galaxies, the only line identified corresponds to [CII], and other lines as CO(6-5), CO(7-6) or CO(8-7) are too faint to be detected by our 3-mm scan.\label{fig:zopt4}}
\end{figure*}

\begin{figure*}[ht]
\centering

\includegraphics[scale=0.65]{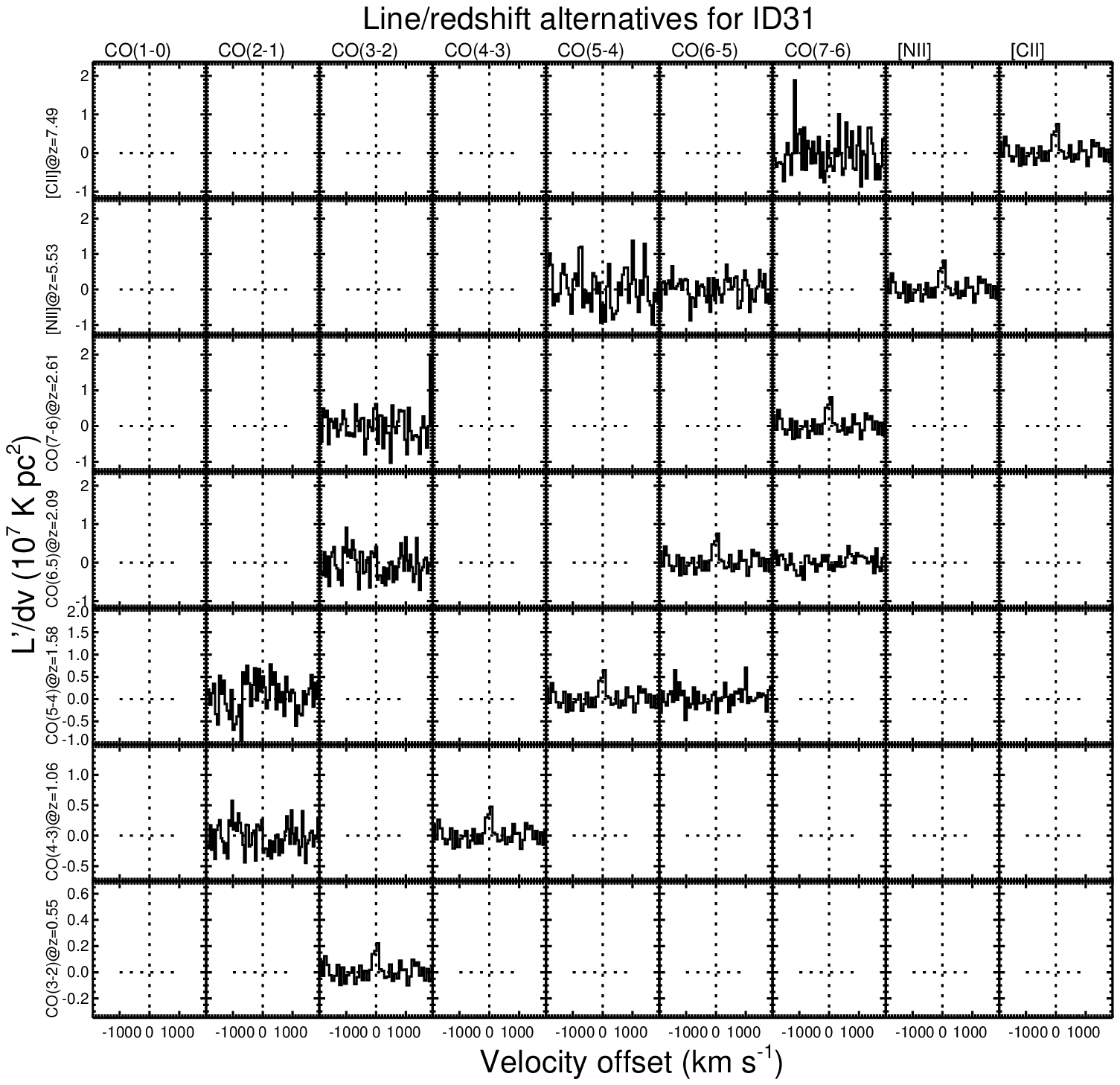}
\includegraphics[scale=0.65]{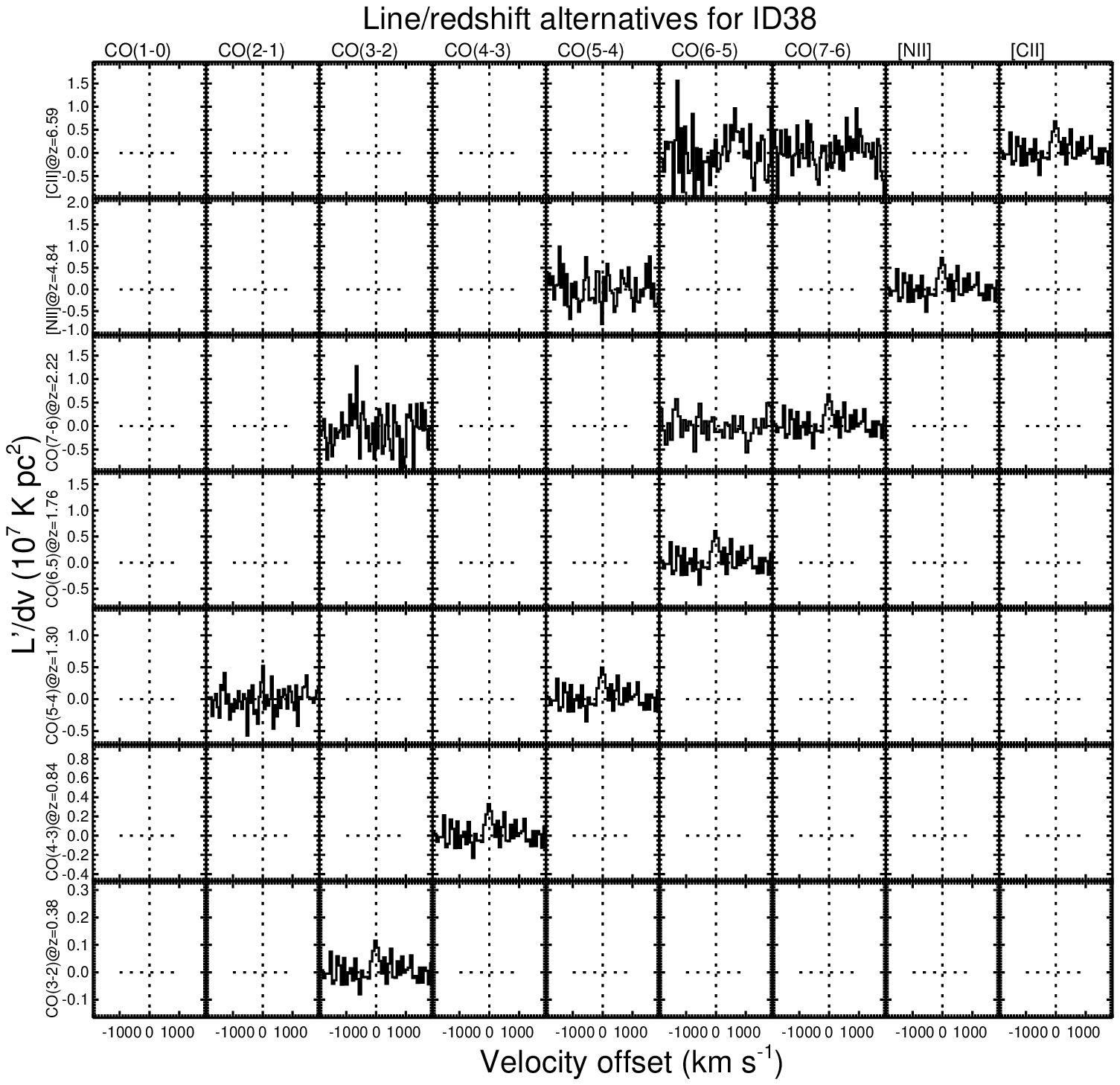}
\caption{Line identification for the optical dropout galaxies associated with a 1-mm line candidate, ID31 and ID38. Each row represents the identification of an emission line and redshift. For each row, the different cells show the spectra around other lines covered by our 3-mm and 1-mm scans at the identified redshift. As expected for distant $z>6$ galaxies, the only line identified corresponds to [CII], and other lines as CO(6-5), CO(7-6) or CO(8-7) are too faint to be detected by our 3-mm scan.\label{fig:zopt5}}
\end{figure*}

\begin{figure*}[ht]
\centering

\includegraphics[scale=0.65]{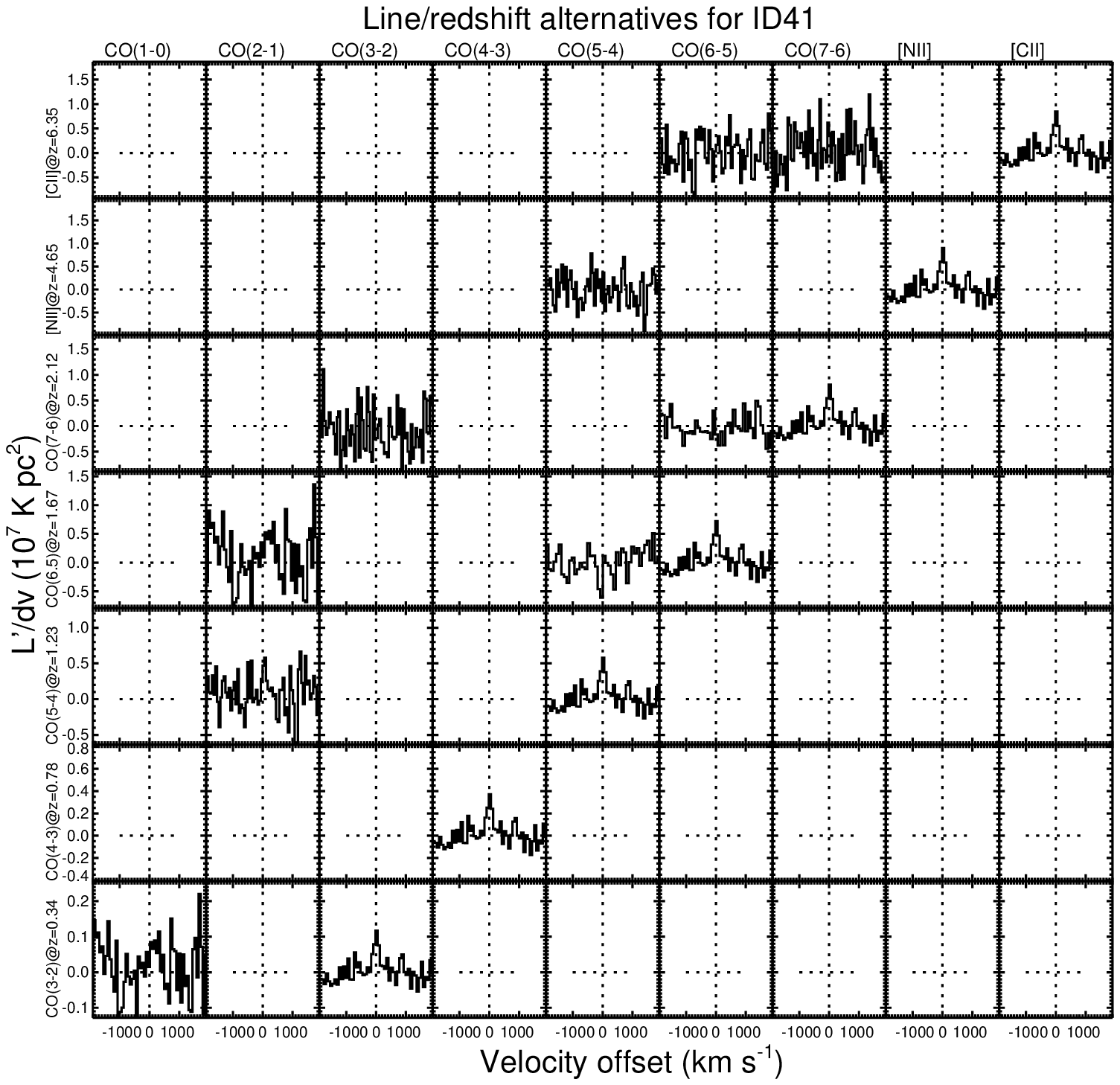}
\includegraphics[scale=0.65]{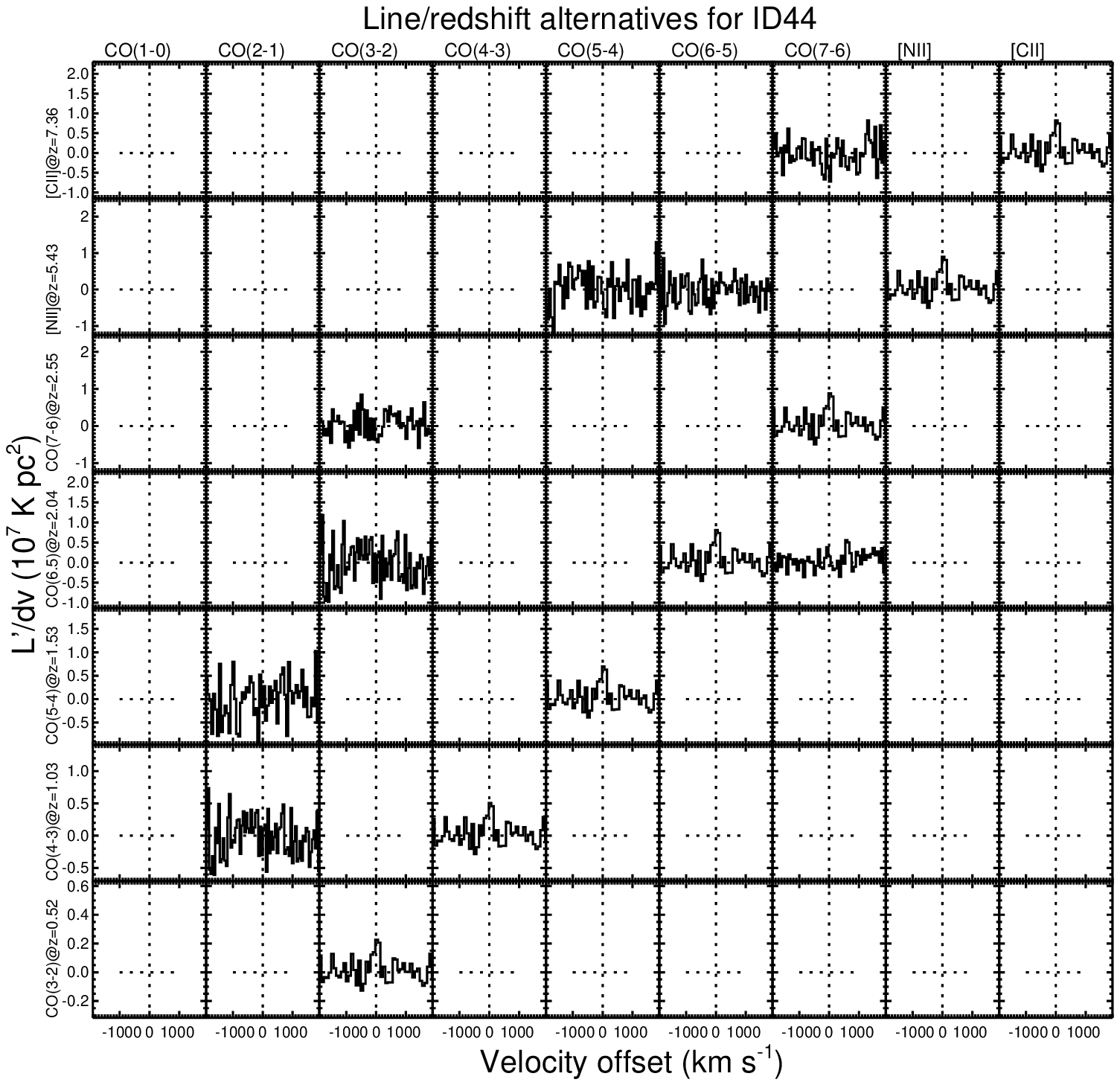}
\caption{Line identification for the optical dropout galaxies associated with a 1-mm line candidate, ID41 and ID44. Each row represents the identification of an emission line and redshift. For each row, the different cells show the spectra around other lines covered by our 3-mm and 1-mm scans at the identified redshift. As expected for distant $z>6$ galaxies, the only line identified corresponds to [CII], and other lines as CO(6-5), CO(7-6) or CO(8-7) are too faint to be detected by our 3-mm scan.\label{fig:zopt6}}
\end{figure*}

\begin{figure*}[ht]
\centering

\includegraphics[scale=0.65]{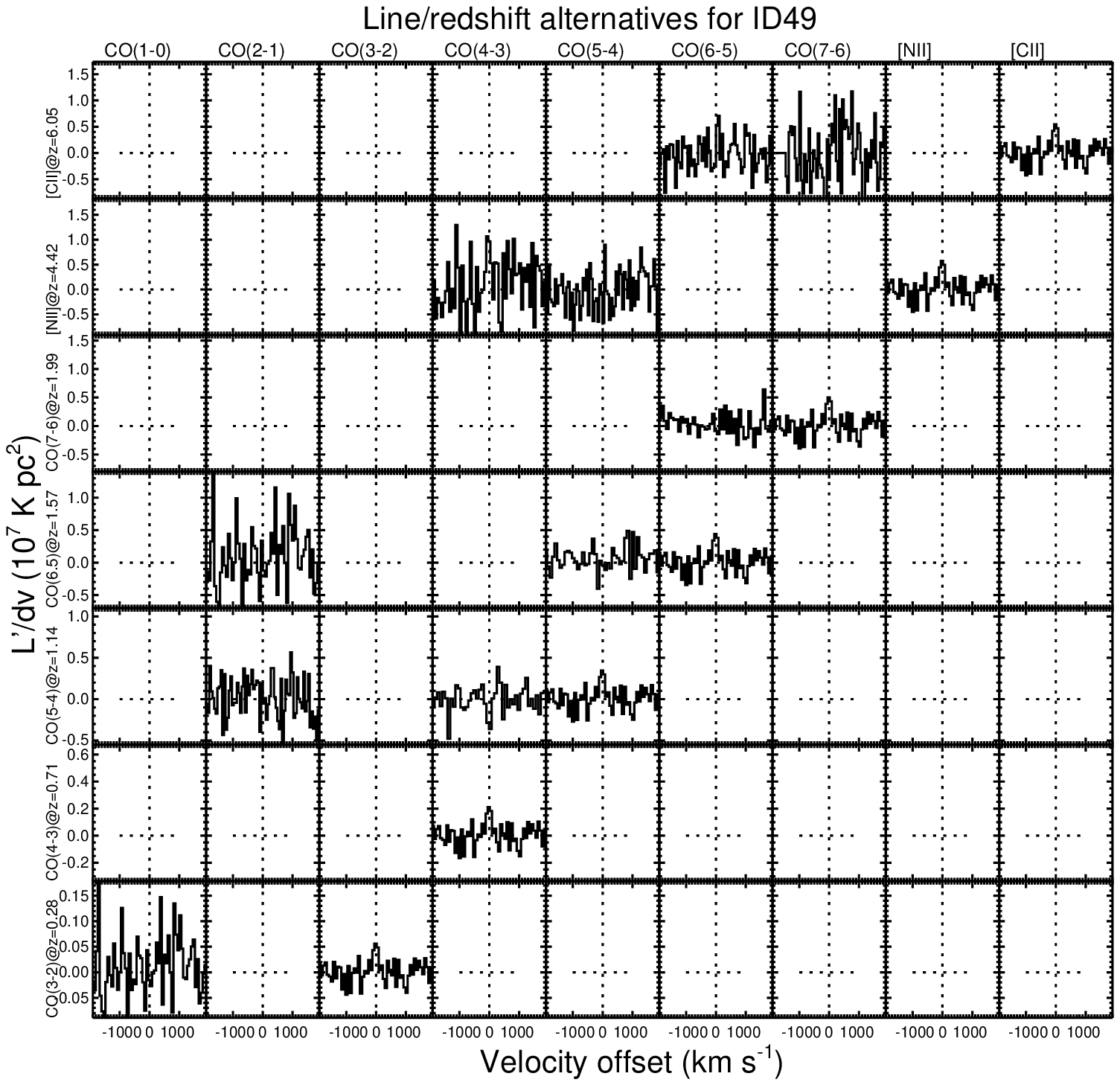}
\caption{Line identification for the optical dropout galaxies associated with a 1-mm line candidate ID49. Each row represents the identification of an emission line and redshift. For each row, the different cells show the spectra around other lines covered by our 3-mm and 1-mm scans at the identified redshift. As expected for distant $z>6$ galaxies, the only line identified corresponds to [CII], and other lines as CO(6-5), CO(7-6) or CO(8-7) are too faint to be detected by our 3-mm scan.\label{fig:zopt7}}
\end{figure*}

\label{lastpage}

\end{document}